\documentclass[twocolumn, appendixfloats, revtex4, numberedappendix, iop, letterpaper]{openjournal}
\usepackage[pass]{geometry}

\usepackage{natbib}

\usepackage[breaklinks,colorlinks,linkcolor=blue,citecolor=blue,urlcolor=blue]{hyperref}

\usepackage[T1]{fontenc}

\usepackage{orcidlink}

\usepackage{xspace}

\usepackage{graphicx}

\usepackage{paralist}

\newcommand{\HI}{\ifmmode \mathrm{H}\,\textsc{i} \else H~{\sc i}\fi\xspace}
\newcommand{\nhi}{\ifmmode \mathrm{N(H}\,\textsc{i}) \else N(H~{\sc i})\fi\xspace}
\newcommand{\mgii}{\ifmmode \mathrm{Mg}\,\textsc{ii} \else Mg~{\sc ii}\fi\xspace}
\newcommand{\civ}{\ifmmode \mathrm{C}\,\textsc{iv} \else C~{\sc iv}\fi\xspace}
\newcommand{\ciii}{\ifmmode \mathrm{C}\,\textsc{iii}] \else C~{\sc iii}]\fi\xspace}
\newcommand{\mgal}{\ifmmode M_* \else $M_*$\fi\xspace}
\newcommand{\modot}{\ifmmode M_\odot \else $M_\odot$\fi\xspace}

\shorttitle{Star formation histories from colour distributions}
\shortauthors{V. Wild et al.}

\begin{document} % OJA

\title{\vspace{-0.8cm}The quenching time and timescale distribution of \lowercase{$z\sim2$} quiescent galaxies from precise colour distribution analysis\vspace{-1.5cm}}

% Authors 
\author{Vivienne Wild\,\orcidlink{0000-0002-8956-7024}$^{1}$\footnote{vw8@st-andrews.ac.uk}, Ho-Hin Leung\,\orcidlink{0000-0003-0486-5178}$^{2}$, Adam Carnall\,\orcidlink{0000-0002-1482-5818}$^{2}$, 
Maya Skarbinski\,\orcidlink{0009-0004-0844-0657}$^{3}$
}
\affiliation{
% List of institutions
$^{1}$School of Physics \& Astronomy, University of St Andrews, North Haugh, St Andrews, KY16 9SS, U.K.\\
$^{2}$Institute for Astronomy, University of Edinburgh, Royal Observatory, Edinburgh, EH9 3HJ, U.K.\\
$^{3}$William H. Miller III Department of Physics and Astronomy, Johns Hopkins University, Baltimore, MD21218, U.S.A.\\
}

% Abstract
\begin{abstract}
Understanding when and how galaxies quench their star formation is crucial for understanding the dominant physical processes at play. The spectral energy distribution (SED) of galaxies encodes significant information on their past histories: the relative importance of different physical processes influences the observed distribution of SED shapes in the galaxy population. We use a simulation based inference (SBI) approach to directly constrain the distribution of formation times, quenching times and quenching timescales within the massive galaxy population at $z \gtrsim 2$ from their broad band photometric colour distribution at $1.7<z<2$.
We demonstrate that a simple distribution of double power-law star formation histories accurately fits the distribution of SED shapes of galaxies with $\log_{10}(M^*/M_\odot)>10.3$.  We measure a quenched galaxy fraction of $0.24\pm0.02$, with the number density of quenched galaxies rising rapidly 2.5\,Gyr after the Big Bang ($z\lesssim2.6$). Galaxies must quench rapidly to achieve the precise bimodal colour distribution: defining the quenching timescale as the time from peak star formation rate (${\rm SFR_{peak}}$) to $0.5\,{\rm SFR_{peak}}$, the quenching timescale distribution has a mode at $97_{-25}^{+31}$\,Myr, a median of $182\pm16$\,Myr and a tail to $\sim700$\,Myr. To achieve full quiescence takes a median time of $\sim400$\,Myr. Comparing to direct number density measurements of quenched galaxies at $z>2$ the combination of recent and rapid quenching inferred from the fossil record suggests a substantial rejuvenation and/or merger rate for quenched galaxies observed directly at $z>3.5$.  

\end{abstract}


\maketitle % OJA

\section{Introduction}\label{sec:intro}

The existence of two distinct populations of galaxies in UV/optical colour distributions leads to one of the most well used galaxy classification systems: red, quiescent and blue, star-forming. At $z<1$ galaxies can be separated into star-forming and quiescent systems based on a single rest-frame optical colour bi-modality \citep[e.g.][]{strateva2001,kauffmann2003,bell2004,baldry2004}, while the bi-modality in the galaxy population is still visible to $z\lesssim5$ when combining two colours \citep[e.g.][]{williams2009,Whitaker2011,leja2019b,Gould2023,vanderWel2025}.  The evolving galaxy stellar mass functions of star-forming and quiescent galaxies, now well-constrained to $z\sim3$ with measurements to $z\sim7$ and beyond improving rapidly, tell us the fraction of galaxies at every redshift and mass that have halted their star formation  \citep[become ``quenched'', e.g.][]{mcleod2021, Santini2021, weaver2023,Weibel2024,Shuntov2026}.

Increasingly wide and deep surveys allow increasingly accurate tracing of demographic trends, which show \emph{how} the local galaxy population was built up. However, it is not straightforward to infer \emph{why}. Aggregated statistics such as stellar mass functions, colour-mass relations and quenched fractions do not uniquely constrain different growth and quenching pathways \citep[e.g.][]{skelton2012,Naab2017,Fu2025}. Additionally, the existence of high-mass quenched systems at the highest observed redshifts challenges our current understanding of pathways to quiescent galaxy formation \citep[e.g.][]{carnall2024,Stevenson2026,Zhang2026a,ji2026}. 

To make further progress on understanding why galaxies quench their star formation, additional constraints must be found. These could involve e.g. the impact of starburst or Active Galactic Nucleus (AGN) phases on growth and quenching \citep[e.g.][]{Zheng2020,Davies2022,Bugiani2025}, the compactness of the morphologies or evidence for recent mergers in recently quenched galaxies \citep[e.g.]{almaini2017,setton2022,Nersesian2026,Ni2025,maltby2026}, the resulting chemical enrichment \citep{leung2024,Gountanis2025}, the level of residual gas and star formation following a quenching process \citep[e.g.][]{rowlands2015,Bezanson2022,lin2026} or the timescales over which growth and quenching happens (see below for references).  Generally, these additional constraints require additional, expensive observations, such as high quality rest-frame optical spectra or deep sub-mm measurements. 

In theory, the timescales over which star-formation quenching occurs should directly relate to the cause of the quenching \citep[see][for a recent comprehensive summary figure]{Iyer2020}. Detailed observational studies of individual objects demonstrate the potential power of measuring timescales for understanding why galaxies stop forming stars \citep{Zheng2020,Khoram2026,Leung2026b}. Ejective AGN feedback, supernova driven winds and galactic mergers are all interrelated physical processes implicated by recent cosmological hydrodynamic simulations as causing rapid ($\lesssim1$\,Gyr) quenching in high mass ($\gtrsim10^{10.3}M_\odot$) galaxies \citep[e.g.][]{trayford2016, nelson2018, Wright2019,Ni2025,Iyer2025,Lagos2025}. In the simplest picture, galactic interactions and mergers are expected to funnel gas into the centre of the galaxies, which is used up rapidly ($\sim100$\,Myr) in a burst of star formation, leading to enhanced feeding of the central super-massive black hole and associated AGN \citep[e.g.][]{DiMAtteo2005,hopkins2008}. Given sufficient coupling of energy from supernovae and black hole accretion to the cold gas, both may expel gas on timescales of 10’s of Myr\footnote{It is worth noting that such short quenching timescales are below the time sampling resolution of most hydrodynamic simulations if merger trees are used to estimate quenching timescales \citep{trayford2016,RodriguezMontero2019,Wright2019,Walters2022}.} or alternatively, may heat the circum-galactic medium (CGM) to starve galaxies on $\sim$Gyr timescales. However, in the past two decades it has become clear that this simple picture is much more complex than first imagined. The exact interplay between each of these processes and quenching are difficult to untangle even within simulations, in part due to their ``sub-grid'' nature \citep[e.g.][]{RodriguezMontero2019,Zheng2022,Davies2022,Lagos2025}.  Halo mass, black hole mass, local environment and galaxy morphology may all work together to provide the conditions needed to induce quenching, potentially resulting in wide ranges of quenching timescales \citep[e.g.][]{Davies2024,Lagos2025,Szpila2025,Kimmig2025,ChandroGomez2026}. 

Unfortunately, extracting star formation quenching timescales from observations of quenched galaxies is challenging, and is typically achieved with high quality spectroscopic observations, which limits sample sizes as well as accessible redshifts and galaxy masses. Although average quenching timescales can be obtained from stacks of either observations or resulting SFHs \citep[e.g.][]{Bravo2023,Pacifici2016,Iyer2025}, or models that track quenched fraction with redshift \citep[e.g.][]{Hahn2017}, these cannot accurately capture the full quenching timescale distribution because stacks and quenched fractions contain no information on the speed of quenching of individual objects, smoothing out any rapid quenching that happens at a wide range of lookback times.  

At higher redshift, observing quenched galaxies closer to their quenching epoch leverages a greater dynamic range in the spectral energy distributions of their stellar populations: A, F and G stars have distinctly different spectral shapes, while G, K and M stars are much more difficult to distinguish in integrated spectra. Typically, rest-frame optical spectra with sufficient signal to observe the stellar continuum are combined with multi-wavelength photometry and analysed with spectral synthesis models to measure star formation histories, including formation and quenching timescales \citep[e.g.][]{CarnallMcLure2019,belli2019,wild2020,Forrest2020,Tacchella2022, Slob2024, park2024, Nanayakkara2025, Hamadouche2026arXiv, Nersesian2026, Leung2026a, skarbinski2026}.  A range of quenching timescales are recovered for quiescent galaxies, depending on redshift and stellar mass of the sample, as well as methodology. At $z\sim1$, $\sim20-50$\% of massive galaxies quenched very rapidly, on timescales of 10's-100's of Myr \citep{wild2016,belli2019,wild2020,Tacchella2022}. By $z\sim2-3$ the fraction of quenched galaxies that are found to have quenched very rapidly increases to $\gtrsim60$\%  \citep{Forrest2020,park2024,skarbinski2026}. 

As well as understanding the physical mechanisms of star-formation quenching, the distribution of quenching timescales is crucial for understanding the implications of early quiescent galaxy formation. Measurement of the full distribution of quenching times and timescales at Cosmic Noon ($1\lesssim z\lesssim 2$) would allow us to predict the number density of quiescent galaxies in the early Universe probed by the James Webb Space Telescope (JWST), where data and models show some tension \citep[e.g.][]{Schreiber2018,Carnall2023Abundance, Gould2023,baker2025,Stevenson2026}. If quenching timescales of $z\sim2$ quenched galaxies are found to have been rapid and recent, this would imply observed high redshift quenched galaxies are not in their final evolutionary stage and will rejuvenate. Alternatively, the fossil record star formation histories measured at a later epoch may agree with those from direct measurements at earlier epochs; currently small spectroscopic sample sizes make this analysis challenging \citep{park2024,Zhang2026b}. 

In this paper we develop a new approach to measuring the distribution of quenching times and timescales in the galaxy population at $z\sim2$. Rather than fitting a small number of individual galaxies with high quality spectra, we make use of the constraints set by the distribution of spectral energy distribution (SED) shapes in the population. Although each galaxy in itself provides less information than would be recoverable from a high quality spectrum, we can include a very large number of galaxies from photometric catalogues. The use of colour distributions to constrain quenching has been tried before \citep{Lian2016, Bravo2023}, providing interesting constraints on average quenching rates from $z\sim0$ observations. Population level modelling is becoming increasingly popular in extragalactic astronomy, as it allows errors to be propagated correctly into scaling relations and degeneracies to be broken. Hierarchical Bayesian methods are one such approach to modelling populations, and have been applied to e.g. integral field datacubes to understand the detailed propagation of star formation and quenching in recent galaxy mergers \citep{Ditrani2024,Leung2026b}, or scaling relations such as the SFR-M* relation to accurately propagate errors and robustly measure the scatter \citep{sandles2022}. However, for spectral fitting in particular, sampling the full likelihood surface is slow for large numbers of data points, models with large numbers of free parameters and complex multi-modal likelihood surfaces. A faster approach can be Simulation Based Inference (SBI) which uses neural networks to map the highly non-linear relationship between population model parameters and data summary statistics via large stochastically sampled model training sets.  This approach has recently been used to fit photometric galaxy catalogues to recover e.g. redshift distributions, evolving stellar mass functions, scaling relations and average star formation histories \citep{alsing2024, Li2024, aufort2025,Deger2026,HarveyLovell2026}.

In Section \ref{sec:data} we present the UKIDSS Ultra-deep survey data that we aim to fit. In Section \ref{sec:methods} we develop and test the SBI method on mock datasets. In Section \ref{sec:results} we present the results of fitting the model to massive galaxies at $1.7<z<2$, and in Section \ref{sec:discussion} we compare our population analysis results to traditional methods to recover quenched fractions and quenching timescales, and discuss the implications of our results for understanding why some galaxies stop forming stars. 

Throughout the paper we assume a flat $\Lambda$CDM cosmology with $\Omega_M=0.3$, $\Omega_\Lambda=0.7$ and H$_0=70$\,km/s/Mpc. Magnitudes are on the AB system and we assume a \citet{Chabrier2003} initial mass function (IMF). Specific star formation rates (sSFR) are calculated from the SFR averaged over the last 100\,Myr since the time of observation, and the current stellar mass. Metallicities are scaled to the default \textsc{bagpipes} value of $Z_\odot=0.02$.

\section{Data}\label{sec:data}

We make use of ground-based imaging data from the the UKIRT Infrared Deep Sky Survey \citep[UKIDSS,][]{lawrence2007} Ultra Deep Survey (UDS) Data Release 11 (DR11)\footnote{\url{https://www.nottingham.ac.uk/astronomy/UDS/index.html}}. Despite the recent arrival of new deep imaging from the James Webb Space Telescope, the UDS remains the largest area imaging dataset with optical and near-infrared (NIR) coverage to a depth and calibration quality required to characterise significant numbers of massive quenched galaxies at cosmic noon. A new analysis of the survey masks for this work gives a survey area of 0.60 square degrees, after masking bright stars and cross talk and combining with optical and mid-IR imaging. UKIRT observations provide 2\arcsec\ diameter aperture $J$, $H$ and $K$ observations to $5\sigma$ limiting depths of 25.4, 24.8 and 25.3 AB magnitudes respectively. Optical observations come from the Subaru XMM-Newton Deep Survey \citep[SXDS, ][]{Furusawa2008}, to depths of 27.2, 27.0, 27.0 and 26.0 in $V$, $R$, $i'$, $z'$ ($5\sigma$, 2\arcsec). $Y$-band coverage with a depth of 23.9 comes from the VISTA-Video survey (P.I. M.~Jarvis). Mid-IR coverage (IRAC 3.6$\mu m$ and 4.5$\mu m$) is provided by the Spitzer UDS Legacy Program (SpUDS, PI. J.~Dunlop) to a depth of 24.2. Photometry was extracted within 2\arcsec\ diameter apertures at the position of the $K$-band sources, with an aperture correction applied for the IRAC 3.6$\mu m$\ and 4.5$\mu m$  images. Details of the photometric redshift estimation are provided in \citet{wilkinson2021}. We use spectroscopic redshifts where available as described in \citet{wild2020}. 

We use an optimal linear combination of weighted observed fluxes, termed super-colours, to parametrise the shapes of the spectral energy distributions (SEDs) of the galaxies, following the method presented in \citet{wild2014}. The application of the method to the UDS DR11 data release is described in \citet{wild2020} and \citet{wilkinson2021} and the catalogue is available as part of the UDS DR11 data release.  Although similar in spirit to traditional colour-colour diagrams, the relative weighting of different bands is optimised to maximise the variance in the dataset using a Principal Component Analysis (PCA) of model SEDs, and the observed data is not forced to fit model SEDs unlike when determining the rest-frame fluxes required for e.g. $UVJ$ diagrams \citep[e.g.][]{williams2009}. The method includes all photometric filters with rest-frame wavelengths $2500$\AA$-1.5\mu m$, with the precise set of filters used dependant on the galaxy redshift. The first and second principal component amplitudes (termed SC1 and SC2) describe the overall red/blue colour of the SED and the strength of the Balmer or 4000\AA\ break respectively. Errors on the super-colours account for the correlated nature of the PCA eigensystem due to the sparse sampling of the observed dataset (i.e. we only observe a small fraction of the full SED), as well as errors on the flux values themselves \citep{connolly1999}.

The SCs are fit to a large suite of \citet{bruzual2003} stochastic burst models in order to determine the median posterior stellar masses of the galaxies \citep{wild2014}. Stellar masses are only used to define the sample and therefore have little impact on the results. However, during the analysis described below we verified that the SC derived stellar masses correspond closely to those determined using the star formation history and dust priors assumed in this paper, with small scatter and no significant offset. 

\begin{figure}
    \centering
    \includegraphics[width=\linewidth]{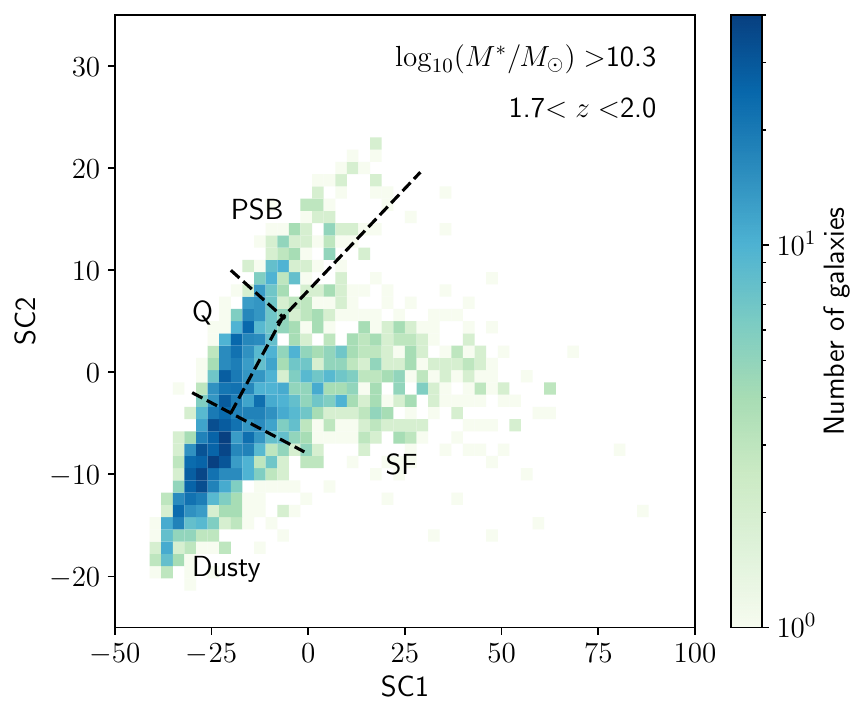}
    \caption{The super-colour distribution of 2,745 galaxies in the UKIDSS Deep Survey with $1.7<z<2.0$ and $\log_{10}(M^*/M_\odot)>10.3$, with logarithmic colour scaling indicating the number of galaxies in each cell as shown by the colour bar. Regions are demarcated by dashed lines for visual purposes only and labels indicate where quiescent (Q), star-forming (SF), dusty and post-starburst (PSB) galaxies are found.}
    \label{fig:SC}
\end{figure} 

For this initial study, we select all 2,745 galaxies from the UDS DR11 catalogue with $1.7<z<2.0$,  $K_{AB}<24$ and $\log_{10}(M^*/M_\odot)>10.3$. Following the completeness calculations in \citet{wilkinson2021} this results in an approximately mass complete sample. This redshift range is optimal because the population of star formation histories is relatively simple to model at this epoch due to the young age of the Universe, and the photometric redshifts and super-colours are well constrained with small errors and minimal bias due to the good coverage of the rest-frame SED by the available broad band filters. Future studies will extend the analysis to different mass and redshift ranges. Fig.~\ref{fig:SC} presents the joint distribution of the first two super-colours for the dataset, with labels to guide the reader in their interpretation in terms of the position of galaxies with different recent star formation histories. This interpretation has been verified with models \citep{wild2014} and comparison to spectroscopy \citep{maltby2016,wild2020}. Demarcation lines match those derived in \citet{wilkinson2021}, but are for visualisation purposes only. It is the distribution in Fig.~\ref{fig:SC} that we aim to model in this paper: the number of galaxies as a function of SC1 and SC2.

\section{Methods}\label{sec:methods}

\begin{figure*}
    \centering
    \includegraphics[width=0.8\linewidth]{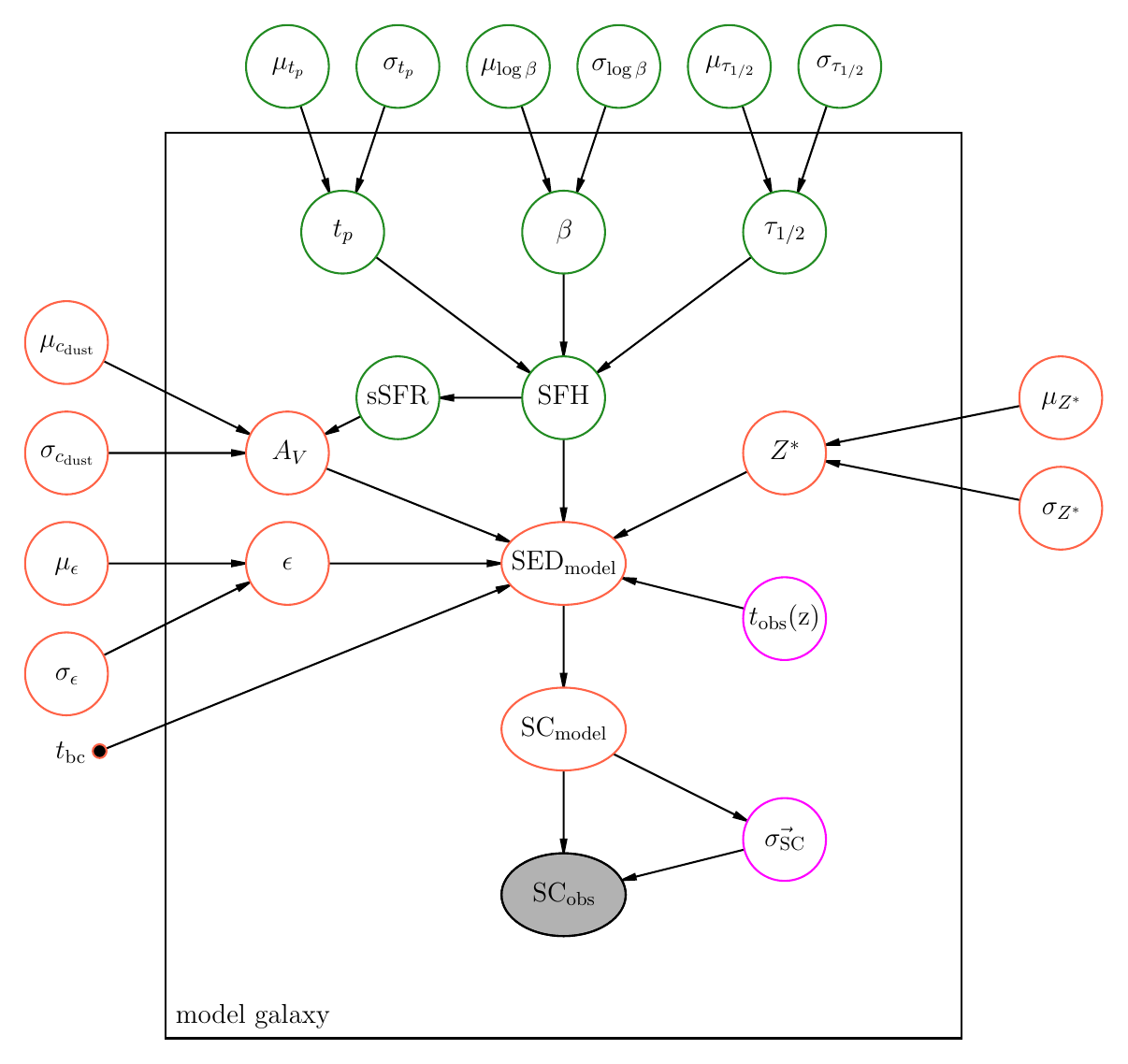}
    \caption{Graphical representation of the model used to construct the colour distribution of a population of galaxies. \emph{Population parameters} (outwith the box) describe the distributions of \emph{model parameters} (inside the box). Parameters that determine the star formation history are in green  (subsection \ref{sec:method:SFH}), while ``nuisance'' parameters are in orange (subsection \ref{sec:method:nuisance}) and observational effects in pink (subsection \ref{sec:method:errors}). For conciseness we have omitted the limits on the truncated Gaussian distributions that additionally describe the populations; these are fixed population parameter variables and would be solid dots outwith the box (i.e. as for the dust birth cloud timescale $t_{bc}$).}
    \label{fig:daft}
\end{figure*}

Simulation-based inference (SBI) allows the estimation of posterior distributions for model parameters given observed data in cases where the likelihood function is analytically intractable, or slow to calculate,  but forward simulations are possible. \citet{Cranmer2020} provide a useful and concise review of the concepts and history behind SBI. In the subsections below we provide precise details of our method; due to the relative novelty of the SBI method to the field we first provide a high-level overview. First we require a generative model, which maps parameters $\theta$ to simulated observables $x$ via a stochastic simulator, assuming a prior distribution $p(\theta)$ which encodes initial parameter beliefs. In the specific case of the rest-frame UV/optical/NIR colour distributions of galaxies, we know that the number of galaxies as a function of colour arises from the underlying distribution of star formation history shapes, alongside the distributions of dust attenuation, stellar metallicity and measurement errors. We can set up a prior that encodes the range of possible star formation histories, metallicities and dust properties of the galaxies. Note that we are not modelling the properties of individual galaxies, but rather the properties of the population, i.e. the distribution of galaxy quenching and formation times, or distribution of quenching rates etc.\footnote{While these population parameters can be helpful in determining the individual properties of galaxies \citep[e.g.][]{thorp2024}, the method is more naturally suited to problems which are focussed on the population parameters themselves.} For this reason, we distinguish \emph{model parameters} that describe the metallicity, dust and SFH of single galaxies and are used to model photometry using stellar population models, and \emph{population parameters} which control the distribution of these model parameters in any particular galaxy population.

Once a model has been defined, a training dataset is constructed by repeatedly sampling the \emph{population parameters} $\theta$ from the prior, which determine the distribution of \emph{model parameters}, and using these model parameters to generate mock datasets $x$, yielding paired samples ${(\theta_i, x_i)}$. In this case, the mock datasets that we wish the model to produce are the normalised number density of galaxies in colour space. These are generated via stellar population synthesis models to create the spectral energy distributions, and therefore colours of mock galaxies (subsections \ref{sec:method:SFH} and \ref{sec:method:nuisance}), with observational errors then applied to the colours (subsection \ref{sec:method:errors}). We compress these image-like colour distributions into summary statistics (``data features'', $x'$) using a principal component analysis (PCA, subsection \ref{sec:method:PCA}). A neural density estimator is then trained to learn the highly non-linear transformations between the population parameters $\theta$ and the summary statistics $x'$, which encapsulate the colour distributions of the galaxies. In this work we use an amortized Neural Posterior Estimation (NPE) algorithm, which is well suited to the image-like data we aim to fit and can be applied directly to different observations without retraining. The resulting trained model can then be applied to observed colour distributions of galaxies, which have been compressed into summary statistics in the same way as the models (subsection \ref{sec:method:inference}). High probability is assigned to population parameters that are consistent with both the data and the prior, resulting in a posterior distribution for the observed data.

The analysis code for this work combines the open source python packages \textsc{sbi} to perform the inference \citep{tejero-cantero2020sbi, BoeltsDeistler_sbi_2025}\footnote{\url{https://github.com/sbi-dev/sbi} (v0.25.0} and \textsc{bagpipes} to model galaxy star formation histories and photometric fluxes \citep{carnall2018}\footnote{\url{https://github.com/ACCarnall/bagpipes} (v1.3.4)}. The full analysis code used in this paper is available on GitHub\footnote{\url{https://github.com/SEDMORPH/bagpus}}. 
In the following sections we describe the model for the star formation history parameters and other parameters relevant to the colour distribution of galaxies, followed by details of the inference process in which we fit the models to the data to determine the distribution of SFH properties.

%%%%%%%%%%%%%%%%%%%%%%%%%%%%%%%%%%%%%%%%%%%%%%%%%%%%%%%%%%%%%%%
\subsection{Modelling the colour distributions of galaxies}\label{sec:method:SSP}
%%%%%%%%%%%%%%%%%%%%%%%%%%%%%%%%%%%%%%%%%%%%%%%%%%%%%%%%%%%%%%%

The rest-frame UV/optical/NIR colour distributions of galaxies are determined by their star formation histories in which we are particularly interested in this paper, as well as their dust contents and, to a lesser extent, stellar metallicities. We address the modelling of each of these in turn in this section. Finally, we address observational errors which are an important  part of the model, as the training datasets will not resemble the true datasets without their inclusion. Fig.~\ref{fig:daft} summarises the parameters in our model, with the relationships between them\footnote{Created by \textsc{daft}. \url{https://docs.daft-pgm.org/en/latest/}.}. 

%-------------------------------------------------------------
\subsubsection{Star formation histories}\label{sec:method:SFH}
%-------------------------------------------------------------
\begin{table*}[]
    \caption{Details of the priors on the population parameters that control the distribution of model parameters. The population parameters $\mu$ and $\sigma$, alongside the allowed limits of the model parameter, describe the truncated Gaussian distribution of model parameters in any given simulation. We assume top hat distribution functions for $\mu$ and $\sigma$ within the limits given. The final column gives a description, including unit where relevant. }
    \label{tab:theta}

    \begin{center}
    \begin{tabular}{c|c|c|c|l}\hline
    & \multicolumn{3}{|c|}{population parameters} & \\
        model parameter & limits & $\mu$ & $\sigma$ & Descriptor\\\hline
        $t_p$ & (0.2,4.84)$^a$& (2,4) & (0.5,2) & approx. SFH peak (Gyr)\\
        $\log_{10}\gamma$ &  (-2,1) & (-2,1) & (0.1,1) & log of quenching timescale (Gyr)\\
        $\log_{10}\beta$ &  (-1,2) & (-1,3) & (0.5,2) & log of SFH rise parameter\\
        $c_{\rm dust}$ & (0.8,2.0) & (1.2,1.5) & (0.1,0.2) & intercept between $\log_{10}(A_V/{\rm mag})$ and $\log_{10}$(sSFR/yr$^{-1}$)\\
        $\epsilon$ & (1,3) & (1,2) & (0.2,2) &  birth cloud dust factor\\
        $Z^*$ & (0.5,2.5) & (0.5,2) & (0.2,2) & stellar metallicity ($Z_\odot$)\\\hline
        
    \end{tabular}       
    \end{center}
    
    \vskip 1pt
    $a:$ {\footnotesize The upper limit on $t_p$ is set to $1.5\times t_{\rm obs}(z=z_{\rm max})$ where $z_{\rm max}$ is the maximum redshift of the observed sample. This matches the limit imposed by the \textsc{bagpipes} double power-law function. We experimented with altering the \textsc{bagpipes} code to extend the limit to a more sensible $z_{\rm min}$, but this did not alter our results. Thus, we keep the limit for consistency with the standard \textsc{bagpipes} distribution.} 
\end{table*}

\begin{figure}
    \centering
    \includegraphics[width=\linewidth]{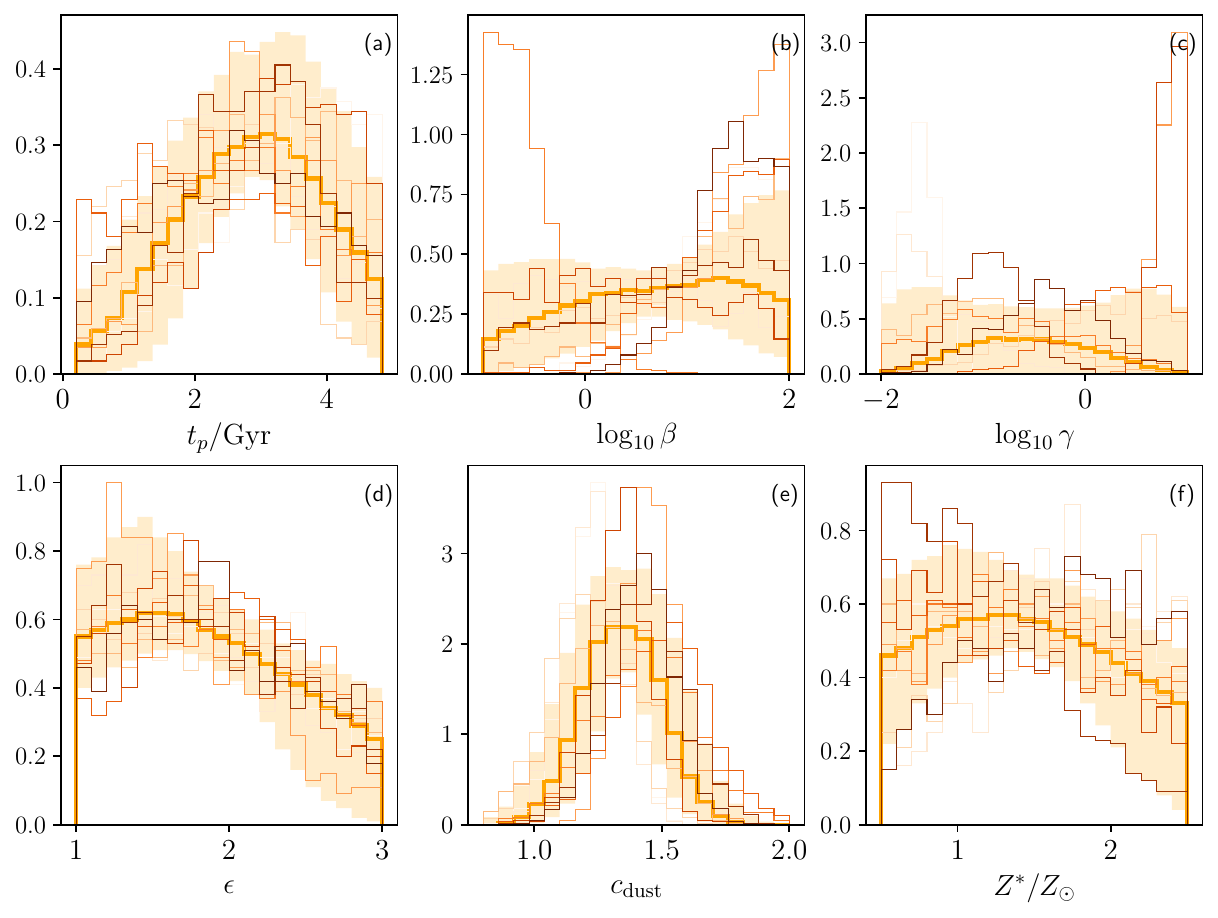}
    \includegraphics[width=\linewidth]{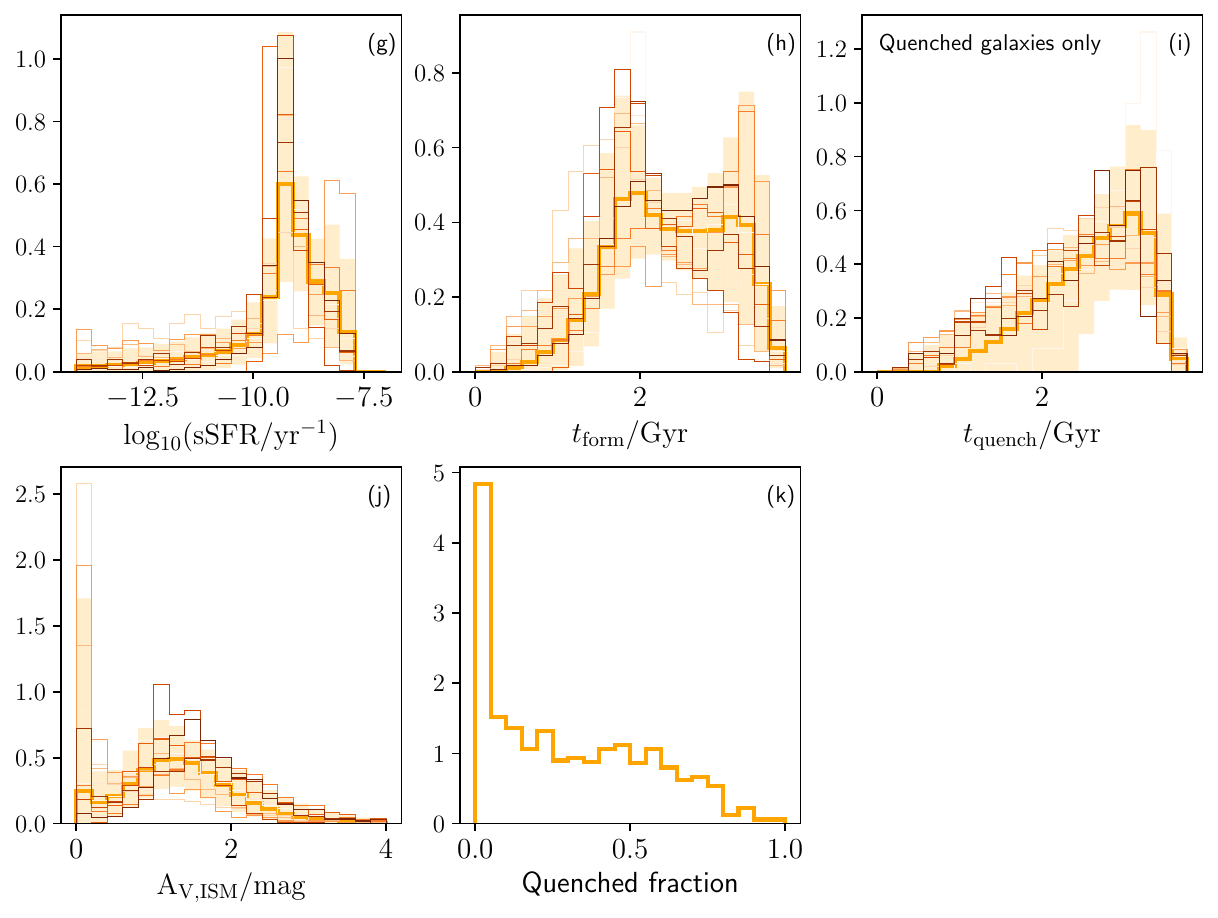}
    \caption{The priors on the distributions of model parameters (top two rows) and derived parameters  (bottom two rows) resulting from draws from the population priors given in Table \ref{tab:theta}. The median and 16th-84th percentile range of the prior distributions are shown as a thick solid line and shaded area. Permissive priors allow a wide range of distribution shapes, and therefore relatively flat median lines and wide percentile ranges. 10 random draws (thin lines) further illustrate some of the range in the shapes of parameter distributions permitted by the population priors. \emph{Top row:} $t_p$, $\beta$ and $\gamma$ describe the shape of the double power-law SFH. \emph{Second row:} $\epsilon$ and $c_{\rm dust}$ describe the dust attenuation, $Z^*$ is the stellar metallicity. \emph{Third row:} the specific star formation rate (with plotting range limited to $>-14$), formation time (mass weighted age) and quenching time (time when SFR decreases to $<10$\% of the average past SFR; includes simulations with $>10$ quenched galaxies only). \emph{Bottom row:} the amplitude of dust attenuation in the $V$ band, and the fraction of quenched galaxies (where $t_{\rm quench}<t_{\rm obs}$).}
    \label{fig:priors}
\end{figure}

Star formation histories (SFHs) of galaxies can be modelled via so-called parametric or non-parametric approaches. In the former, the SFHs are parametrised into simple functional forms, while in the latter the SFHs are parametrised by (usually top-hat) fractions of stars formed in fixed or variable time bin widths. In both cases it is important to compare posterior distributions to those directly or indirectly provided by the priors, in order to ensure that the data is truly constraining the quantities of interest. For the purposes of this paper we focus on the former, which has fewer free parameters. We demonstrate in Section \ref{sec:results} that this approach adequately describes the galaxy population at these masses and redshifts. We model galaxy SFHs as a double power-law functional form:
\begin{equation}\label{eqn:sfh}
    {\rm SFR}(t) \propto \left[ \left( \frac{t}{t_{p}}\right)^{\alpha} + \left( \frac{t}{t_p}\right)^{-\beta} \right]^{-1}
\end{equation}
where SFR$(t)$ is the star formation rate at cosmic time $t$, $t_p$ controls the time at which the star formation peaks, $\alpha$ controls the rate of the decline in SFR and $\beta$ controls the rate of increase. Unless explicitly noted, all times in this paper are measured forwards from the Big Bang. Large values of $\alpha$ and $\beta$ indicate fast rates of change. The double power-law SFH has been shown previously to fit simulated galaxy SFHs well, including the rapid decline in SFR experienced by many quiescent galaxies \citep{Diemer2017,carnall2018, Mosleh2025, Iyer2025}, although can lead to narrower specific star formation rate distributions for star forming galaxies than more complex SFH models \citep{leja2019,CarnallLeja2019}. We keep in mind the potential limitations of this model when presenting our results in Section \ref{sec:results} below.

As we were particularly interested in the quenching timescale distribution, we found it more convenient to directly fit for a parameter related to this, rather than the parameter $\alpha$ in the double power-law. We define $\gamma$ as the time taken for the SFR to drop from the value at $t=t_p$ to half of this value. $\gamma$ and $\alpha$ are related via:
\begin{equation}\label{eqn:tauhalftoalpha}
    \alpha = \frac{\log_{10}\left(4 - \left(1+\frac{\gamma}{t_p}\right)^{-\beta}\right)} {\log_{10}(1+\frac{\gamma}{t_p})}.
\end{equation}
We note that while $\gamma$ can be thought of as a quenching timescale, and has the unit of time, it is not equivalent to other timescales used in the literature to measure galaxy quenching because $t_p$ can be before or after the peak SFR; it only corresponds to the peak SFR when $\alpha$ and $\beta$ are equal. Therefore, we will define other more intuitive quenching timescales below, that are also more appropriate for comparisons to other work. 

As we are interested in fitting the observed distribution of SFHs, rather than individual galaxy SFHs, our model does not directly involve the SFH parameters $t_p$, $\beta$ and $\gamma$, but rather \emph{population parameters that describe the distribution of each of these SFH parameters}. For simplicity we assume the model parameter distributions are described by truncated Gaussians. Population parameters are the means and standard deviations of these Gaussians (e.g. $\mu_{t_p}$ and $\sigma_{t_p}$) which describe the distribution of model parameter values in any given simulation. It is these means and standard deviations that we fit for, and we set flat priors for each of them (i.e. top hat distribution functions). Both the population parameters and the respective model parameters are allowed to vary within fixed ranges, which we set via experimentation to ensure the prior space is small enough to allow efficient training sample generation, while large enough to allow the true data to not be overly constrained by the priors. 

The SFH population parameters and their limits are provided in the top half of Table \ref{tab:theta}. The priors on the population parameters directly determine the distributions of model parameters, which are illustrated in the top row (panels $a,b,c$) of Fig.~\ref{fig:priors}. In order to make this figure we generate 1000 random galaxy populations, each containing 1000 random galaxies with the same redshift distribution as the UDS dataset described in Section \ref{sec:data}. The population models have population parameters that are drawn from the top hat priors given in Table \ref{tab:theta}; these population parameters in turn determine the truncated Gaussian distributions of model parameters for the galaxies in that particular population. In Fig.~\ref{fig:priors} we then take the median, 16th and 84th percentiles of the distributions, thus a broad orange shaded region indicates a wide range of possible distribution shapes allowed by the priors. The distributions of 10 random galaxy populations are over-plotted, to also help visualise the possible range in distribution shapes. Note that the median shown in these plots is not in itself a distribution, this is the median value of all the distributions at a given model value. For all of the SFH parameters, we set permissive priors that allow a wide range of distributions. 

Taking $\log_{10}\gamma$ as an example in panel $c$, the prior given in Table \ref{tab:theta} allows $\mu_{\log\gamma}$ to vary between -2 (rapid quenching) and 1 (slow quenching) and $\sigma_{\log\gamma}$ between 0.1 (narrow distribution) and 1 (wide distribution). The fixed limits then truncate the Gaussian at -2 and 1. This results in a wide range of different possible distributions for $\log_{10}\gamma$, from a population that predominantly quenches slowly, on timescales of $\sim10$\,Gyr, to a population that predominantly quenches fast (timescales of $\sim10$\,Myr), or a population with a quenching time distribution that is relatively flat in $\log_{10}\gamma$. This can be seen either by looking at the 10 random draws from the prior, or from the rather flat median and wide shaded areas. We set the limits on $\log_{10}\gamma$ to encode our prior knowledge about likely quenching timescales, while ensuring that when we fit the model to real data our results are not strongly constrained by the prior assumptions.   

While the model SFH parameters shown in the top row describe the model directly, it is useful to derive further quantities that capture the SFH of the galaxies. For each model we calculate the specific star formation rates (sSFR = SFR/M$^*$), time of formation ($t_{\rm form}$) and time of quenching ($t_{\rm quench}$), following \citet{carnall2018}. As per the default in \textsc{bagpipes}, sSFRs are calculated from the SFR averaged over the last 100\,Myr since the time of observation, and the current stellar mass. The time of formation is defined as the mass-weighted age (although is expressed as a time forward from the Big Bang) and quantifies when the bulk of the stellar mass is being assembled. The time of quenching, $t_{\rm quench}$, is defined as when the SFR decreases to 10\% of the average SFR across the past history of the galaxy:
\begin{equation}
{\rm SFR}(t_{\rm quench}) = 0.1 \frac{{{\rm M_{formed}}(t_{\rm quench})}}{t_{\rm quench}}
\end{equation}
where SFR and $\rm M_{formed}$ are the SFR and total formed mass at the time of quenching.

Although $t_{\rm quench}$ is formally defined for all SFHs, we follow the \textsc{bagpipes} convention and measure it only when it is smaller than the age of the Universe at the time of observation, i.e. only for quenched galaxies at the time of observation. \citet{carnall2018} note that $t_{\rm quench}$ defined in this way is broadly equivalent to a UVJ selection, as well as the common sSFR criterium of ${\rm 0.2/t_{U}(z)}$ (where $t_{U}(z)$ is the age of the Universe at the time of observation). They also note that it is preferable to both of these criteria, as it does not depend on exact filter definitions, nor mass recycling rates. We verified that the quenched galaxies in the dataset fitted in Section \ref{sec:results} have $\log_{10}({\rm sSFR/yr^{-1}})<-10.28$, which is marginally more conservative than $-10.27<{\rm \log_{10}(0.2/t_{U}(z)/ yr^{-1}})<-10.21$ at the redshift limits of the dataset. 

The priors on the derived SFH parameter distributions are visualised in panels $g,h,i$ of Fig.~\ref{fig:priors}\footnote{Note that $t_{\rm quench}$ and $t_{\rm form}$ are calculated at the point of observation, thus these distributions depend on the redshift distribution of the galaxies in our model.}.  It is important to remember that derived (or indirect) priors are a result of the priors on the population parameters, rather being directly imposed by the model. In some cases indirect priors can be surprisingly peaked e.g. for the sSFR of star-forming galaxies \citep[panel $g$, see also][]{CarnallLeja2019}. In panel $h$ we see the model allows for both predominantly early or late formation time distributions, together with flatter distributions\footnote{The apparently bimodal prior in $t_{\rm form}$ is artificial, and caused by the fact that the $t_{\rm form}$ parameter is calculated only up to the time of observation. Galaxies which will in fact form a significant fraction of their mass in the future all have $t_{\rm form}\sim t_{\rm obs}$.}. The prior distribution of quenching times of quenched galaxies (panel $i$) peaks at later times in our model due to the number of quenched galaxies naturally increasing over time in a double power-law SFH. However, the prior includes models where galaxies quench only $>2$\,Gyr after the Big Bang, or models where galaxies begin to quench immediately. The final panel $k$ plots the prior on the quenched fraction, i.e. the fraction of galaxies with $t_{\rm quench}<t_{obs}$, which we see allows for a galaxy population with anything from 0 to 100\% quenched galaxies. Unlike the other distributions, panel $k$ is a simple histogram, as each of our randomly generated populations results in a single value for the fraction of quenched galaxies, rather than a distribution of properties as in the other panels. 

%-------------------------------------------------------------
\subsubsection{Dust and metallicity}\label{sec:method:nuisance}
%-------------------------------------------------------------

There are additional galaxy physical parameters that impact the UV/optical/NIR colour distribution of galaxies for which we must fit to avoid biasing the SFH parameters, but in which we are less interested for our results (so-called ``nuisance'' parameters). The most important of these is dust attenuation, which we model with a \citet{calzetti2000} dust law normalised in the $V$ band ($A_V$/mag). We allow the attenuation of light from stars younger than $t_{bc}=10^7$ years to be larger by a factor $1<\epsilon<3$ than from old stars, i.e. $\epsilon=1$ implies equal amounts of dust attenuation for old and young stars, while $\epsilon=3$ implies three times the amount of attenuation for young stars in the $V$ band to account for their stellar birth clouds \citep{CharlotFall2000}. We fit for the distribution of $\epsilon$ using truncated Gaussian priors, as for the SFH parameters. 

Dust presents an interesting complication for modelling galaxy colour distributions as the mixing of different populations of stars with different dust quantities and geometries means that attenuation impacting the stellar continuum varies systematically with a galaxy's stellar population, and therefore position in colour space \citep[e.g.][]{CharlotFall2000,wild2011,Chevallard2013,Price2014}. To accurately reproduce the colour distribution of galaxies we must capture this mutual dependency. In order to reveal the functional form of the tied variation that is most appropriate to the high mass dataset being fitted, we used \textsc{bagpipes} to fit the photometry of $\sim300$ individual galaxies in the UDS data set described in Section \ref{sec:data}. We assumed SFH model and priors identical to those used in the full population model (i.e. double power-law star formation histories, a \citealt{calzetti2000} dust law and two-component dust attenuation). We found that the data was well fit by a linear relation with sSFR, with little additional mass or redshift dependence within the ranges used in this paper:
\begin{equation}
\log_{10}A_V = c_{\rm dust} + m\times \log_{10}{\rm sSFR}
\end{equation}
where $A_V$ is the attenuation in the $V$ band in the interstellar medium (in magnitudes), sSFR is the specific star formation rate in $yr^{-1}$, and the slope ($m$) and intercept ($c_{\rm dust}$) of the relation were fitted to the median posterior values output by \textsc{bagpipes}. We further found, via fitting galaxies at a wider range of redshifts and masses, that the slope of this relation did not vary substantially with galaxy sample, although the intercept and scatter on the relation did. We therefore fix the slope in our model to $m=0.13$ and fit for the shape parameters that describe the distribution of the intercept using truncated Gaussian distributions. The $\mu_{c_{\rm dust}}$ of this distribution can be taken as the true intercept of the relation, while the $\sigma_{c_{\rm dust}}$ can be taken as the scatter in $A_V$ around the relation.

Stellar metallicity is poorly constrained by the colours that we use in this paper (as we will see below), but we allow it to vary to ensure its unknown distribution is accounted for in the errors on the parameters of interest. We do not allow metallicity to vary with time \citep[see][]{leung2024}, as the impact of this is not expected to be significant for broadband photometric colour distributions. We use the default \textsc{bagpipes} metallicity scale of $Z_\odot=0.02$. 

The dust and metallicity population parameters and their limits are provided in the bottom half of Table \ref{tab:theta} and the priors visualised in terms of model parameter distributions (panels $d,e,f$) and derived $A_V$ distributions (panel $j$) in Fig.~\ref{fig:priors} using the same random sampling as described above. The distribution of $c_{\rm dust}$ is very tightly constrained by the colour distribution of galaxies; after experimentation we decided on relatively restrictive limits for $\mu_{c_{\rm dust}}$ and $\sigma_{c_{\rm dust}}$, in order to avoid the computation of large numbers of unrealistic simulations (panel $e$). The priors allow metallicity distributions which are either predominantly sub-solar, super-solar or relatively flat (panel $f$). Dust attenuation in the ISM varies predominantly between $1<A_V/{\rm mag}<2.5$ (panel $j$), with the spike at $A_V=0$ caused by quiescent galaxies with very low ISM attenuations and positivity constraint on $A_V$.

%-------------------------------------------------------------
\subsubsection{Observational effects}\label{sec:method:errors}
%-------------------------------------------------------------
The final important parameters that impact the colour distribution of galaxies are observational: the redshift distribution and the errors on the observations. The redshift distribution impacts the colour distribution in two ways. Firstly, the distribution of observation time impacts at what point we catch the galaxies in the evolution of the SFHs, e.g. for declining SFHs a later distribution of observation times results in lower sSFRs for the same distribution of SFH parameters, and therefore shifts galaxies to lower SC1. Secondly, the super-colours show some trends with redshift which result from different filters moving in and out of the rest-frame wavelength range used, as well as different sampling of the Balmer/4000\AA\ break region \citep[see Fig. 6 in][]{wild2014}. For both these reasons, we build model sets with a matching redshift distribution to the data, by stochastically sampling from the observed redshift distribution each time we build a model set.

We use the errors on the super-colours to model the error distribution in the data. Like with dust, the SC errors vary systematically with galaxy colour, and we must therefore take this into account in our model rather than simply ensuring a matching overall error distribution as we do for the redshifts. In bins of SC1 we calculate the mean and standard deviation of the errors on SC1 and SC2 in the dataset described in Section \ref{sec:data}.  However, the distribution of errors in bins of SC1 is not perfectly Gaussian: an error floor is observed in the data which increases linearly with SC1, which we fit for using \textsc{scipy}'s \textsc{curvefit} routine. The errors are then applied to the model SC1 and SC2 values as a Gaussian random distribution with mean, variance and error floor appropriate for the model SC1 value\footnote{A feature of PCA is that it does not see Gaussian errors on the input values, therefore it is not possible to model errors on the SCs by including Gaussian errors on the individual input fluxes.}.

A limitation of using the observed redshift distribution and error properties to define the model is that the model becomes specific to the dataset, and a new training sample must be generated for datasets with different redshift or noise properties. This reduces flexibility, but avoids a large number of additional free parameters and potential for mistaking noise-driven scatter for physical scatter in the parameters of interest. Our light-weight approach to modelling galaxy colour distributions means that it is inexpensive to generate new training sets with different redshift and observed error properties, re-train the neural network, and obtain posterior distributions for a new dataset. We therefore prefer this simpler approach compared to more complete and flexible approaches which aim to model full photometric datasets, but require a significantly larger training set and more analysis to understand potential covariances between fitted parameters \citep[e.g.][]{alsing2024, Li2024, aufort2025}.

%%%%%%%%%%%%%%%%%%%%%%%%%%%%%%%%%%%%%%%%%%%%%%%%%%%%%%%%%%%%%%%
\subsection{Building, training and fitting the model}
%%%%%%%%%%%%%%%%%%%%%%%%%%%%%%%%%%%%%%%%%%%%%%%%%%%%%%%%%%%%%%%

Now that the model is described, we detail the process of fitting it to the data. We generate 10,000 training datasets of 7,000 galaxies each by drawing stochastically from the population priors given in Table \ref{tab:theta} and the redshift distribution of the dataset in Section \ref{sec:data}, which jointly determine the distribution of SFH, metallicity and dust parameters for the galaxies. We experimented with different numbers of training datasets and galaxies within them to ensure that the final results were robust. The observed photometric fluxes for each model galaxy are calculated using the \textsc{bagpipes} code, employing the \textsc{update} method of the \textsc{model\_galaxy} class to ensure rapid generation of the models. We changed the default \textsc{bagpipes} stellar population models to the latest 2019 version (CB19) of the \citet[][BC03]{bruzual2003} models\footnote{\url{https://www.bruzual.org/CB19/}}, which were originally published in \citet{plat2019} and described further in \citet{sanchez2022}. To match these new spectral synthesis models we update the default \textsc{cloudy} grids using the 2025 \textsc{cloudy} software \citep{Cloudy2025}, which \textsc{bagpipes} uses to calculate nebular line and continuum emission.  We fix the ionisation parameter for these nebular grids to $\log U=-3$, and verified that allowing this to vary does not impact our results. Early on in the project we compared the resulting SC colour distributions between the default 2016 version of the BC03 models used in \textsc{bagpipes}, the CB19 models used here, BPASS \citep{stanway2018} and FSPS \citep{conroy2009}, finding both the CB19 and FSPS models provided the best overall fit to the super-colour data. 

To calculate the super-colours of the models, the photometric fluxes are projected onto the same super-colour eigenbasis set as used for the data, and errors are added to match the dataset that we are modelling. The number of mock galaxies as a function of SC1 and SC2 is calculated employing the \textsc{fast histogram} mini package\footnote{\url{https://pypi.org/project/fast-histogram/}}. We bin the super-colour values into a 50x55 density array with $-50<{\rm SC1}<100$ and $-25<{\rm SC2}<35$, resulting in an image-like array of 2750 numbers describing the density of galaxies in each super-colour cell. The generation of a single SC density image with 7000 galaxies takes $\sim8$ seconds on a single core of a high-end Apple Mac, and the generation of 10,000 models can be achieved in a few hours on a standard HPC utilising only 24 cores.

%-------------------------------------------------------------
\subsubsection{Compressing the super-colour distributions}\label{sec:method:PCA}
%-------------------------------------------------------------

While we could apply the SBI method to the full SC image array, this is not efficient as much of the array contains empty space, as well as shot noise in the outer regions where the colour space is poorly sampled (in models and data alike). Our aim is to achieve a method that does not take up unnecessary amounts of computing power, therefore we compress the density arrays using a principal component analysis (PCA). While other more complex methods are available that may result in fewer final summary statistics describing the colour distribution (e.g. convolutional neural network [CNN] embedding nets, as demonstrated in the \textsc{sbi} package documentation), the images we are fitting here are not complex, and the simplicity of the easily visualisable PCA method outweighed any additional benefits of more advanced methods. 

\begin{figure}
    \centering
    \includegraphics[width=\linewidth]{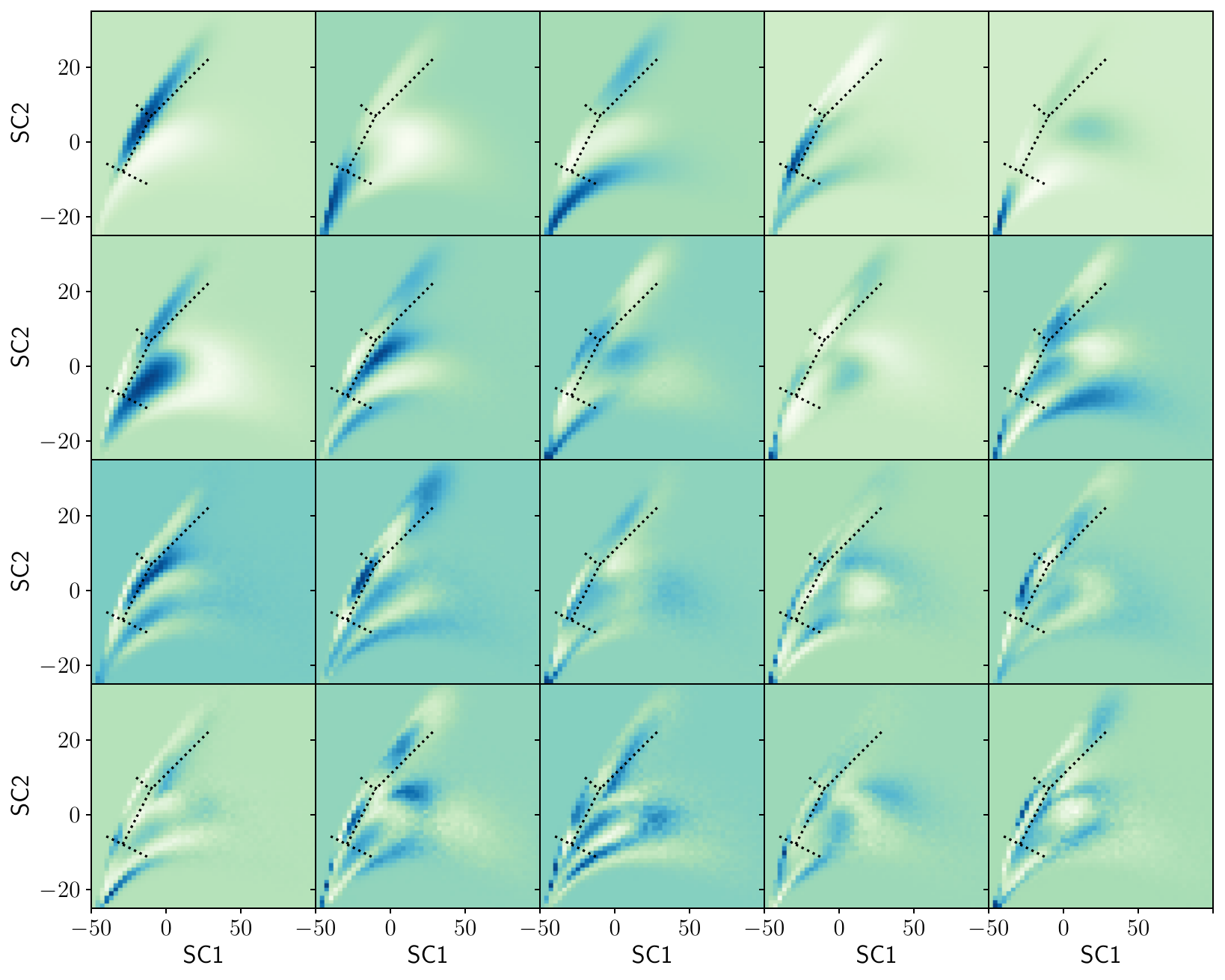}
    \caption{The PCA components used to reduce the dimensionality of the super-colour distributions. Colour scaling is logarithmic number density with a floor at a probability of 0.0001 per cell, equivalent to 0.7 galaxies in our training datasets of 7000 galaxies and each image scaled individually to best visualise the patterns in each component. The dotted lines show the observationally motivated spectral class separation from Fig.~\ref{fig:SC}.}
    \label{fig:PCAcomponents}
\end{figure}

We first place a floor at a probability density of 0.0001 per cell and take the $\log_{10}$ of the image to ensure both the lower and higher density regions contribute similarly to the variance. In our training dataset which uses super-colour distributions of 7000 galaxies, this corresponds to ignoring cells with fewer than 0.7 galaxies in them. Due to the larger number of simulated galaxies compared to galaxies in the UDS dataset that we aim to fit, this floor removes only cells which are unlikely to contain any observed galaxies. We use the \textsc{scikit learn} PCA package\footnote{\url{https://scikit-learn.org/stable/modules/generated/sklearn.decomposition.PCA.html}} to reduce the dimensionality of the 50x55 image-like supercolour arrays into 20 PCA components.  Fig. \ref{fig:PCAcomponents} presents these components, which explain 82\% of the variance in the training image arrays, with the 20th component explaining only 0.01\% of the variance. The number of components was decided on by visually inspecting the components to ensure they still had recognisable features, alongside plotting the explained variance vs. component number which drops rapidly for low component numbers and then flattens off at around 20 for this model set. We also experimented with simulations of star formation histories that were outwith the prior model distribution, and found that the 20 component PCA was still able to produce an accurate image compression. This is important, as we do not want the image compression step to inadvertently introduce model-data mismatch. The final results are not dependent on the exact number of components used, and the appropriate number of components will always depend on the complexity of the model. 

Although it is not necessary to interpret these components, they do provide some insight into the main regions of SC space that vary between models in our training sample, and for this reason we have included on Fig.~\ref{fig:PCAcomponents} the observationally motivated demarcation lines between galaxies of different spectral classes. The first component in the top left of Fig.~\ref{fig:PCAcomponents} alters the relative fraction of quiescent (left branch) to star-forming (right branch) galaxies; the second component alters the age and dust distribution in the star-forming population from older / high dust (more galaxies on the left) to younger / low dust (more galaxies on the right); the third alters the importance of starbursts and associated rapid quenching which introduce more galaxies at low and high SC2. The final row demonstrates the key limitation of PCA - each component must be orthogonal to all others, which introduces interference-like patterns into the components. This does not impact the ability of PCA to reconstruct images, but rather makes the components of limited use for physical interpretation, a problem common to many data compression methods.

\begin{figure}
    \centering
    \includegraphics[width=\linewidth]{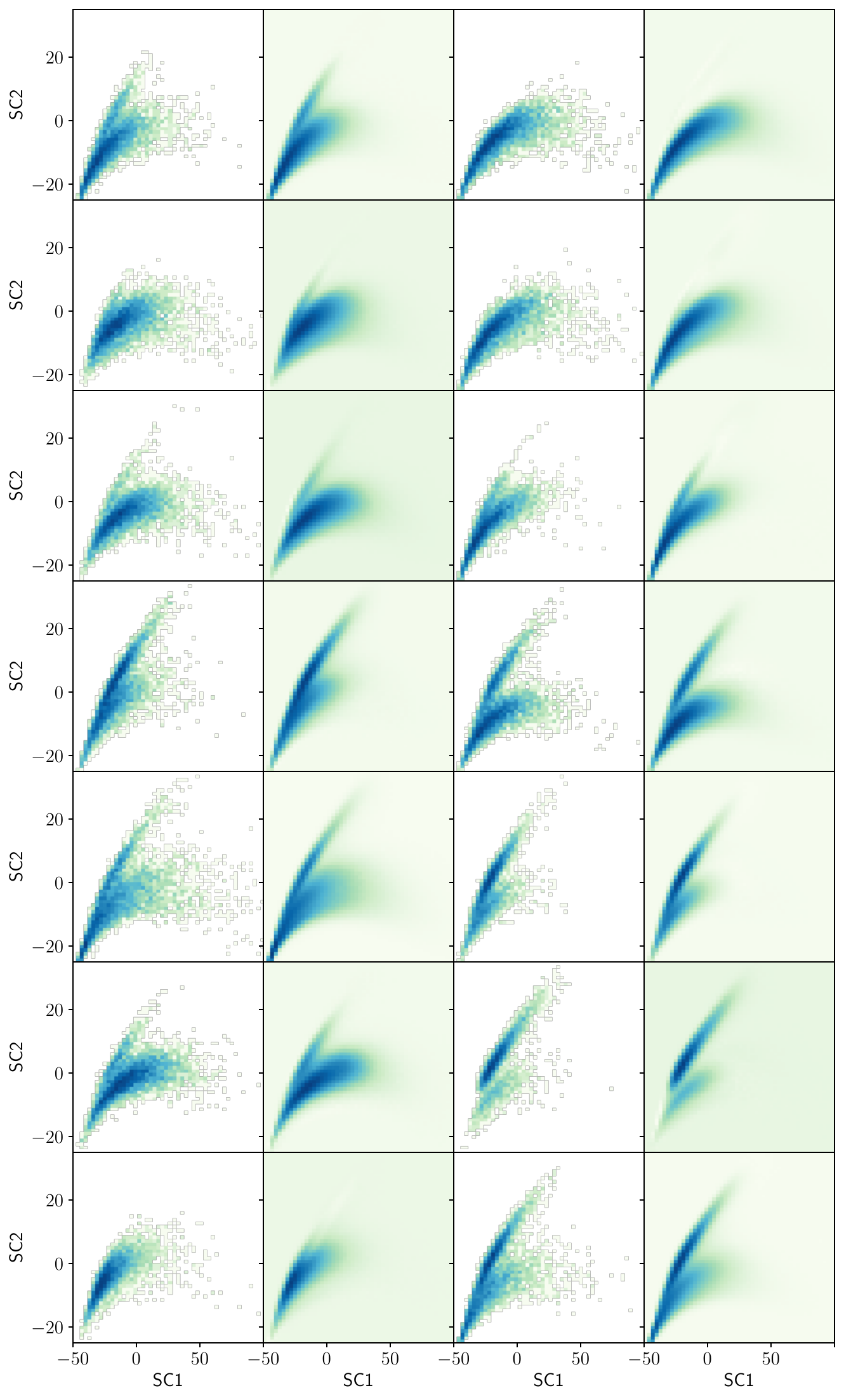}
    \caption{Example model super-colour distributions drawn from the prior population parameters, with their PCA reconstructions to the right of each draw. The colour scaling is logarithmic number density with a floor at a probability density of 0.0001 per cell (equivalent to 0.7 galaxies in the training dataset, or 0.27 galaxies for the UDS dataset) for the PCA reconstructions. We see that the colour distributions vary widely between the different models extracted from the prior population parameters, and the PCA reconstructions capture the wide range of model colour distributions very well.}
    \label{fig:PCArecon}
\end{figure}

Fig. \ref{fig:PCArecon} presents example super-colour distributions drawn from the model prior, alongside their PCA compressed images. This demonstrates the ability of the 20 PCA components to capture the full diversity of the super-colour distributions in the training set, while reducing dimensionality and shot noise at the edges of the distribution. These simulations cover the full range of the model prior, and we can see how substantially the colour distribution of galaxies can change depending on the distribution of SFHs, dust and metallicity. We will return to this point in Section \ref{sec:discussion} when we look at the sensitivity of the colour distribution to parameters of particular interest in this paper.   

%-------------------------------------------------------------
\subsection{Fitting the model to data (inference)}\label{sec:method:inference}
%-------------------------------------------------------------
The next stage is to train the neural density estimator to learn the association between the 20 PCA amplitudes that describe each 50x55 super-colour image, and the underlying population parameters that were used to create it. We input the training set of model parameters and PCA amplitudes into the neural density estimator, which takes only $\sim1$ minute to learn the associations. Given a new observation (either a new mock, or a real dataset), we can then extract the posterior from the neural posterior estimation, and sample from this posterior to obtain the constraints on the model and derived parameters that we are interested in. Typically we draw 1000 samples which takes around one second. In the following subsection we demonstrate this process with a mock dataset. 

%-------------------------------------------------------------
\subsection{Mock parameter recovery tests}\label{sec:methods:mocks}
%-------------------------------------------------------------
\begin{figure}
    \centering
    \includegraphics[width=\linewidth]{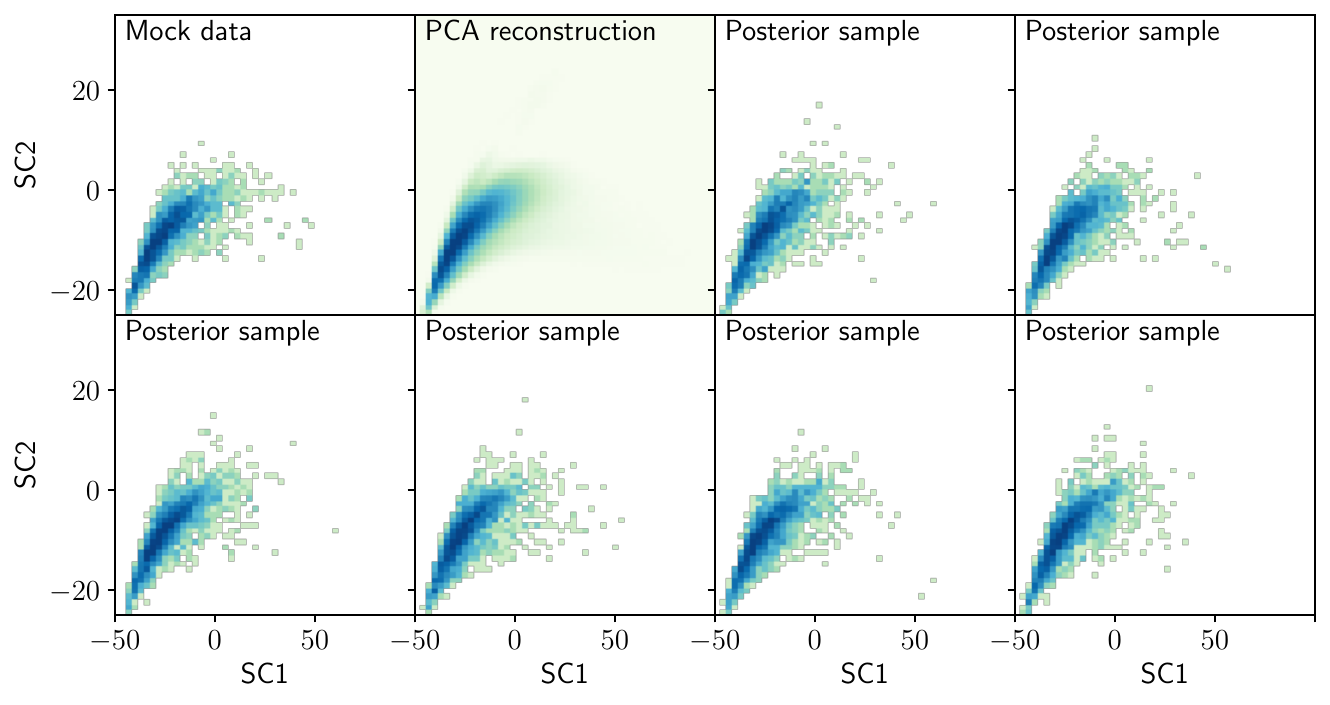}
    \caption{An input mock super-colour distribution with 2745 galaxies (top left), PCA compressed image (top second from left) and 6 samples from the fitted posterior. We can see that all the posterior samples look similar to the input mock. Compared to the wide range of colour distributions allowed by the prior (Fig.~\ref{fig:PCArecon}) this demonstrates that the SBI has correctly fitted the data. }
    \label{fig:mockposterior}
\end{figure}

We draw a new mock dataset from the prior distribution, this time with 2745 galaxies to match the UDS dataset we will be fitting in the next section, and use the method described above to recover its input parameters. Fig.~\ref{fig:mockposterior} shows the mock dataset, alongside the PCA compressed image and 6 images drawn from the posterior. When we compare the images drawn from the posterior to the wide range of super-colour distributions allowed by the prior in Fig.~\ref{fig:PCArecon}, it is clear that the near-identical images extracted by the SBI in the posterior sampling demonstrates that the input data was correctly fit. We calculate the reduced $\chi^2_\nu$ between the mock and posterior models as represented by the PCA reconstruction, summing only over cells with one or more observed or model galaxy, finding $\chi^2_\nu\sim1.6$ and that 59\% of these cells lie within $1\sigma$ of the expected values. We note that the $\chi^2_\nu$ is sensitive to both the number of galaxies in the mock dataset, and the cells over which we sum, so should not necessarily be treated as an absolute indication of goodness of model fit. However, it provides a useful metric when comparing to the fit to data below. The resulting corner plot constraints on the population parameters are shown in Appendix Fig.~\ref{fig:mockcorners}. 

To visualise the ability of the SC distribution to constrain each of the model population parameters over the full prior space, we generate 500 new models with 7000 galaxies each - a test dataset. We then fit these 500 models to extract the posterior distributions for each of the parameters and calculate the posterior median value minus the input value for each population parameter. The resulting distributions of differences between true and fitted values are shown in Appendix Fig.~\ref{fig:mockrecovery}. Both the result for the single mock and these distributions demonstrate that the distribution of dust attenuation ($\mu_{\rm c_{dust}}$ and $\sigma_{\rm c_{dust}}$), quenching timescale distribution ($\mu_{\gamma}$ and $\sigma_{\gamma}$) and mean of the double power-law peak time ($\mu_{t_p}$) are best constrained by the observed distribution of super-colours among the physical properties. Parameters that are poorly constrained are the width of the distribution of the SFH increase rate $\sigma_{\log\beta}$, the width of the metallicity distribution $\sigma_{Z^*}$ and the distribution of the birth-cloud to ISM dust ratio ($\mu_{\epsilon}$ and $\sigma_{\epsilon}$). The limited ability of broad band photometry to recover metallicities is well known \citep[e.g.][]{cheng2025}. Similarly the fraction of dust in birth clouds requires spectroscopic observations to constrain well \citep[e.g.][]{daCunha2008_magphys}. The difficulty in recovering the distribution of the rising SFH rates is related to the well known ``outshining'' problem in SED analysis, where the brighter younger stellar populations make it difficult to obtain detailed information about the fainter older populations \citep[see e.g.][for a recent study]{narayanan2024}.

Improved data, such as spectroscopy, will be able to constrain these parameters better. This section demonstrates the sort of information that can be reliably extracted from the broad-band SED distributions of galaxies, and we use this to focus our results and discussion below.

\section{Results}\label{sec:results}

We apply the same super-colour image creation and PCA image compression to the UDS dataset described in Section \ref{sec:data} to generate the 20 summary statistics used to describe its colour distribution. We provide these summary statistics to the neural posterior estimation to calculate the posterior for this dataset. We then draw 1000 samples from the posterior population model to determine the posterior model and derived parameters. 

%-----------------------------------------------------------------
\subsection{Super-colour distributions and residuals}
%-----------------------------------------------------------------

\begin{figure*}
    \centering
    \includegraphics[width=\linewidth]{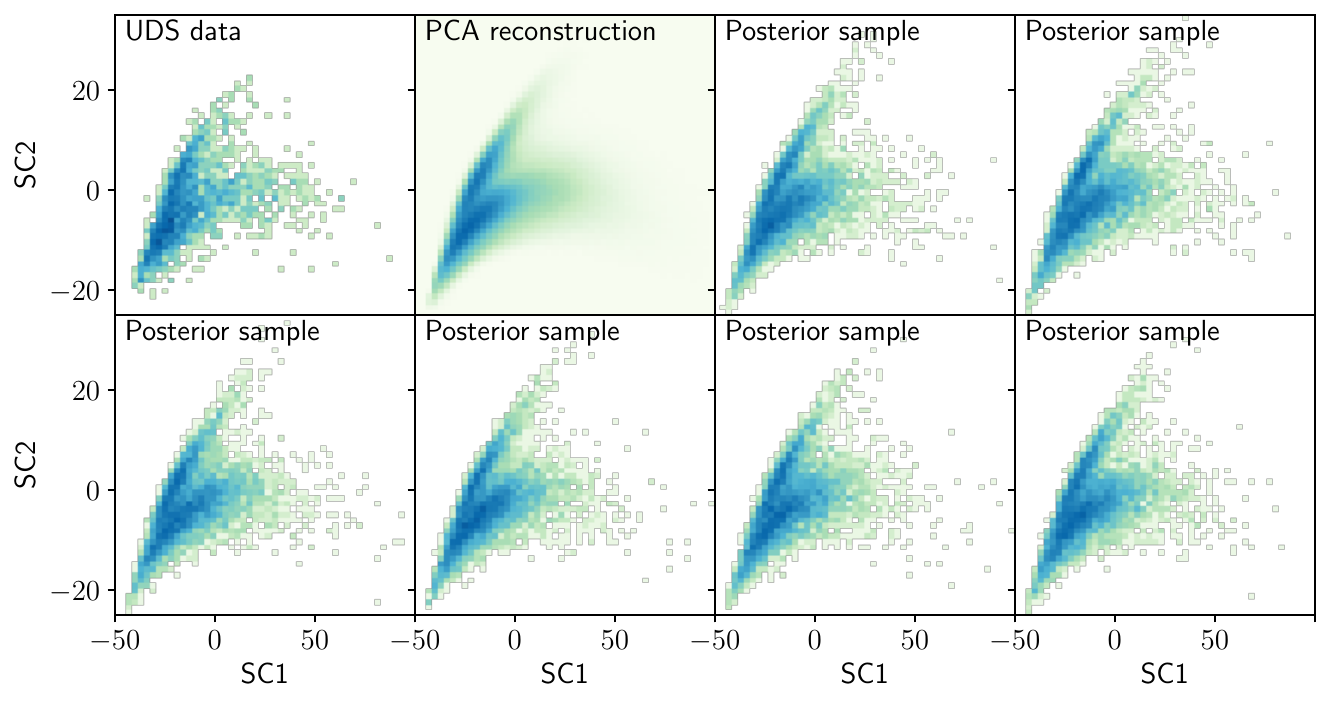}
    
    \caption{The UDS data super-colour distribution for galaxies with $1.7<z<2.0$ and $\log_{10}(M^*/M_\odot)>10.3$ (top left), the PCA compressed image (top second from left) and 6 samples from the fitted posterior. We can see that all the posterior samples look similar, and similar to the observed data. The posteriors have a median $\chi^2_\nu=1.05$, assuming Poisson errors on each SC1/2 cell. Compared to the wide range of colour distributions allowed by the prior (Fig.~\ref{fig:PCArecon}) this demonstrates that the SBI has correctly fitted the data. }
    \label{fig:dataposterior}
\end{figure*}

In Fig. \ref{fig:dataposterior} we show the super-colour distribution of the data in the top left, and PCA reconstruction second from left. The PCA reconstruction resembles the data distribution well. This tells us that the model can capture the main aspects of the data. In itself, this is non-trivial as the PCA eigenbasis was constructed from models with an arguably rather simple SFH and dust prescription. Additionally, the stellar population synthesis models are not infallible and reproducing colour distributions of galaxies from physically consistent population models is typically challenging. 

The following six panels in  Fig. \ref{fig:dataposterior} show random samples from the SBI fitted posterior, i.e. population model fits to the data. There is a strong resemblance between the colour distribution of the data and the population model fits, with little variation between the different posterior samples, indicating that the SBI has correctly identified the primary features of the dataset. This result tells us that a smooth double power-law SFH, with no stochasticity in the SFR, and a simple scaling relation between dust attenuation strength and sSFR, is sufficient to quite accurately produce the SED shape distribution of galaxies at $z\sim2$. It does not, however, tell us that SFHs are in fact smooth - only that the distribution of these two super-colours, based on broad band NUV-NIR SEDs, does not provide information on stochasticity. 

\begin{figure}
    \centering
    \includegraphics[width=\linewidth]{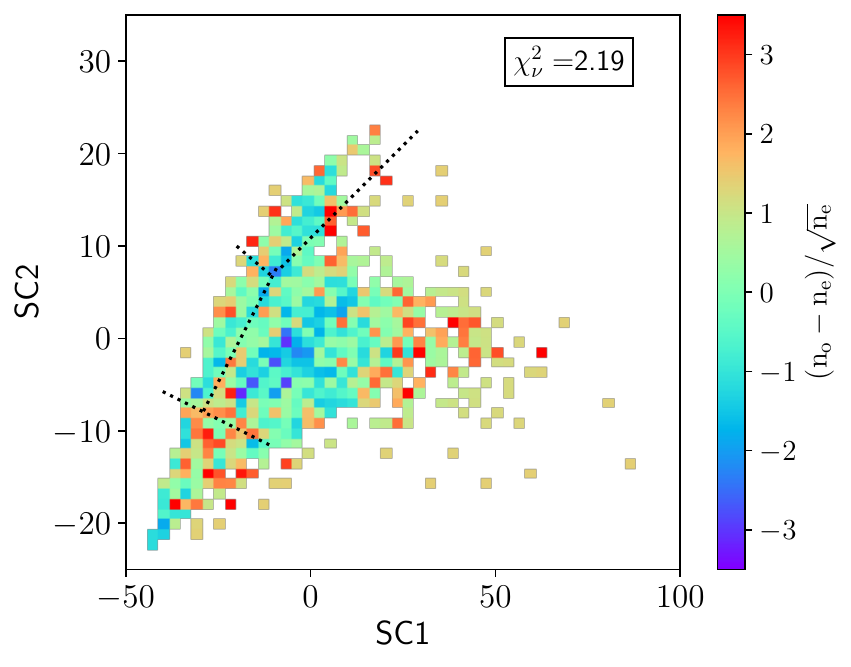}
    \caption{The stacked residuals between the UDS data super-colour distribution ($\rho$) and 100 posterior samples from the fitted model ($\rho_m$), normalised by the estimated Poisson error from the expected number of galaxies in each cell based on the PCA reconstructed posterior image. The resulting colour scale thus represents how many $\sigma$ the data is away from the fitted model. The median reduced $\chi^2_\nu$ of all the posterior samples is given in the top right. The dotted lines show the observationally motivated spectral class separation from Fig.~\ref{fig:SC}. In general the model fits the data well, with observed number of galaxies within $3\sigma$ from the model value. Regions where the model fits less well are visible in purple (more model galaxies) or red (more real galaxies). These are still mostly within $\sim3.5\sigma$, but their clustering in certain regions of SC1/2 space tells us about regions where the model might be improved.} 
    \label{fig:residuals}
\end{figure}

In Fig. \ref{fig:residuals} we show the stacked residuals between the data and 100 draws from the fitted posterior, normalised by the error given the finite number of galaxies expected in each cell. To avoid the effect of shot noise in the model posteriors, we use the PCA reconstruction for both the model and expected error.  We calculate the reduced $\chi^2_\nu$ between the data and posterior distributions, summing over all cells with at least one observed or model galaxy, finding a median value of $\chi^2_\nu=2.19$. We find 97\% of these cells have galaxy number counts within 3$\sigma$ of the model and 45\% within 1$\sigma$. Comparing both the $\chi^2_\nu$ and residual distribution to the mock dataset in subsection \ref{sec:methods:mocks} shows that the fit to the model is less good than in the perfect case of where the data is drawn from the same parameter space as the model. We can conclude the model is a good, but not perfect, description of the data. In a forthcoming paper, we investigate the model fits to cosmological simulations finding that the residuals between fitted posterior and model provide a very good indication of whether the population model is a good match to the data. Thus, the relatively low level of residuals found here provides confidence in our model choices.

There are a few contiguous regions in Fig. \ref{fig:residuals} that are consistently $\sim3\sigma$ away: there are slightly too few model galaxies in the dusty left-hand-side and bluer centre of the star-forming sequence, with slightly too many at $-30\lesssim{\rm SC1}\lesssim-20$ and SC1$>25$. These differences can also be seen from close inspection of the number density distributions in Fig.~\ref{fig:dataposterior}. This could either be due to a slight mismatch with the assumed attenuation curve and/or the correspondence between $A_V$ and sSFR, or the star formation histories of the star-forming population are not allowing enough breadth in the sSFR distribution, a known feature of the double power-law SFH model. The other most notable region with too many model galaxies is along the ridge of the post-starburst branch, with too few either side of it, which may indicate we have under-estimated the errors on the super-colours in this region of the diagram, or again that our SFHs are too simplistic. However, the overall fit is very good, and we would require better data to constrain a more extensive model with a more flexible dust and SFH prescription. 

%-----------------------------------------------------------------
\subsection{Posterior model parameter distributions}\label{sec:results:post}
%-----------------------------------------------------------------

\begin{table}
\centering
\caption{Median, 16th and 84th percentile values of the population model parameters fitted to the  super-colour distribution of galaxies in the UDS with $1.7<z<2.0$ and $\log_{10}(M^*/M_\odot)>10.3$. }\label{tab:popparams}
\begin{tabular}{cccc}\hline
Parameter & 16th & median & 84th \\\hline
$\mu_{t_p}$ & 3.44 & 3.53 & 3.61 \\
$\sigma_{t_p}$ & 0.63 & 0.71 & 0.81 \\
$\mu_{\log\beta}$ & -0.15 & 0.24 & 0.53 \\
$\sigma_{\log\beta}$ & 1.11 & 1.41 & 1.68 \\
$\mu_{\log\gamma}$ & -1.01 & -0.93 & -0.84 \\
$\sigma_{\log\gamma}$ & 0.36 & 0.45 & 0.55 \\
$\mu_\epsilon$ & 1.49 & 1.73 & 1.89 \\
$\sigma_\epsilon$ & 0.78 & 1.27 & 1.67 \\
$\mu_{c_{\rm dust}}$ & 1.27 & 1.28 & 1.29 \\
$\sigma_{c_{\rm dust}}$ & 0.14 & 0.15 & 0.15 \\
$\mu_{Z^*}$ & 0.97 & 1.20 & 1.46 \\
$\sigma_{Z^*}$ & 0.47 & 0.78 & 1.19 \\
\hline
\end{tabular}
\end{table}

\begin{figure}
    \centering
    \includegraphics[width=\linewidth]{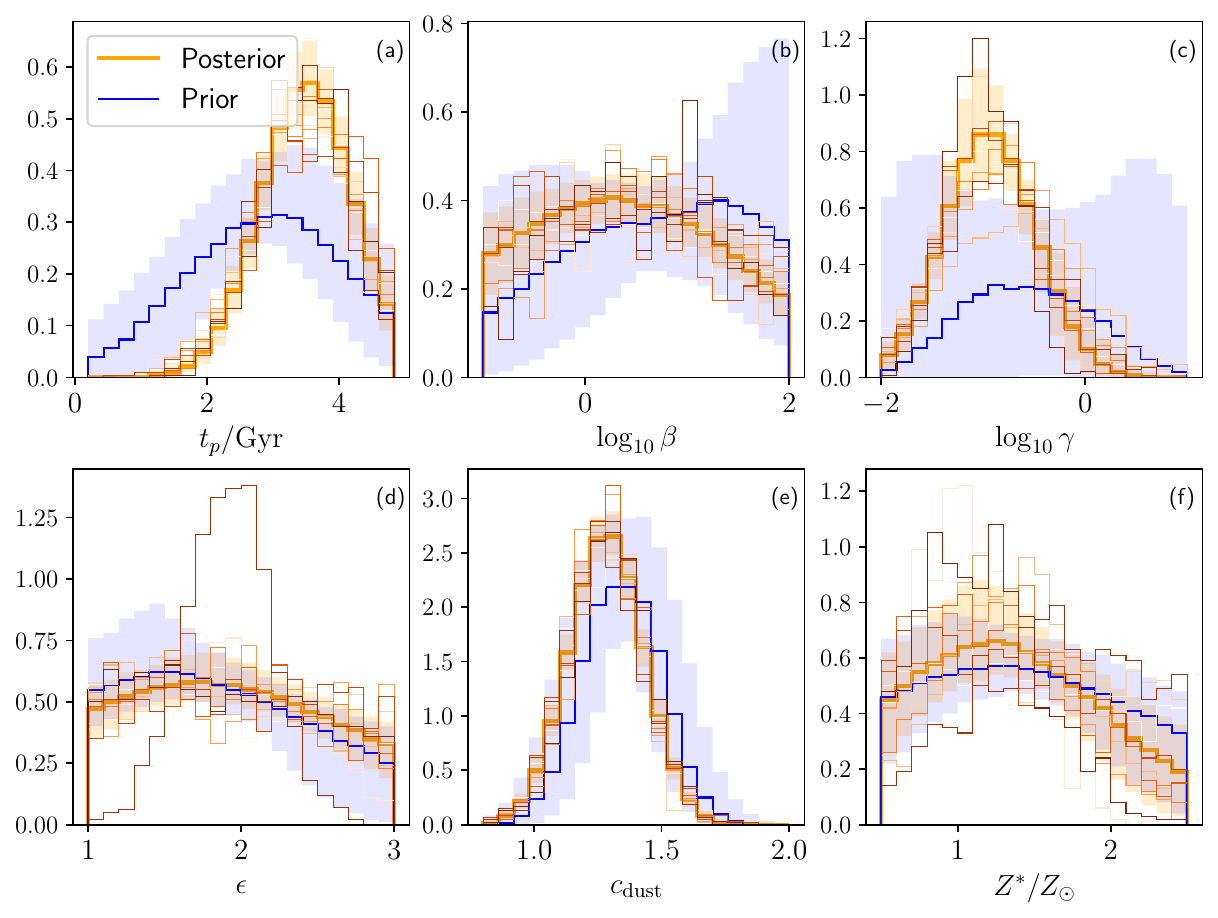}
    \includegraphics[width=\linewidth]{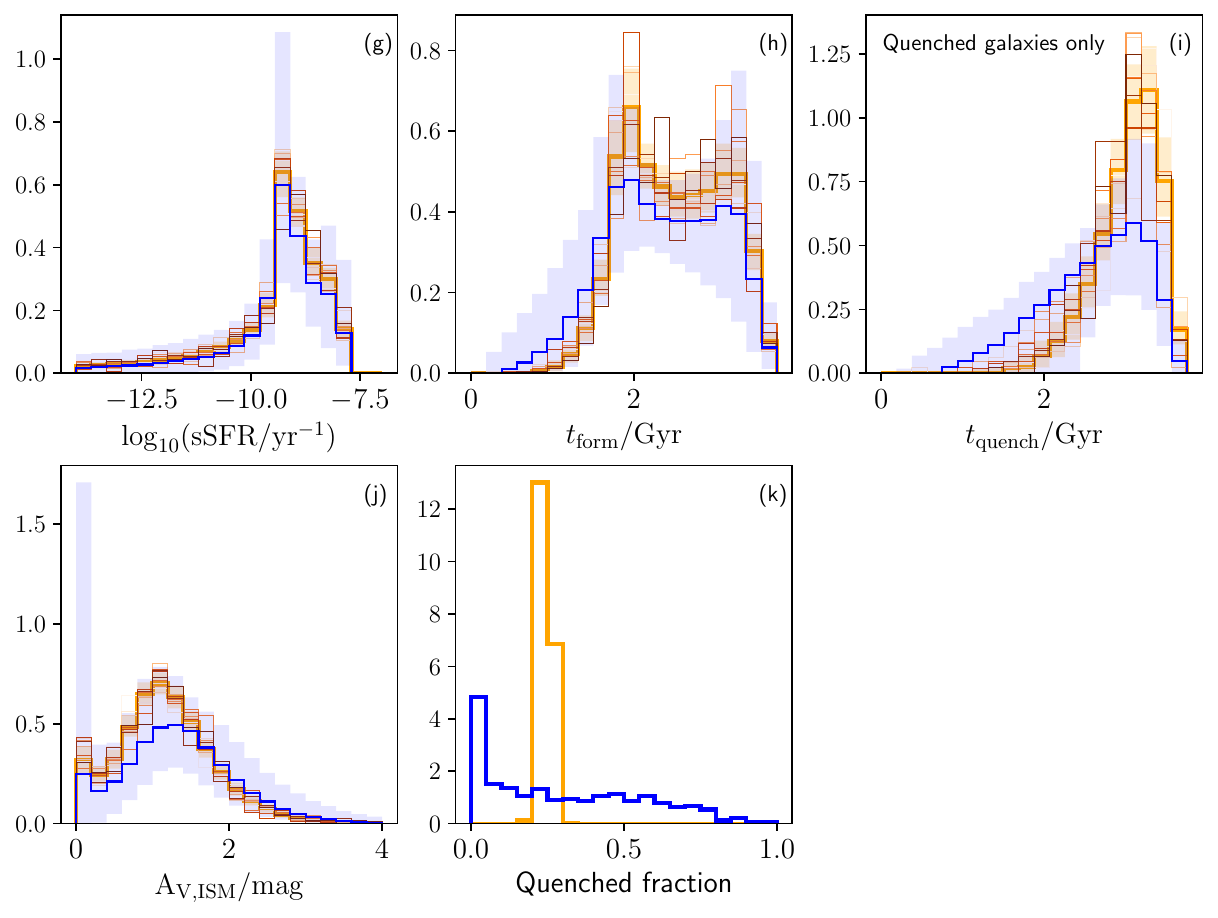}
    \caption{The inferred galaxy model parameter distributions (orange) with prior range (blue) also shown for ease of comparison. Panels, lines and shading as for the prior model parameter distributions shown in Fig.~\ref{fig:priors}. The top two rows show fitted model parameters, while the lower two rows show derived properties. By comparing the narrow orange shaded regions to the broad blue shaded regions, we can see how tightly the data has constrained the model for most parameters.}
    \label{fig:dataparams}
\end{figure}

\begin{figure*}
    \centering
    \includegraphics[width=0.31\linewidth]{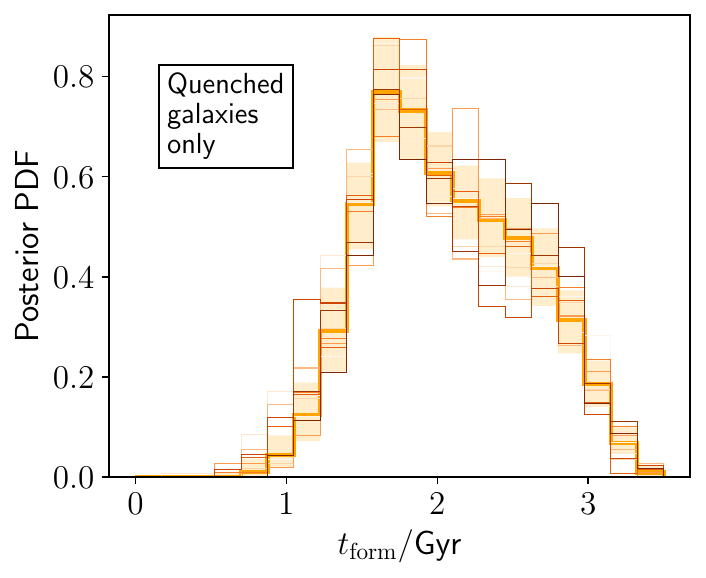}
    \includegraphics[width=0.31\linewidth]{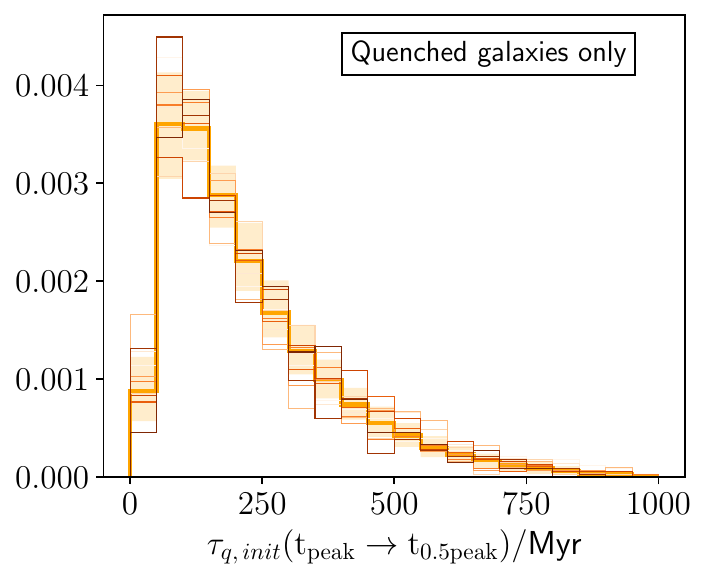}
    \includegraphics[width=0.31\linewidth]{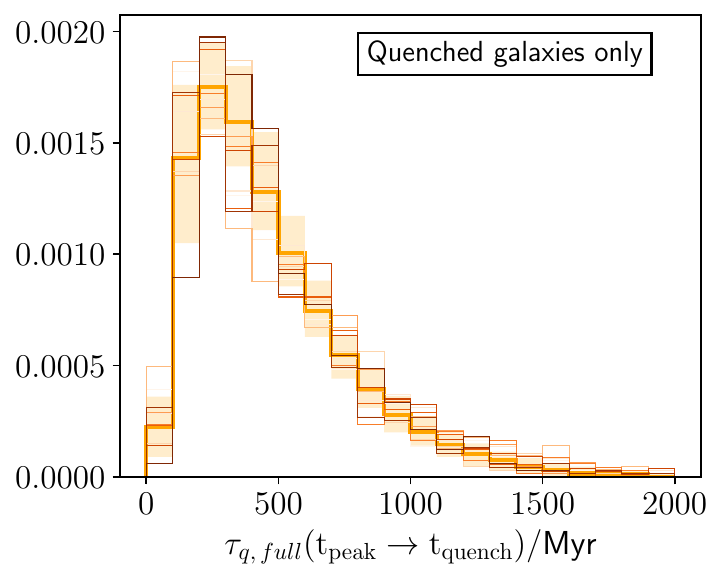}
    
    \caption{The inferred distributions of the time of formation (mass weighted age, left) and quenching timescales (centre and right) for quenched galaxies only (those with $t_{quench}<t_{obs}$) in the UDS dataset. The central panel captures the timescale of the initial fall in SFR from peak to 50\% of the peak, which is relevant for initial quenching mechanisms such as ejective AGN feedback. The right hand panel captures the time taken for the galaxy to reach complete quiescence, which is relevant for quantifying the build up of the quiescent population. Combined with the time of quenching shown in panel $i$ of Fig.~\ref{fig:dataparams}, these distributions imply a steady formation, followed by late and rapid ($\lesssim 1$\,Gyr) quenching for the majority of quenched galaxies observed at $1.7<z<2.0$ with $\log_{10}(M^*/M_\odot)>10.3$.}
    \label{fig:quenched}
\end{figure*}

In order to visualise the impact of the fitted population parameters on the distribution of parameters describing the SFH and dust attenuation of the galaxies in the UDS dataset, we sample the posterior population model 1000 times. In Appendix Fig.~\ref{fig:datacorners} we show the corner plot for the fitted population parameters and the median, 16th and 84th percentiles are given in Table \ref{tab:popparams}.

For each posterior population model we draw 1000 galaxies, in the same way as for the prior distributions in Fig.~\ref{fig:priors}, and  plot the resulting posterior distributions of the model parameters in the top two rows of Fig.~\ref{fig:dataparams},  including 10 random draws from the posterior to further demonstrate the tightness of the constraints on the distributions in most cases. As expected from the mock test and visualised in Fig.~\ref{fig:datacorners}, the $t_p$, $\gamma$ and $c_{\rm dust}$ distributions (panels $a,c,e$) are the best constrained: the orange shaded posterior error bounds are tight and random draws do not show a lot of variation between distributions. When we compare the fitted posterior to the prior distributions in blue we see that most parameters are well constrained by the data rather than the prior; either the shape of the median posterior distribution is different, or the scatter (orange shaded region) is significantly reduced compared to the blue shaded regions of the prior. 

The bottom two rows of Fig.~\ref{fig:dataparams} show the posteriors on the distributions of derived model quantities compared to the input prior distributions. Again, the data constrains these distributions to be tighter than the prior distribution.

Panel $a$ of Fig.~\ref{fig:dataparams} shows that the data strongly prefers a SFH distribution that peaks late ($t_p$), at around the age of the Universe at the time of observation ($3.2<t_{\rm obs}/{\rm Gyr}<3.8$ for the galaxies in the dataset). This tells us that a large fraction of the star-forming galaxies in our sample are still increasing their SFR at the time of observation and results in a later formation time distribution than the prior median (panel $h$). The $t_{\rm quench}$ distribution of the quiescent galaxies (panel $i$) shows that most quenched galaxies quenched late, $>2$\,Gyr after the Big Bang. The distribution of SFH rising rates ($\log_{10}\beta$) in panel $b$ indicates SFHs with very rapid rises (large $\beta$) are not favoured by the data. The parameter controlling the quenching timescale ($\log_{10}\gamma$) in panel $c$ shows that the data disfavours distributions with significant numbers of slower quenching galaxies (large $\gamma$).

The amount of dust in the ISM ($A_{\rm V,ISM}$, panel $j$) is distributed with a peak of around 1.2 magnitude, with a tail to up to $\sim3$ magnitudes. The additional dust in the birth clouds ($\epsilon$, panel $d$) is not well constrained for this particular dataset, although models with a wide range of $\epsilon$ values are preferred. The metallicity distribution is also poorly constrained for this population (panel $f$). 

We measure the quenched fraction from each posterior sample as the fraction of galaxies with time of quenching less than time of observation, $t_{quench}<t_{obs}$. This is strongly constrained at $0.24\pm0.02$ (panel $k$), where we quote the median of the quenched fraction in each of the posterior samples, with errors from the 16th and 84th percentiles. Note that the constraints on the quenched fraction are presented as a simple histogram, as each model galaxy population leads to a single quenched fraction value, rather than a distribution of values such as e.g. sSFR. The value we measure from this probabilistic method is $\sim25$\% higher than that determined from the demarcation lines in Fig.~\ref{fig:SC} of 0.18 (i.e. summing all galaxies classified as quiescent and post-starburst), reflecting the conservative positioning of these lines in order to achieve samples with minimal contamination in \citet{wilkinson2021}. We will return to these points in detail in the discussion section below. 

While the model parameter prior distributions are generally broad, ensuring the data is not too constrained by the prior, this does not guarantee broad derived parameter distributions and the choices of our model are reflected in their posterior distributions to some extent. The sSFR distribution is well known for being constrained by the assumed form for the SFH -- in this case the double power-law SFHs strongly informs its shape. The shape of the $t_{\rm form}$ and $t_{\rm quench}$ distributions are equally shaped by the SFH prior, e.g. as the Universe ages, more galaxies can become quenched when a double power-law SFH is assumed. As noted above, the dust parameter priors were kept narrow to improve efficiency of the training set generation, as they were found to be well constrained by the data for any assumed model -- this results in a relatively tight prior on the $A_V$ distribution.

%-----------------------------------------------------------------
\subsection{The formation times, quenching times and quenching timescales of quenched galaxies}
%-----------------------------------------------------------------

\begin{figure}
    \centering
    \includegraphics[width=\linewidth]{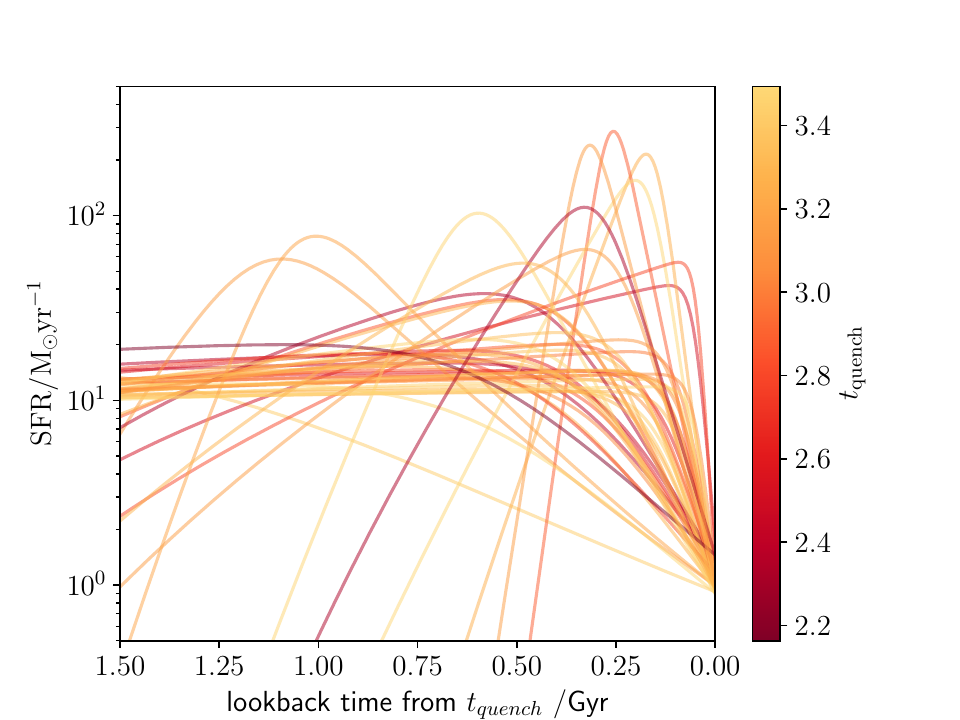}
    \caption{Example star formation histories of 55 quenched galaxies drawn randomly from the posterior model distribution fit to the UDS data. The $x$-axis is shifted such that all galaxies quench at the rightmost edge of the plot, and the $y$-axis is normalised for galaxies with a formed mass of $\log_{10}(M/M_\odot)=10.5$. The colour scale indicates time of quenching.}
    \label{fig:SFH}
\end{figure}
Finally, in Fig.~\ref{fig:quenched} we show the posterior distributions for the formation time (mass-weighted age) and two different definitions of quenching timescale, for the quenched population only. We find that the quenched galaxies observed at $1.7<z<2.0$ formed the bulk of their stellar mass between 1 and 3\,Gyr after the Big Bang, which corresponds to $5.5>z>2.15$. This is not significantly different from the overall population shown in panel $h$ of Fig.~\ref{fig:dataparams}, aside from the relative lack of very recently formed galaxies, showing that many of the quiescent galaxies start out growing like normal star-forming galaxies. The time of quenching for this population, shown in panel $i$ of Fig.~\ref{fig:dataparams}, is late, with the bulk of quiescent galaxies becoming quiescent $\gtrsim2.5$\,Gyr after the Big Bang, or at $z\lesssim2.6$. 

The quenching timescales are defined as the time from the peak of the SFR (calculated numerically, i.e. not $t_p$) to half of the peak SFR (central panel), or to $t_{\rm quench}$ (right hand panel, close to ${\rm sSFR}=0.2/t_U(z)$). The ``initial'' quenching timescale is relevant to the feedback processes that initiate shut-off of star formation in the galaxy, while the ``full'' quenching timescale tells us how quickly galaxies then traverse the green-valley and enter the red sequence. With a double power-law parametric SFH, these two timescales are interrelated, but we show both due to their different physical interpretations and to aid comparison with the literature. We find that the initial quenching timescale distribution for the quenched population peaks at $97_{-25}^{+31}$\,Myr, has a median quenching timescale of $182\pm16$\,Myr and a tail to $\sim700$\,Myr. The full quenching timescale distribution peaks at $256_{-75}^{+87}$\,Myr, has a median quenching timescale of $399\pm27$\,Myr and a tail to $\sim1.5$\,Gyr. 

To give a better intuition for what this distribution of SFHs look like, in Fig.~\ref{fig:SFH} we show a sample of randomly drawn SFHs for the quenched galaxies in our posterior model fit to the UDS dataset, with the $x$-axis shifted such that $t_{\rm quench}$ is at the right most edge for all galaxies. It is important to note that we do not expect individual galaxy SFHs to be smooth, or even uni-modal. Cosmological simulations demonstrate how stochastic SFHs are expected to be \citep[e.g.][]{Lagos2025}. However, the overall shape of the SFH, and quenching timescale, must be approximately correct to lead to the correct distribution of colours. We can see a wide range of SFH shapes: most quiescent galaxies form steadily and quench late, while a minority form late in significant starbursts. We also see the higher fraction of rapidly than slowly quenched galaxies.  We note that there is plenty of opportunity for galaxies to form their mass earlier and quench more slowly, the fact that they do not must be a result of the processes responsible for the quenching, rather than the time at which they quench. We explore the connection between SFH population parameters and colour distribution further in the following section.

\section{Discussion}\label{sec:discussion}
\begin{figure*}
    \centering
    \includegraphics[width=\linewidth]{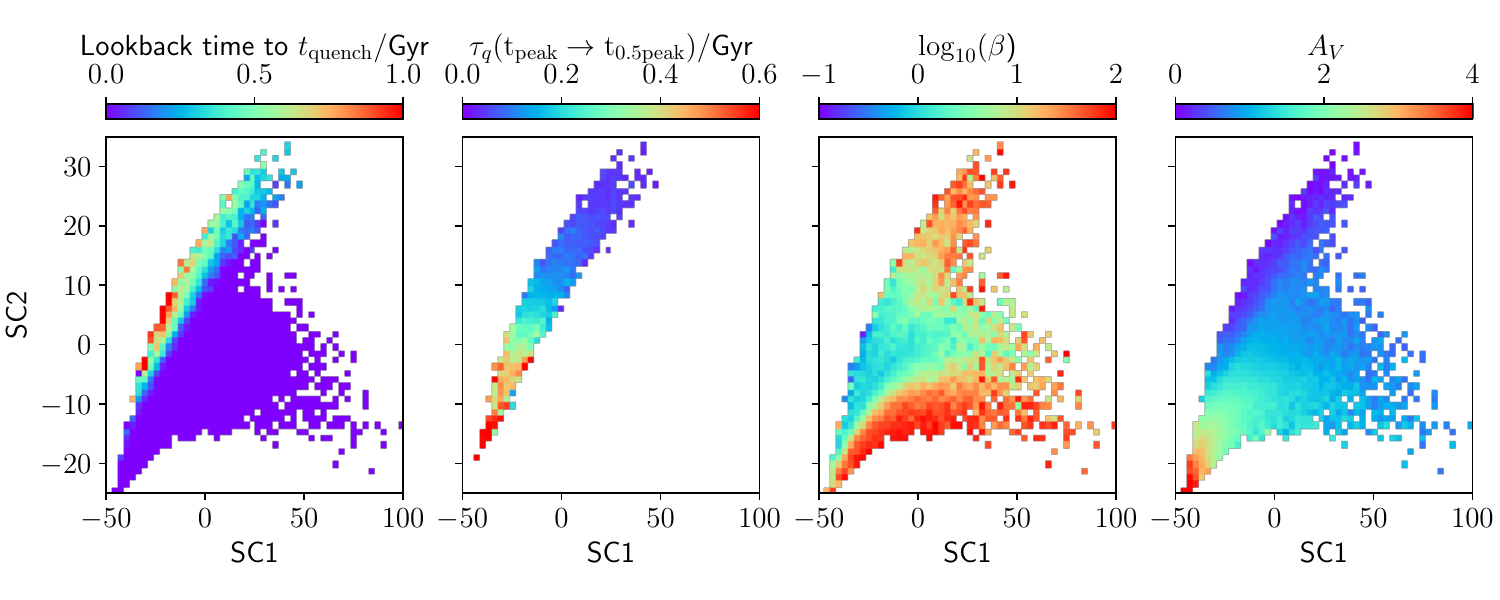}
    \caption{The mean physical parameters from posterior models fitted to the UDS galaxies. We calculate the mean physical parameter for each model draw in each SC1/2 bin, and then average over 100 draws from the population posterior. Note that the colour scale is set to best illustrate the parameter range of the dataset being fit, rather than the full range allowed by the priors. } 
    \label{fig:params_fitted_model}
\end{figure*}

\begin{figure}
    \centering
    \includegraphics[width=\linewidth]{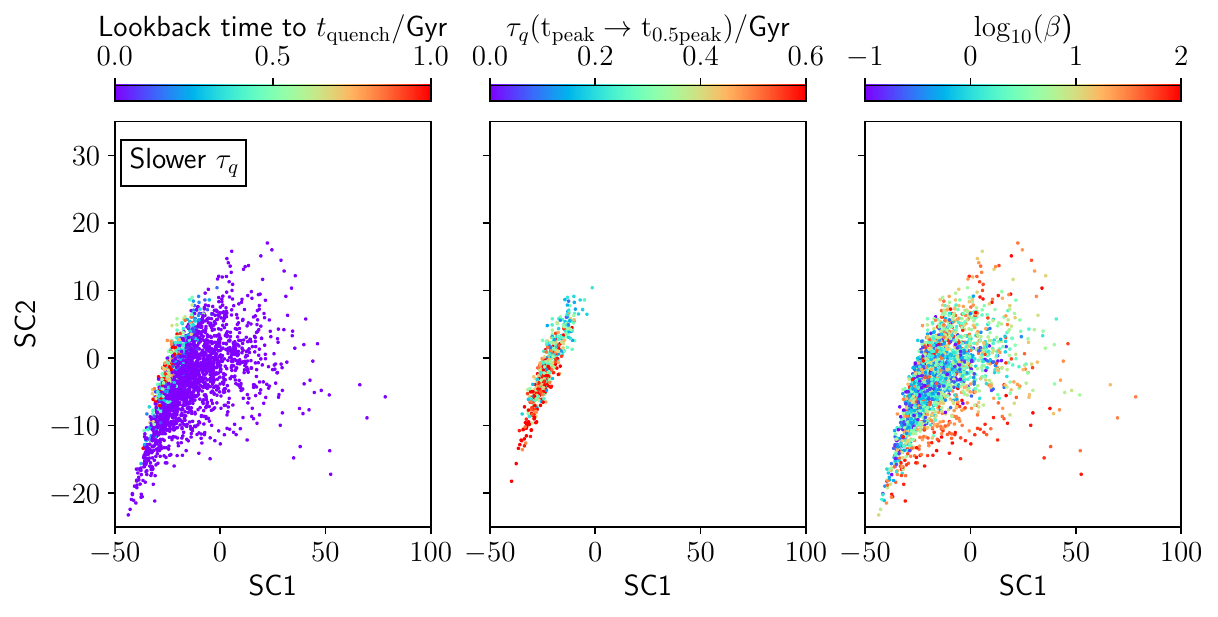}
    \includegraphics[width=\linewidth]{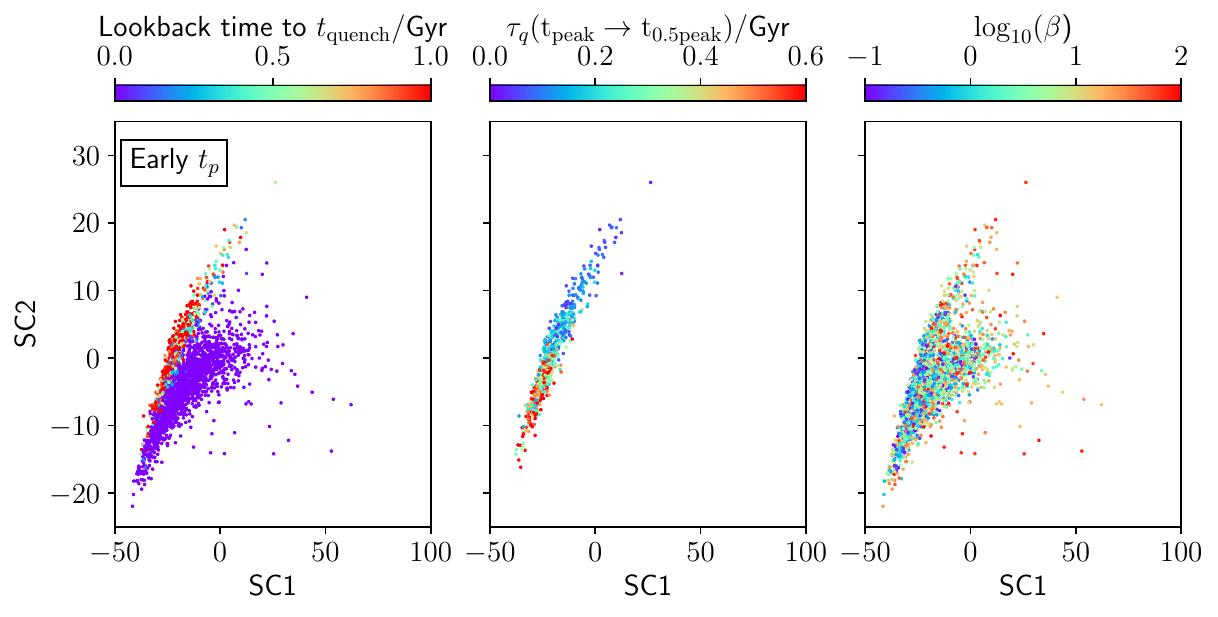}
    \caption{Two alternative models to demonstrate the impact of setting a slightly slower quenching timescale (top panels) or slightly earlier quenching time (bottom panels) as described in the text. In both cases the models have a similar quenched fraction to that observed in the data. }
    \label{fig:params_other_models}
\end{figure}

In the following sections we compare this rapidly and relatively late quenching population to massive quenched galaxies directly observed in the first half of Cosmic time. We start by exploring which features of the colour distribution constrain these parameters, and therefore which of our results are most robust to observational and model uncertainties. 

\subsection{The link between colour distribution and star formation histories}\label{sec:discussion:colourSFH}

In Fig.~\ref{fig:params_fitted_model} we show the average physical parameters of the posterior models fitted to the UDS galaxies on the SC1-SC2 plane, averaging over 100 draws from the posterior.  Note that the number of galaxies contributing to each cell in SC1/2 varies, and not all models will contribute to all cells, particularly in the outer regions where number densities are very low - number density distributions are shown in Fig.~\ref{fig:dataposterior}. The top left panel shows the lookback time to $t_{\rm quench}$ (i.e. $t_{\rm obs}-t_{\rm quench}$), with a value of zero (purple) implying that the galaxy is not yet quenched. We can clearly identify the quenched population to the left edge of the distribution, with those that quenched longest ago lying in a narrow strip just above SC2=0. The second panel shows the initial quenching timescale of the quenched galaxies: the upper wing of the SC1/2 distribution clearly contains the rapidly quenching galaxies, as expected from toy models and spectroscopic comparison, while quiescent galaxies that experienced slower quenching appear lower down on the quiescent branch. The third panel shows the SFH rise timescale ($\beta$), with high values indicating a rapid rise. The posterior distribution shown in the top centre panel of Fig.~\ref{fig:dataparams} shows that the data prefers a distribution with relatively steady rise rates, and this figure explains why this is - a rapid rise creates strong starbursts and post-starbursts (at low and high SC2, in red), and relatively few of these are observed in the data. The final panel shows the dust distribution, which makes it clear that the maximum dust attenuation allowed is strongly constrained by the bottom left point of the distribution. It can also be seen that the dusty star-forming galaxies lie in a distinctly different part of the super-colour diagram compared to quiescent galaxies, thus confusion between these two populations is not expected to impact our results.

For comparison, Fig.~\ref{fig:params_other_models} shows two example models with different values for the distribution of quenching timescale ($\gamma$) and peak of the SFH ($t_p$), while fixing all other parameters to the median of the posterior fit to the UDS galaxies. In each case we adjust the other parameter to result in a similar quenched fraction as observed in the data. In the top panels we show a ``slower $\tau_q$'' model by setting $\mu_{\log\gamma}=-0.6$ and $\sigma_{\log\gamma}=0.2$. To match the fraction of quenched galaxies found in the data, we decrease the mean peak star formation $\mu_{t_p}$ slightly to 3.2\,Gyr. Despite the presence of high $\beta$ models with rapid SFH increases, the upper branch of the SC1/2 distribution and bimodality in the colours has been lost, with galaxies quenched typically on $\gtrsim300$\,Myr timescales. This demonstrates why the data constrains initial quenching to have typical timescales of $\lesssim200$\,Myr.

In the lower panels of Fig.~\ref{fig:params_other_models} we create an ``early $t_p$'' model by setting $\mu_{t_p}=2.3$\,Gyr while retaining the same standard deviation. To match the quenched fraction we increase $\mu_{\log\gamma}=0$. This results in a shift in the quenching time of the quenched population to earlier cosmic times as shown in the lower left hand panel. However, the quenched population retains the same predominantly rapid quenching timescales as found in the posterior fits to the data. 

In this ``early $t_p$'' model the slightly older quenched galaxies fall at slightly lower SC2 values on average. However, the difference is subtle and the exact colour of red sequence galaxies is likely sensitive to a few challenging aspects of the modelling e.g. any amount of dust still existing in quenched galaxies will push them to lower SC2, stochasticity in the SFH would cause galaxies to be on average bluer than modelled with a smooth double power-law SFH, and these colours may also be sensitive to non-solar abundances caused by rapid formation. We conclude that the difference between this ``early $t_p$'' model and the fitted model is much more subtle than for the ``slower $\tau_{\rm quench}$'' model and may not be robust to changes in modelling assumptions. We take this into account in our discussion below.  

Of course, many of these results depend at least in part on the appropriateness of the model choices that we have made, i.e. a double power-law SFH and two component dust attenuation linked to the sSFR of the galaxy. The framework presented here provides a new way to compare the impact of different population model assumptions on the observational properties of galaxies, and future work will investigate different assumed priors.  

\subsection{Quenched galaxy number densities}

\begin{figure}
    \centering
    \includegraphics[width=\linewidth]{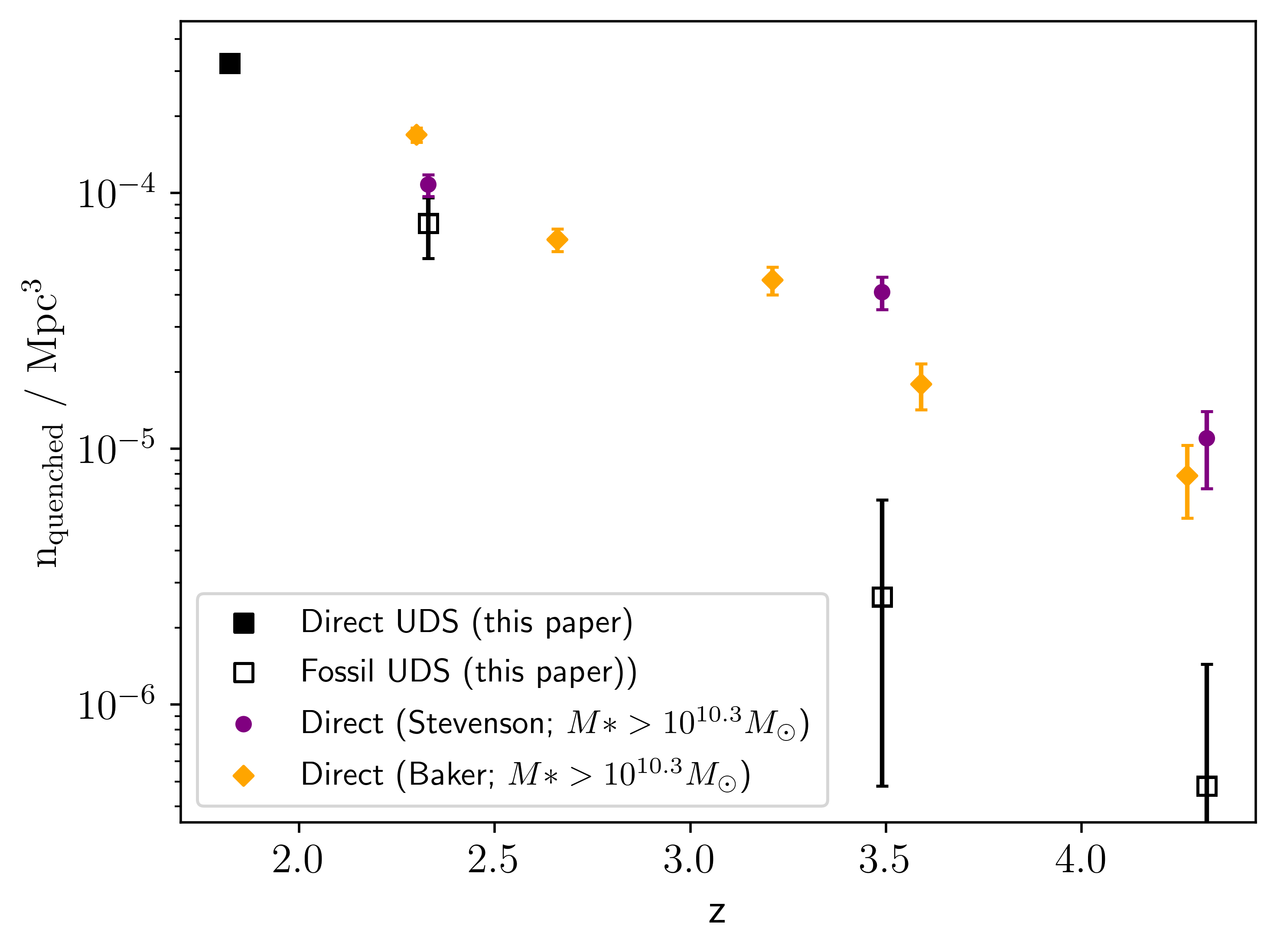}
    \caption{The quenched galaxy number density as a function of redshift. The filled square shows the median posterior value measured from the colour distribution fit to the UDS data, with errors from the 16th and 84th percentiles of the posterior distribution. The open squares show the fossil record number density of quenched galaxies inferred from the fitted SFH distribution at different redshifts. These are compared to values measured directly in JWST and HST fields by \citet{Stevenson2026} as filled circles and \citet{baker2025} as filled diamonds. In both cases, number densities have been recalculated for our mass limit of $\log_{10}(M^*/M_\odot)>10.3$. The difference between direct and fossil measurements can be interpreted as evidence for rejuvenation and/or merging of quenched galaxies between higher redshift and $z\sim2$.}
    \label{fig:numberdensity}
\end{figure}

From the fraction of quenched galaxies ($f_{\rm quench} = 0.24\pm0.02$, Fig.~\ref{fig:dataparams}, panel $k$) and a knowledge of the UDS survey area we can calculate the number density of quenched galaxies at $1.7<z<2.0$ and $\log_{10}(M^*/M_\odot)>10.3$ to be $n_q = 32.0^{+1.8}_{-2.0}\times10^{-5}\,{\rm Mpc}^{-3}$, where we quote the median posterior value and errors from the 16th and 84th percentiles of the posterior distributions. The definition of quiescence used in this paper was calibrated by \citet{carnall2018} to match the $UVJ$-selected quiescent population, and is very close to the popular sSFR cut of $0.2/t_U(z)$ where $t_U(z)$ is the age of the Universe at the redshift of observation. 

Our number density compares very well to the number density of 14 $\log_{10}(M^*/M_\odot)>10.3$ quiescent galaxies with $1.7<z<3.5$ in the spectroscopic JWST Blue Jay survey of $n_q = 27.2\pm0.64\times10^{-5}\,{\rm Mpc}^{-3}$\citep{park2024}; we note that while it is reassuring that such different methods obtain the same answer, it is better to compare in the exact redshift range due to the rapid evolution in quenched fraction with cosmic time at these epochs. Following \citet{wild2016} we therefore calculate the number density of galaxies classified as quiescent and post-starburst in the UDS dataset studied here, using the demarcation lines of \citet{wilkinson2021} shown in Fig.~\ref{fig:SC}, obtaining $n_q = 24\times10^{-5}\,{\rm Mpc}^{-3}$. As noted in Section \ref{sec:results:post} this 25\% lower number density (and similarly quenched fraction) likely reflects the conservative positioning of the demarcation lines in order to achieve a pure sample of quiescent galaxies in \citet{wilkinson2021} vs. the probabilistic approach taken here. 

A similarly lower value of $n_q = 19.0\times10^{-5}\,{\rm Mpc}^{-3}$ is obtained from integrating the Schechter function fit for $\log_{10}(M^*/M_\odot)>10.3$ quiescent galaxies with $1.75<z<2.25$ selected from rest-frame UVJ colours from \citet{mcleod2021}, or $23.0\times10^{-5}\,{\rm Mpc}^{-3}$ for quiescent galaxies with $1.5<z<2$ selected from rest-frame NUVrJ colours from \citet{weaver2023}, showing that the hard demarcation lines in SC, UVJ and NUVrJ space all result in values that are $\sim25$\% lower than the probabilistic method here \citep[see e.g.][for a comparison of quiescent fractions using different selection methods]{Sherman2020}. The fact that the probabilistic method matches better to the spectroscopic result, rather than photometric selections is relevant however, as  the additional constraints afforded by fitting the full colour distribution, rather than individual galaxies, should allow us to get closer to the true answer.  In the future it would be useful to compare with other probabilistic photometric methods such as presented in \citet{Gould2023}, although the same datasets should ideally be used to understand the impact of different methods on absolute number densities.

\subsection{Fossil record vs. direct number densities}

In Fig.~\ref{fig:quenched} we showed the posterior distributions for the quenching timescales of galaxies in the UDS at $1.7<z<2$, showing that most quiescent galaxies quench rapidly. In panel $i$ of Fig.~\ref{fig:dataparams} we see the quenching time distribution for the quenched population, showing that this quenching happens relatively late, with the number density of quenched galaxies rising rapidly 2.5\,Gyr after the Big Bang ($z\lesssim2.6$). Here we investigate the implications of a rapid and late quenching population for the past history of red sequence build up.  

From our fitted model we can calculate the number density of quenched galaxies as a function of redshift, which we find to be 7.6, 0.24 and $0.05\times10^{-5}$Mpc$^{-3}$ at redshifts of 2.3, 3.5 and 4.3 respectively, and is shown in Fig.~\ref{fig:numberdensity}. We compare to direct observations of quenched galaxies in fields with JWST (and HST) coverage \citep{baker2025,Stevenson2026}, with data points plotted at the median redshifts of galaxies in each bin, and number densities adjusted to include only galaxies with  $\log_{10}(M^*/M_\odot)>10.3$ for the purpose of this plot (W. Baker, S. Stevenson priv. comm.)\footnote{The error bars on the points from \citet{baker2025} include only the Poisson contribution, i.e. without the additional cosmic variance component.}. 
Comparing directly to the points from \citet{Stevenson2026} at their sample median redshifts we find lower number densities of quenched galaxies from the fossil record method compared to direct observations by factors of 1.4, 16 and 23 at redshifts of 2.3, 3.5 and 4.3. These factors increase slightly at lower redshifts and decrease slightly at higher redshifts if we instead compare to the results from \citet{baker2025}, but the overall pattern of lower fossil record number densities remains. If correct, this can be interpreted as evidence for rejuvenation and/or merging. Rejuvenation would cause galaxies that are observed at higher redshifts as quenched to become star-forming at later times. These would therefore not be identified as quenched in the colour-colour diagram of $z\sim2$ galaxies. Merging of quiescent galaxies between the higher redshift observations and $z\sim2$ would equally cause their number density measured at $z\sim2$ to be reduced. 

Constraining rejuvenation fractions is crucial for understanding the formation of high redshift quiescent galaxies. Given the high gas contents and chaotic environments expected in the early Universe, high levels of rejuvenation might be one area by which agreement could be improved between observed and simulated numbers of quenched galaxies. \citet{Remus2025} find that only 30\% of massive quiescent galaxies at $z=3.4$ remain fully quenched until $z=2$ in the \textsc{magneticum} simulation, although another 40\% maintain low SFRs. \citet{ChandroGomez2026} find that $\sim$50\% of quiescent galaxies at $z=3$ that survive without merging to $z=2$ rejuvenate in \textsc{colibre}, although comparisons with the observed age-mass relation of high redshift quenched galaxies suggests this may be an over-estimation \citep{Leung2026a}. On the other hand, \citet{Szpila2025} find that a lower fraction of $\sim$10\% of quiescent galaxies at $z\sim2.5$ rejuvenate for longer than $\sim$200\,Myr in \textsc{simba}, with most rejuvenation events being short lived. Our results imply that $\sim$30\% of quiescent galaxies at $z=2.3$ rejuvenate by $1.7<z<2$ which, given the errors on the observations and difference between simulations, is certainly broadly in line with the simulation results. At higher redshifts, the observations imply closer to 90\% rejuvenation rates between $z>3$ and $z\sim1.8$, however with simulations struggling to generate enough quiescent galaxies at $z\gtrsim3$ it becomes more difficult to know whether this is plausible, or related to the increasing difficulty of measuring quenching times from the stellar population fossil record as the time since quenching increases. 

In particular, it is worth cautioning that the exact time of quenching measured in this work depends sensitively on the precise positioning of the oldest red sequence galaxies in colour-colour space, as we discussed in Section \ref{sec:discussion:colourSFH}. If the stellar population synthesis models have a very slightly too weak Balmer/4000\AA\ break for galaxies with ages greater than 1\,Gyr then this would result in an age estimate that is too young. This might happen if the expected different chemical abundances or stellar initial mass function of these rapidly formed and quenched galaxies impact this particular part of the SED. A careful analysis of different spectral synthesis and modelling assumptions would be required to be sure of the accuracy of this result and progress is currently hampered by varying agreement between $\alpha$ enhanced spectral synthesis models \citep{Leung2026a}. Our predicted number densities at $z=3.5$ and $z=4.3$ are slightly lower than those predicted by \citet{park2024} from a similar fossil record analysis of 14 $z\sim2$ quenched galaxy spectra of $n=(1.5-6.0) \times10^5$Mpc$^{-3}$ at $z=4-6$. \citet{Zhang2026b} used a small number of quenched galaxy spectra at $z\sim2.5$ to perform the same experiment; our predictions are lower than their fiducial model at $z\sim3.5$, but higher at $z\sim4.5$. We refer the reader to \citet{Zhang2026b} for further discussion on the uncertainties involved in this calculation. 

\subsection{Quenching timescale distributions}

\begin{figure}
    \centering
    \includegraphics[width=\linewidth]{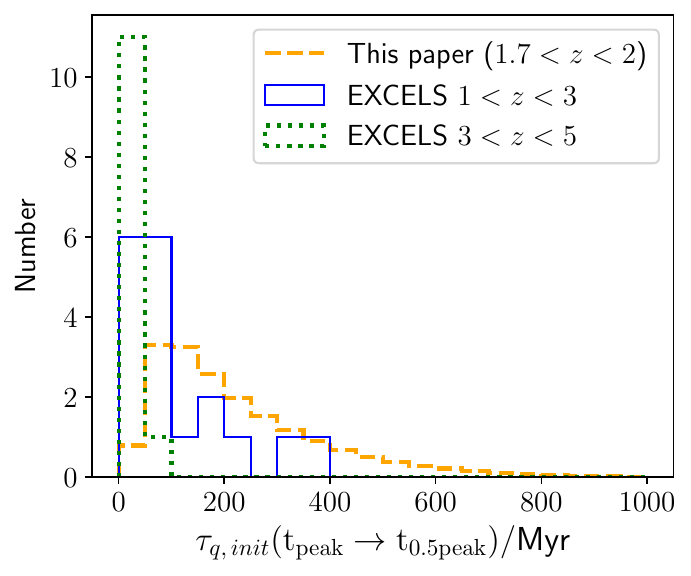}
    \caption{The quenching timescale distribution measured in this paper is compared to values measured directly from JWST spectroscopy of 18 quenched galaxies at $1<z<3$ \citep{skarbinski2026} and 12 at $3<z<5$ \citep{Leung2026a}. The distribution from this paper has been normalised to the equivalent of 18 galaxies to match the $1<z<3$ sample. There is broad agreement between the direct measurements from a small number of galaxies with high quality spectra and the statistical measurement from a large number of galaxies with broad-band photometry, although the spectral analyses prefer even shorter quenching timescales, particularly for quenched galaxies observed at $z>3$. }
    \label{fig:tauhalf}
\end{figure}

In Fig.~\ref{fig:tauhalf} we compare the initial quenching timescale distribution recovered from the broad band photometric analysis in this paper, to direct measurements from a small number of good quality galaxy spectra in the JWST EXCELS survey \citep{carnall2024}. \citet{Leung2026a} used double power-law SFHs to fit the JWST spectra and multi-wavelength photometry of 14 quiescent galaxies with $3<z<5$. \citet{skarbinski2026} fit the JWST spectra and multi-wavelength photometry of 24 quiescent galaxies at $1<z<3$; in order to match the method used in this paper, we take their double power-law fitted SFHs, rather than the exponential+burst SFH used in \citet{skarbinski2026}. We calculate the median $\tau_q$ values from the posterior fits to all galaxies with well constrained SFH parameters (13 at $3<z<5$ and 18 at $1<z<3$). The galaxies in these samples have stellar masses $\log_{10}(M^*/M_\odot)\gtrsim10.5$ at $3<z<5$ and $\log_{10}(M^*/M_\odot)\gtrsim10.2$ at $1<z<3$.

While we cannot draw too firm conclusions from the small number of objects in these studies, the rapid quenching timescale distribution measured from individual galaxies is clear. A larger fraction of the spectroscopic observations than measured in this paper have very short timescales, which may reflect sample selection biases or the spectra providing finer-grained timescale information. There is some evidence for a trend with redshift, with the EXCELS galaxies at $3<z<5$ having very short quenching timescales, while the two galaxies with the longest quenching timescales are found at $z\sim1.5$. A similar result is found when using the full quenching timescales ($t_{\rm peak}\rightarrow t_{\rm quench}$, not shown). This shift to shorter quenching timescales at higher redshifts may relate to the younger age of the Universe when the galaxies quench, i.e. a selection effect whereby shorter quenching times are required at higher redshift in order to find quenched galaxies. This effect is implicitly included in our model, as quenched galaxies with early $t_{\rm quench}$ will only exist if they have rapid quenching times. However,  the model does not allow for inter-relatedness or evolution in the SFH parameters with cosmic time, e.g. the quenching timescale can not evolve with time, or be tied to the SFR rise time etc., beyond the requirement of increasing maximal quenching timescale with cosmic time. Fig.~\ref{fig:tauhalf} suggests that such trends may be required to accurately reproduce the quenched galaxy population, and could be explored in more detail with population-level spectroscopic studies in the future. 

In Fig.~\ref{fig:params_other_models} we set the ``early $t_p$'' model to not only have a similar quenched fraction to the data, but also to have similar  quenched number densities as a function of redshift to those observed directly by \citet{Stevenson2026}. As a reminder, this model is ruled out by the data, but is worth considering given the small difference in colour distribution that it results in, and possible imperfections in spectral synthesis modelling. The earlier peak SFH allows this particular model to have a distribution of quenching timescales that is slightly slower, with a median initial $\tau_{q}({\rm t_{peak}\rightarrow t_{0.5peak}})= 264$\,Myr instead of 182\,Myr for the fitted model. However, the shape of the distribution is the same, with a peak close to 100\,Myr and tail to $\sim800$\,Myr. Thus the predominantly rapid quenching timescale distribution is robust, even if the quenching happens a little earlier than measured here. The spectroscopic results in Fig.~\ref{fig:tauhalf} favour more rapid quenching, suggesting that perhaps this concern is not relevant. 

Direct comparison with other results in the literature is challenging, due to the wide range of definitions of $\tau_q$: some authors start the clock on or above the star-forming main sequence (MS), while others start the clock once the galaxy has already fallen significantly below the MS, others calculate a gradient in SFR. Additionally, when calculating sSFR, SFR can be averaged over 10 vs. 100\,Myr, while stellar mass can either be current or formed over the galaxy's entire history, making direct comparisons even more difficult. A combined analysis of all available spectroscopic datasets at these redshifts using consistent definitions for quenching timescales would clearly be beneficial, but is beyond the scope of this paper. 

Overall there is a growing consensus that rapid quenching routes are dominant prior to Cosmic Noon \citep[e.g.][]{Forrest2020, Slob2024,park2024,Nanayakkara2025,skarbinski2026}. Fig.~\ref{fig:tauhalf} suggests that quenching timescales may become even more rapid in the early Universe \citep{Leung2026a}.  These rapid timescales are significantly more rapid than the age of the Universe, and thus relate directly to the \emph{processes} causing the quenching, not simply when the galaxies are quenching. By $z\sim1$ a wider range of quenching timescales are apparent, with $\lesssim50\%$ of quenched galaxies having quenched rapidly, many of which are identified as post-starburst galaxies \citep[e.g.][]{wild2016,belli2019,wild2020,Tacchella2022,Nersesian2026}. How this diversification of quenching timescales relates to the decreasing gas content of galaxies, or the greater range of processes available to actively quench galaxies, requires further study.

\section{Summary}\label{sec:summary}

We present a simulation-based inference (SBI) framework to infer population-level star formation history (SFH) distributions directly from galaxy colour distributions, without fitting individual galaxies.

\medskip

\begin{itemize}
    
    \item Applying this method to 2,745 massive galaxies ($\log_{10} M_*/M_\odot > 10.3$) at $1.7 < z < 2.0$ in the UKIDSS Ultra-Deep Survey, we show that a simple double power-law SFH model, combined with a scaling relation between dust attenuation and specific star formation rate, reproduces the observed super-colour distribution accurately (Figs.~\ref{fig:dataposterior} and \ref{fig:residuals}).

    \item The inferred galaxy population exhibits a clear bimodality in colour space, including a distinct post-starburst population, without requiring stochastic star formation or complex SFH parametrisations.

    \item We measure a quenched galaxy fraction of $0.24 \pm 0.02$ at $1.7<z<2$ (Fig.~\ref{fig:dataparams}, panel $k$), with the majority of quenched galaxies having ceased star formation $\gtrsim2.5$\,Gyr after the Big Bang ($z\lesssim2.6$, Fig.~\ref{fig:dataparams}, panel $i$)).

    \item Quenching must be rapid to reproduce the observed colour distribution: the distribution of timescales taken for the SFR to fall from peak to 50\% of peak (the initial quenching timescale) peaks at $97^{+31}_{-25}$\,Myr, has a median of $182\pm16$\,Myr and a tail to $\sim700$\,Myr (Fig.~\ref{fig:quenched}, central panel). The distribution of timescales taken to reach full quiescence has a median of $\sim400$\,Myr (Fig.~\ref{fig:quenched}, right panel). 

    \item Direct measurements of the number density of quenched galaxies at $z > 2.5$ are higher than those inferred from the star formation histories measured here, implying significant rejuvenation and/or merging of quenched systems between $z \sim 5$ and $z \sim 2$ (Fig.~\ref{fig:numberdensity}).

    \item While the quenching timescale distribution is robust to modelling uncertainties, the exact quenching time is likely to depend on details of the stellar population, star formation and dust models (Fig.~\ref{fig:params_other_models}). 
    
\end{itemize}

The precise form of the bimodal colour distribution of massive galaxies at $z \sim 2$ requires that star-formation quenching is both common and rapid, with $24\pm2\%$ of galaxies already quenched, predominantly within the last Gyr from the time of observation. While this result has been achieved using novel statistical methods applied to broad band photometric data, the addition of spectroscopic observations of large numbers of galaxies at cosmic noon would allow for greater insights. Spectroscopic surveys with near-infrared instruments such as MOONS and PFS could easily adapt the summary statistics used here for application to spectroscopy, thus combining the benefits of tighter constraints on the SFH of individual systems with the population level constraints of spectral shape distributions. With improved data, additional parameters could be included in the model, such as a variation in quenching timescale with time of quenching, bimodal quenching timescale distributions, evolving metallicities or initial mass functions, each leading to interesting new information on how galaxies form, evolve and quench.

\section*{Data Availability}
UKIDSS data, including super-colours, are available from \url{https://www.nottingham.ac.uk/astronomy/UDS/index.html}. The code \textsc{bagpus} (Bayesian Analysis of Galaxy PopUlations with Spectra) is available at \url{https://github.com/SEDMORPH/bagpus}. This will allow you to reproduce the main paper figures, as well as create your own galaxy population fits. The training and test datasets, the SBI generated posterior, and the CB19 stellar population grids with associated Cloudy grids for use with \textsc{bagpus} and \textsc{bagpipes} are available at \url{https://doi.org/10.5281/zenodo.22670811}.

% Acknowledgements
\section*{Acknowledgements}
We would like to thank both OJA referees for the very thoughtful and careful refereeing of the paper. VW acknowledges the support of the Science and Technologies Facilities Council (ST/Y00275X/1) and Leverhulme Research Fellowship (RF-2024-589/4). HL and ACC acknowledge support from a UKRI Frontier Research Grantee Grant (PI Carnall; EP/Y037065/1). MS acknowledges support for Program number JWST-GO-03543.014 that was provided through a grant from the STScI under NASA contract NAS5-03127.

\textit{Software:} \textsc{Astropy} \citep{astropy}, \textsc{Bagpipes} \citep{carnall2018,CarnallMcLure2019}, \textsc{SBI} \citep{tejero-cantero2020sbi}, \textsc{Daft} \citep{daft}, \textsc{Matplotlib} \citep{matplotlib}, \textsc{Numpy} \citep{numpy}, \textsc{Scipy} \citep{scipy}.

\bibliography{biblist}

@ARTICLE{astropy,
       author = {{Astropy Collaboration} and {Robitaille}, Thomas P. and {Tollerud}, Erik J. and {Greenfield}, Perry and {Droettboom}, Michael and {Bray}, Erik and {Aldcroft}, Tom and {Davis}, Matt and {Ginsburg}, Adam and {Price-Whelan}, Adrian M. and {Kerzendorf}, Wolfgang E. and {Conley}, Alexander and {Crighton}, Neil and {Barbary}, Kyle and {Muna}, Demitri and {Ferguson}, Henry and {Grollier}, Fr{\'e}d{\'e}ric and {Parikh}, Madhura M. and {Nair}, Prasanth H. and {Unther}, Hans M. and {Deil}, Christoph and {Woillez}, Julien and {Conseil}, Simon and {Kramer}, Roban and {Turner}, James E.~H. and {Singer}, Leo and {Fox}, Ryan and {Weaver}, Benjamin A. and {Zabalza}, Victor and {Edwards}, Zachary I. and {Azalee Bostroem}, K. and {Burke}, D.~J. and {Casey}, Andrew R. and {Crawford}, Steven M. and {Dencheva}, Nadia and {Ely}, Justin and {Jenness}, Tim and {Labrie}, Kathleen and {Lim}, Pey Lian and {Pierfederici}, Francesco and {Pontzen}, Andrew and {Ptak}, Andy and {Refsdal}, Brian and {Servillat}, Mathieu and {Streicher}, Ole},
        title = "{Astropy: A community Python package for astronomy}",
      journal = {\aap},
         year = 2013,
        month = oct,
       volume = {558},
          eid = {A33},
        pages = {A33},
          doi = {10.1051/0004-6361/201322068},
archivePrefix = {arXiv},
       eprint = {1307.6212},
 primaryClass = {astro-ph.IM},
       adsurl = {https://ui.adsabs.harvard.edu/abs/2013A&A...558A..33A}
}

@ARTICLE{matplotlib,
       author = {{Hunter}, John D.},
        title = "{Matplotlib: A 2D Graphics Environment}",
      journal = {Computing in Science and Engineering},
         year = 2007,
        month = may,
       volume = {9},
       number = {3},
        pages = {90-95},
          doi = {10.1109/MCSE.2007.55},
       adsurl = {https://ui.adsabs.harvard.edu/abs/2007CSE.....9...90H}
}

@ARTICLE{numpy,
       author = {{Harris}, Charles R. and {Millman}, K. Jarrod and {van der Walt}, St{\'e}fan J. and {Gommers}, Ralf and {Virtanen}, Pauli and {Cournapeau}, David and {Wieser}, Eric and {Taylor}, Julian and {Berg}, Sebastian and {Smith}, Nathaniel J. and {Kern}, Robert and {Picus}, Matti and {Hoyer}, Stephan and {van Kerkwijk}, Marten H. and {Brett}, Matthew and {Haldane}, Allan and {del R{\'\i}o}, Jaime Fern{\'a}ndez and {Wiebe}, Mark and {Peterson}, Pearu and {G{\'e}rard-Marchant}, Pierre and {Sheppard}, Kevin and {Reddy}, Tyler and {Weckesser}, Warren and {Abbasi}, Hameer and {Gohlke}, Christoph and {Oliphant}, Travis E.},
        title = "{Array programming with NumPy}",
      journal = {\nat},
         year = 2020,
        month = sep,
       volume = {585},
       number = {7825},
        pages = {357-362},
          doi = {10.1038/s41586-020-2649-2},
archivePrefix = {arXiv},
       eprint = {2006.10256},
 primaryClass = {cs.MS},
       adsurl = {https://ui.adsabs.harvard.edu/abs/2020Natur.585..357H}
}

@ARTICLE{scipy,
       author = {{Virtanen}, Pauli and {Gommers}, Ralf and {Oliphant}, Travis E. and {Haberland}, Matt and {Reddy}, Tyler and {Cournapeau}, David and {Burovski}, Evgeni and {Peterson}, Pearu and {Weckesser}, Warren and {Bright}, Jonathan and {van der Walt}, St{\'e}fan J. and {Brett}, Matthew and {Wilson}, Joshua and {Millman}, K. Jarrod and {Mayorov}, Nikolay and {Nelson}, Andrew R.~J. and {Jones}, Eric and {Kern}, Robert and {Larson}, Eric and {Carey}, C.~J. and {Polat}, {\.I}lhan and {Feng}, Yu and {Moore}, Eric W. and {VanderPlas}, Jake and {Laxalde}, Denis and {Perktold}, Josef and {Cimrman}, Robert and {Henriksen}, Ian and {Quintero}, E.~A. and {Harris}, Charles R. and {Archibald}, Anne M. and {Ribeiro}, Ant{\^o}nio H. and {Pedregosa}, Fabian and {van Mulbregt}, Paul and {SciPy 1. 0 Contributors}},
        title = "{SciPy 1.0: fundamental algorithms for scientific computing in Python}",
      journal = {Nature Methods},
         year = 2020,
        month = feb,
       volume = {17},
        pages = {261-272},
          doi = {10.1038/s41592-019-0686-2},
archivePrefix = {arXiv},
       eprint = {1907.10121},
 primaryClass = {cs.MS},
       adsurl = {https://ui.adsabs.harvard.edu/abs/2020NatMe..17..261V}
}

@software{daft,
       author = {{Foreman-Mackey}, Dan and {Hogg}, David W. and {Fulford}, David S. and {Daft-Bot} and {Dobos}, L{\'a}szl{\'o} and {McFee}, Brian and {Murphy}, Kevin P and {Lindemann}, Oliver and {Gerold}, Pierre and {Agrawal}, Varun},
        title = "{daft-dev/daft}",
         year = 2021,
        month = mar,
          eid = {10.5281/zenodo.4615289},
          doi = {10.5281/zenodo.4615289},
      version = {v0.1.2},
    publisher = {Zenodo},
       adsurl = {https://ui.adsabs.harvard.edu/abs/2021zndo...4615289F}
}

@article{tejero-cantero2020sbi,
  doi = {10.21105/joss.02505},
  url = {https://doi.org/10.21105/joss.02505},
  year = {2020},
  publisher = {The Open Journal},
  volume = {5},
  number = {52},
  pages = {2505},
  author = {Alvaro Tejero-Cantero and Jan Boelts and Michael Deistler and Jan-Matthis Lueckmann and Conor Durkan and Pedro J. Gonçalves and David S. Greenberg and Jakob H. Macke},
  title = {sbi: A toolkit for simulation-based inference},
  journal = {Journal of Open Source Software}
}

@article{BoeltsDeistler_sbi_2025,
  doi = {10.21105/joss.07754},
  url = {https://doi.org/10.21105/joss.07754},
  year = {2025},
  publisher = {The Open Journal},
  volume = {10},
  number = {108},
  pages = {7754},
  author = {Jan Boelts and Michael Deistler and Manuel Gloeckler and Álvaro Tejero-Cantero and Jan-Matthis Lueckmann and Guy Moss and Peter Steinbach and Thomas Moreau and Fabio Muratore and Julia Linhart and Conor Durkan and Julius Vetter and Benjamin Kurt Miller and Maternus Herold and Abolfazl Ziaeemehr and Matthijs Pals and Theo Gruner and Sebastian Bischoff and Nastya Krouglova and Richard Gao and Janne K. Lappalainen and Bálint Mucsányi and Felix Pei and Auguste Schulz and Zinovia Stefanidi and Pedro Rodrigues and Cornelius Schröder and Faried Abu Zaid and Jonas Beck and Jaivardhan Kapoor and David S. Greenberg and Pedro J. Gonçalves and Jakob H. Macke},
  title = {sbi reloaded: a toolkit for simulation-based inference workflows},
  journal = {Journal of Open Source Software}
}

@article{Cranmer2020,
author = {Kyle Cranmer  and Johann Brehmer  and Gilles Louppe },
title = {The frontier of simulation-based inference},
journal = {Proceedings of the National Academy of Sciences},
volume = {117},
number = {48},
pages = {30055-30062},
year = {2020},
doi = {10.1073/pnas.1912789117},
URL = {https://www.pnas.org/doi/abs/10.1073/pnas.1912789117},
eprint = {https://www.pnas.org/doi/pdf/10.1073/pnas.1912789117}
}

@ARTICLE{Shuntov2026,
       author = {{Shuntov}, Marko and {Ilbert}, Olivier and {Lagos}, Claudia del P. and {Toft}, Sune and {Valentino}, Francesco and {Mercier}, Wilfried and {Akins}, Hollis B. and {Binh}, Nguyen and {Brinch}, Malte and {Casey}, Caitlin M. and {Franco}, Maximilien and {Gentile}, Fabrizio and {Gozaliasl}, Ghassem and {Haghjoo}, Aryana and {Harish}, Santosh and {Hirschmann}, Michaela and {Huertas-Company}, Marc and {Jin}, Shuowen and {Kartaltepe}, Jeyhan S. and {Koekemoer}, Anton M. and {Laigle}, Clotilde and {Lewis}, Joseph S.~W. and {Magdis}, Georgios E. and {Joy McCracken}, Henry and {Mobasher}, Bahram and {Moutard}, Thibaud and {Oesch}, Pascal A. and {Paquereau}, Louise and {Renzini}, Alvio and {Rich}, Michael R. and {Sanders}, David B. and {Toni}, Greta and {Tresse}, Laurence and {Weibel}, Andrea and {Weaver}, John R. and {Yang}, Lilan},
        title = "{The stellar mass function of quiescent and star-forming galaxies and its dependence on morphology in COSMOS-Web}",
      journal = {\aap},
         year = 2026,
        month = mar,
       volume = {707},
          eid = {A391},
        pages = {A391},
          doi = {10.1051/0004-6361/202558022},
archivePrefix = {arXiv},
       eprint = {2511.05259},
 primaryClass = {astro-ph.GA},
       adsurl = {https://ui.adsabs.harvard.edu/abs/2026A&A...707A.391S}
}

@ARTICLE{Weibel2024,
       author = {{Weibel}, Andrea and {Oesch}, Pascal A. and {Barrufet}, Laia and {Gottumukkala}, Rashmi and {Ellis}, Richard S. and {Santini}, Paola and {Weaver}, John R. and {Allen}, Natalie and {Bouwens}, Rychard and {Bowler}, Rebecca A.~A. and {Brammer}, Gabe and {Carnall}, Adam C. and {Cullen}, Fergus and {Dayal}, Pratika and {Dickinson}, Mark and {Donnan}, Callum T. and {Dunlop}, James S. and {Giavalisco}, Mauro and {Grogin}, Norman A. and {Illingworth}, Garth D. and {Koekemoer}, Anton M. and {Labbe}, Ivo and {Marchesini}, Danilo and {McLeod}, Derek J. and {McLure}, Ross J. and {Naidu}, Rohan P. and {P{\'e}rez-Gonz{\'a}lez}, Pablo G. and {Shuntov}, Marko and {Stefanon}, Mauro and {Toft}, Sune and {Xiao}, Mengyuan},
        title = "{Galaxy build-up in the first 1.5 Gyr of cosmic history: insights from the stellar mass function at z   4-9 from JWST NIRCam observations}",
      journal = {\mnras},
         year = 2024,
        month = sep,
       volume = {533},
       number = {2},
        pages = {1808-1838},
          doi = {10.1093/mnras/stae1891},
archivePrefix = {arXiv},
       eprint = {2403.08872},
 primaryClass = {astro-ph.GA},
       adsurl = {https://ui.adsabs.harvard.edu/abs/2024MNRAS.533.1808W}
}

@ARTICLE{Lian2016,
       author = {{Lian}, Jianhui and {Yan}, Renbin and {Zhang}, Kai and {Kong}, Xu},
        title = "{The Quenching Timescale and Quenching Rate of Galaxies}",
      journal = {\apj},
         year = 2016,
        month = nov,
       volume = {832},
       number = {1},
          eid = {29},
        pages = {29},
          doi = {10.3847/0004-637X/832/1/29},
archivePrefix = {arXiv},
       eprint = {1609.04805},
 primaryClass = {astro-ph.GA},
       adsurl = {https://ui.adsabs.harvard.edu/abs/2016ApJ...832...29L}
}

@ARTICLE{wild2011,
       author = {{Wild}, Vivienne and {Charlot}, St{\'e}phane and {Brinchmann}, Jarle and {Heckman}, Timothy and {Vince}, Oliver and {Pacifici}, Camilla and {Chevallard}, Jacopo},
        title = "{Empirical determination of the shape of dust attenuation curves in star-forming galaxies}",
      journal = {\mnras},
         year = 2011,
        month = nov,
       volume = {417},
       number = {3},
        pages = {1760-1786},
          doi = {10.1111/j.1365-2966.2011.19367.x},
archivePrefix = {arXiv},
       eprint = {1106.1646},
 primaryClass = {astro-ph.CO},
       adsurl = {https://ui.adsabs.harvard.edu/abs/2011MNRAS.417.1760W}
}

@ARTICLE{Price2014,
       author = {{Price}, Sedona H. and {Kriek}, Mariska and {Brammer}, Gabriel B. and {Conroy}, Charlie and {F{\"o}rster Schreiber}, Natascha M. and {Franx}, Marijn and {Fumagalli}, Mattia and {Lundgren}, Britt and {Momcheva}, Ivelina and {Nelson}, Erica J. and {Skelton}, Rosalind E. and {van Dokkum}, Pieter G. and {Whitaker}, Katherine E. and {Wuyts}, Stijn},
        title = "{Direct Measurements of Dust Attenuation in z \raisebox{-0.5ex}\textasciitilde 1.5 Star-forming Galaxies from 3D-HST: Implications for Dust Geometry and Star Formation Rates}",
      journal = {\apj},
         year = 2014,
        month = jun,
       volume = {788},
       number = {1},
          eid = {86},
        pages = {86},
          doi = {10.1088/0004-637X/788/1/86},
archivePrefix = {arXiv},
       eprint = {1310.4177},
 primaryClass = {astro-ph.CO},
       adsurl = {https://ui.adsabs.harvard.edu/abs/2014ApJ...788...86P}
}

@ARTICLE{Chevallard2013,
       author = {{Chevallard}, J. and {Charlot}, S. and {Wandelt}, B. and {Wild}, V.},
        title = "{Insights into the content and spatial distribution of dust from the integrated spectral properties of galaxies}",
      journal = {\mnras},
         year = 2013,
        month = jul,
       volume = {432},
       number = {3},
        pages = {2061-2091},
          doi = {10.1093/mnras/stt523},
archivePrefix = {arXiv},
       eprint = {1303.6631},
 primaryClass = {astro-ph.CO},
       adsurl = {https://ui.adsabs.harvard.edu/abs/2013MNRAS.432.2061C}
}

@ARTICLE{Zheng2022,
       author = {{Zheng}, Yirui and {Dave}, Romeel and {Wild}, Vivienne and {Montero}, Francisco Rodr{\'\i}guez},
        title = "{Rapidly quenched galaxies in the SIMBA cosmological simulation and observations}",
      journal = {\mnras},
         year = 2022,
        month = jun,
       volume = {513},
       number = {1},
        pages = {27-41},
          doi = {10.1093/mnras/stac905},
archivePrefix = {arXiv},
       eprint = {2110.01935},
 primaryClass = {astro-ph.GA},
       adsurl = {https://ui.adsabs.harvard.edu/abs/2022MNRAS.513...27Z}
}

@ARTICLE{RodriguezMontero2019,
       author = {{Rodr{\'\i}guez Montero}, Francisco and {Dav{\'e}}, Romeel and {Wild}, Vivienne and {Angl{\'e}s-Alc{\'a}zar}, Daniel and {Narayanan}, Desika},
        title = "{Mergers, starbursts, and quenching in the SIMBA simulation}",
      journal = {\mnras},
         year = 2019,
        month = dec,
       volume = {490},
       number = {2},
        pages = {2139-2154},
          doi = {10.1093/mnras/stz2580},
archivePrefix = {arXiv},
       eprint = {1907.12680},
 primaryClass = {astro-ph.GA},
       adsurl = {https://ui.adsabs.harvard.edu/abs/2019MNRAS.490.2139R}
}

@ARTICLE{vanderWel2025,
       author = {{van der Wel}, A. and {Martorano}, M. and {Marchesini}, D. and {Wuyts}, S. and {Bell}, E.~F. and {Meidt}, S.~E. and {Gebek}, A. and {Brammer}, G.~B. and {Whitaker}, K.~E. and {Bezanson}, R. and {Nelson}, E.~J. and {Rudnick}, G.~H. and {Kriek}, M. and {Leja}, J. and {Dunlop}, J.~S. and {Casey}, C.~M. and {Kartaltepe}, J.~S.},
        title = "{Even redder than we knew: Color and A$_{V}$ evolution up to z = 2.5 from JWST/NIRCam photometry}",
      journal = {\aap},
         year = 2025,
        month = sep,
       volume = {701},
          eid = {A30},
        pages = {A30},
          doi = {10.1051/0004-6361/202555488},
archivePrefix = {arXiv},
       eprint = {2506.23669},
 primaryClass = {astro-ph.GA},
       adsurl = {https://ui.adsabs.harvard.edu/abs/2025A&A...701A..30V}
}

@ARTICLE{DiMAtteo2005,
       author = {{Di Matteo}, Tiziana and {Springel}, Volker and {Hernquist}, Lars},
        title = "{Energy input from quasars regulates the growth and activity of black holes and their host galaxies}",
      journal = {\nat},
         year = 2005,
        month = feb,
       volume = {433},
       number = {7026},
        pages = {604-607},
          doi = {10.1038/nature03335},
archivePrefix = {arXiv},
       eprint = {astro-ph/0502199},
 primaryClass = {astro-ph},
       adsurl = {https://ui.adsabs.harvard.edu/abs/2005Natur.433..604D}
}

@ARTICLE{Davies2022,
       author = {{Davies}, Jonathan J. and {Pontzen}, Andrew and {Crain}, Robert A.},
        title = "{Galaxy mergers can initiate quenching by unlocking an AGN-driven transformation of the baryon cycle}",
      journal = {\mnras},
         year = 2022,
        month = sep,
       volume = {515},
       number = {1},
        pages = {1430-1443},
          doi = {10.1093/mnras/stac1742},
archivePrefix = {arXiv},
       eprint = {2203.08157},
 primaryClass = {astro-ph.GA},
       adsurl = {https://ui.adsabs.harvard.edu/abs/2022MNRAS.515.1430D}
}

@ARTICLE{Davies2024,
       author = {{Davies}, Jonathan J. and {Pontzen}, Andrew and {Crain}, Robert A.},
        title = "{Are the fates of supermassive black holes and galaxies determined by individual mergers, or by the properties of their host haloes?}",
      journal = {\mnras},
         year = 2024,
        month = jan,
       volume = {527},
       number = {3},
        pages = {4705-4716},
          doi = {10.1093/mnras/stad3456},
archivePrefix = {arXiv},
       eprint = {2301.04145},
 primaryClass = {astro-ph.GA},
       adsurl = {https://ui.adsabs.harvard.edu/abs/2024MNRAS.527.4705D}
}

@ARTICLE{Chabrier2003,
       author = {{Chabrier}, Gilles},
        title = "{Galactic Stellar and Substellar Initial Mass Function}",
      journal = {\pasp},
         year = 2003,
        month = jul,
       volume = {115},
       number = {809},
        pages = {763-795},
          doi = {10.1086/376392},
archivePrefix = {arXiv},
       eprint = {astro-ph/0304382},
 primaryClass = {astro-ph},
       adsurl = {https://ui.adsabs.harvard.edu/abs/2003PASP..115..763C}
}

@ARTICLE{Nanayakkara2025,
       author = {{Nanayakkara}, Themiya and {Glazebrook}, Karl and {Schreiber}, Corentin and {Chittenden}, Harry and {Brammer}, Gabriel and {Esdaile}, James and {Jacobs}, Colin and {Kacprzak}, Glenn G. and {Kawinwanichakij}, Lalitwadee and {Kimmig}, Lucas C. and {Labbe}, Ivo and {Lagos}, Claudia and {Marchesini}, Danilo and {Mart{\`\i}nez-Mar{\`\i}n}, M. and {Marsan}, Z. Cemile and {Oesch}, Pascal A. and {Papovich}, Casey and {Remus}, Rhea-Silvia and {Tran}, Kim-Vy H.},
        title = "{The Formation Histories of Massive and Quiescent Galaxies in the 3 < z < 4.5 Universe}",
      journal = {\apj},
         year = 2025,
        month = mar,
       volume = {981},
       number = {1},
          eid = {78},
        pages = {78},
          doi = {10.3847/1538-4357/ada6ac},
archivePrefix = {arXiv},
       eprint = {2410.02076},
 primaryClass = {astro-ph.GA},
       adsurl = {https://ui.adsabs.harvard.edu/abs/2025ApJ...981...78N}
}

@ARTICLE{wild2020,
       author = {{Wild}, Vivienne and {Taj Aldeen}, Laith and {Carnall}, Adam and {Maltby}, David and {Almaini}, Omar and {Werle}, Ariel and {Wilkinson}, Aaron and {Rowlands}, Kate and {Bolzonella}, Micol and {Castellano}, Marco and {Gargiulo}, Adriana and {McLure}, Ross and {Pentericci}, Laura and {Pozzetti}, Lucia},
        title = "{The star formation histories of z {\textasciitilde} 1 post-starburst galaxies}",
      journal = {\mnras},
         year = 2020,
        month = may,
       volume = {494},
       number = {1},
        pages = {529-548},
          doi = {10.1093/mnras/staa674},
archivePrefix = {arXiv},
       eprint = {2001.09154},
 primaryClass = {astro-ph.GA},
       adsurl = {https://ui.adsabs.harvard.edu/abs/2020MNRAS.494..529W}
}

@ARTICLE{Zheng2020,
       author = {{Zheng}, Yirui and {Wild}, Vivienne and {Lah{\'e}n}, Natalia and {Johansson}, Peter H. and {Law}, David and {Weaver}, John R. and {Jimenez}, Noelia},
        title = "{Comparison of stellar populations in simulated and real post-starburst galaxies in MaNGA}",
      journal = {\mnras},
         year = 2020,
        month = oct,
       volume = {498},
       number = {1},
        pages = {1259-1277},
          doi = {10.1093/mnras/staa2358},
archivePrefix = {arXiv},
       eprint = {2005.14112},
 primaryClass = {astro-ph.GA},
       adsurl = {https://ui.adsabs.harvard.edu/abs/2020MNRAS.498.1259Z}
}

@ARTICLE{leja2019,
       author = {{Leja}, Joel and {Carnall}, Adam C. and {Johnson}, Benjamin D. and {Conroy}, Charlie and {Speagle}, Joshua S.},
        title = "{How to Measure Galaxy Star Formation Histories. II. Nonparametric Models}",
      journal = {\apj},
         year = 2019,
        month = may,
       volume = {876},
       number = {1},
          eid = {3},
        pages = {3},
          doi = {10.3847/1538-4357/ab133c},
archivePrefix = {arXiv},
       eprint = {1811.03637},
 primaryClass = {astro-ph.GA},
       adsurl = {https://ui.adsabs.harvard.edu/abs/2019ApJ...876....3L}
}

@ARTICLE{bruzual2003,
       author = {{Bruzual}, G. and {Charlot}, S.},
        title = "{Stellar population synthesis at the resolution of 2003}",
      journal = {\mnras},
         year = 2003,
        month = oct,
       volume = {344},
       number = {4},
        pages = {1000-1028},
          doi = {10.1046/j.1365-8711.2003.06897.x},
archivePrefix = {arXiv},
       eprint = {astro-ph/0309134},
 primaryClass = {astro-ph},
       adsurl = {https://ui.adsabs.harvard.edu/abs/2003MNRAS.344.1000B}
}

@ARTICLE{nelson2018,
       author = {{Nelson}, Dylan and {Pillepich}, Annalisa and {Springel}, Volker and {Weinberger}, Rainer and {Hernquist}, Lars and {Pakmor}, R{\"u}diger and {Genel}, Shy and {Torrey}, Paul and {Vogelsberger}, Mark and {Kauffmann}, Guinevere and {Marinacci}, Federico and {Naiman}, Jill},
        title = "{First results from the IllustrisTNG simulations: the galaxy colour bimodality}",
      journal = {\mnras},
         year = 2018,
        month = mar,
       volume = {475},
       number = {1},
        pages = {624-647},
          doi = {10.1093/mnras/stx3040},
archivePrefix = {arXiv},
       eprint = {1707.03395},
 primaryClass = {astro-ph.GA},
       adsurl = {https://ui.adsabs.harvard.edu/abs/2018MNRAS.475..624N}
}

@ARTICLE{carnall2018,
       author = {{Carnall}, A.~C. and {McLure}, R.~J. and {Dunlop}, J.~S. and {Dav{\'e}}, R.},
        title = "{Inferring the star formation histories of massive quiescent galaxies with BAGPIPES: evidence for multiple quenching mechanisms}",
      journal = {\mnras},
         year = 2018,
        month = nov,
       volume = {480},
       number = {4},
        pages = {4379-4401},
          doi = {10.1093/mnras/sty2169},
archivePrefix = {arXiv},
       eprint = {1712.04452},
 primaryClass = {astro-ph.GA},
       adsurl = {https://ui.adsabs.harvard.edu/abs/2018MNRAS.480.4379C}
}

@ARTICLE{CarnallMcLure2019,
       author = {{Carnall}, A.~C. and {McLure}, R.~J. and {Dunlop}, J.~S. and {Cullen}, F. and {McLeod}, D.~J. and {Wild}, V. and {Johnson}, B.~D. and {Appleby}, S. and {Dav{\'e}}, R. and {Amorin}, R. and {Bolzonella}, M. and {Castellano}, M. and {Cimatti}, A. and {Cucciati}, O. and {Gargiulo}, A. and {Garilli}, B. and {Marchi}, F. and {Pentericci}, L. and {Pozzetti}, L. and {Schreiber}, C. and {Talia}, M. and {Zamorani}, G.},
        title = "{The VANDELS survey: the star-formation histories of massive quiescent galaxies at 1.0 < z < 1.3}",
      journal = {\mnras},
         year = 2019,
        month = nov,
       volume = {490},
       number = {1},
        pages = {417-439},
          doi = {10.1093/mnras/stz2544},
archivePrefix = {arXiv},
       eprint = {1903.11082},
 primaryClass = {astro-ph.GA},
       adsurl = {https://ui.adsabs.harvard.edu/abs/2019MNRAS.490..417C}
}

@ARTICLE{CarnallLeja2019,
       author = {{Carnall}, Adam C. and {Leja}, Joel and {Johnson}, Benjamin D. and {McLure}, Ross J. and {Dunlop}, James S. and {Conroy}, Charlie},
        title = "{How to Measure Galaxy Star Formation Histories. I. Parametric Models}",
      journal = {\apj},
         year = 2019,
        month = mar,
       volume = {873},
       number = {1},
          eid = {44},
        pages = {44},
          doi = {10.3847/1538-4357/ab04a2},
archivePrefix = {arXiv},
       eprint = {1811.03635},
 primaryClass = {astro-ph.GA},
       adsurl = {https://ui.adsabs.harvard.edu/abs/2019ApJ...873...44C}
}

@ARTICLE{wild2016,
       author = {{Wild}, Vivienne and {Almaini}, Omar and {Dunlop}, Jim and {Simpson}, Chris and {Rowlands}, Kate and {Bowler}, Rebecca and {Maltby}, David and {McLure}, Ross},
        title = "{The evolution of post-starburst galaxies from z=2 to 0.5}",
      journal = {\mnras},
         year = 2016,
        month = nov,
       volume = {463},
       number = {1},
        pages = {832-844},
          doi = {10.1093/mnras/stw1996},
archivePrefix = {arXiv},
       eprint = {1608.00588},
 primaryClass = {astro-ph.GA},
       adsurl = {https://ui.adsabs.harvard.edu/abs/2016MNRAS.463..832W}
}

@ARTICLE{Whitaker2011,
       author = {{Whitaker}, Katherine E. and {Labb{\'e}}, Ivo and {van Dokkum}, Pieter G. and {Brammer}, Gabriel and {Kriek}, Mariska and {Marchesini}, Danilo and {Quadri}, Ryan F. and {Franx}, Marijn and {Muzzin}, Adam and {Williams}, Rik J. and {Bezanson}, Rachel and {Illingworth}, Garth D. and {Lee}, Kyoung-Soo and {Lundgren}, Britt and {Nelson}, Erica J. and {Rudnick}, Gregory and {Tal}, Tomer and {Wake}, David A.},
        title = "{The NEWFIRM Medium-band Survey: Photometric Catalogs, Redshifts, and the Bimodal Color Distribution of Galaxies out to z \raisebox{-0.5ex}\textasciitilde 3}",
      journal = {\apj},
         year = 2011,
        month = jul,
       volume = {735},
       number = {2},
          eid = {86},
        pages = {86},
          doi = {10.1088/0004-637X/735/2/86},
archivePrefix = {arXiv},
       eprint = {1105.4609},
 primaryClass = {astro-ph.CO},
       adsurl = {https://ui.adsabs.harvard.edu/abs/2011ApJ...735...86W}
}

@ARTICLE{belli2019,
       author = {{Belli}, Sirio and {Newman}, Andrew B. and {Ellis}, Richard S.},
        title = "{MOSFIRE Spectroscopy of Quiescent Galaxies at 1.5 < z < 2.5. II. Star Formation Histories and Galaxy Quenching}",
      journal = {\apj},
         year = 2019,
        month = mar,
       volume = {874},
       number = {1},
          eid = {17},
        pages = {17},
          doi = {10.3847/1538-4357/ab07af},
archivePrefix = {arXiv},
       eprint = {1810.00008},
 primaryClass = {astro-ph.GA},
       adsurl = {https://ui.adsabs.harvard.edu/abs/2019ApJ...874...17B}
}

@ARTICLE{stanway2018,
       author = {{Stanway}, E.~R. and {Eldridge}, J.~J.},
        title = "{Re-evaluating old stellar populations}",
      journal = {\mnras},
         year = 2018,
        month = sep,
       volume = {479},
       number = {1},
        pages = {75-93},
          doi = {10.1093/mnras/sty1353},
archivePrefix = {arXiv},
       eprint = {1805.08784},
 primaryClass = {astro-ph.GA},
       adsurl = {https://ui.adsabs.harvard.edu/abs/2018MNRAS.479...75S}
}

@ARTICLE{HarveyLovell2026,
       author = {{Harvey}, Thomas and {Lovell}, Christopher C. and {Newman}, Sophie and {Conselice}, Christopher J. and {Austin}, Duncan and {Roper}, William J. and {Vijayan}, Aswin P. and {Wilkins}, Stephen M. and {Iglesias-Navarro}, Patricia and {Rusakov}, Vadim and {Li}, Qiong and {Adams}, Nathan and {Magdwick}, Kai and {Goolsby}, Caio M. and {Huertas-Company}, Marc and {Ho}, Matthew},
        title = "{Flexible simulation-based inference for galaxy photometric fitting with synthesizer}",
      journal = {\mnras},
         year = 2026,
        month = mar,
       volume = {547},
       number = {1},
          eid = {stag282},
        pages = {stag282},
          doi = {10.1093/mnras/stag282},
archivePrefix = {arXiv},
       eprint = {2511.10640},
 primaryClass = {astro-ph.GA},
       adsurl = {https://ui.adsabs.harvard.edu/abs/2026MNRAS.547ag282H}
}

@ARTICLE{aufort2025,
       author = {{Aufort}, G. and {Laigle}, C. and {McCracken}, H.~J. and {Le Borgne}, D. and {Arango-Toro}, R. and {Ciesla}, L. and {Ilbert}, O. and {Tresse}, L. and {Dubois}, Y.},
        title = "{Reconstructing galaxy star formation histories from COSMOS2020 photometry using simulation-based inference}",
      journal = {\aap},
         year = 2025,
        month = jul,
       volume = {699},
          eid = {A328},
        pages = {A328},
          doi = {10.1051/0004-6361/202452457},
archivePrefix = {arXiv},
       eprint = {2410.00795},
 primaryClass = {astro-ph.GA},
       adsurl = {https://ui.adsabs.harvard.edu/abs/2025A&A...699A.328A}
}

@ARTICLE{Gould2023,
       author = {{Gould}, Katriona M.~L. and {Brammer}, Gabriel and {Valentino}, Francesco and {Whitaker}, Katherine E. and {Weaver}, John. R. and {Lagos}, Claudia del P. and {Rizzo}, Francesca and {Franco}, Maximilien and {Hsieh}, Bau-Ching and {Ilbert}, Olivier and {Jin}, Shuowen and {Magdis}, Georgios and {McCracken}, Henry J. and {Mobasher}, Bahram and {Shuntov}, Marko and {Steinhardt}, Charles L. and {Strait}, Victoria and {Toft}, Sune},
        title = "{COSMOS2020: Exploring the Dawn of Quenching for Massive Galaxies at 3 < z < 5 with a New Color-selection Method}",
      journal = {\aj},
         year = 2023,
        month = jun,
       volume = {165},
       number = {6},
          eid = {248},
        pages = {248},
          doi = {10.3847/1538-3881/accadc},
archivePrefix = {arXiv},
       eprint = {2302.10934},
 primaryClass = {astro-ph.GA},
       adsurl = {https://ui.adsabs.harvard.edu/abs/2023AJ....165..248G}
}

@ARTICLE{park2024,
       author = {{Park}, Minjung and {Belli}, Sirio and {Conroy}, Charlie and {Johnson}, Benjamin D. and {Davies}, Rebecca L. and {Leja}, Joel and {Tacchella}, Sandro and {Mendel}, J. Trevor and {Benton}, Chlo{\"e} and {Bugiani}, Letizia and {Emami}, Razieh and {Khoram}, Amir H. and {Li}, Yijia and {Maheson}, Gabriel and {Mathews}, Elijah P. and {Naidu}, Rohan P. and {Nelson}, Erica J. and {Terrazas}, Bryan A. and {Weinberger}, Rainer},
        title = "{Widespread Rapid Quenching at Cosmic Noon Revealed by JWST Deep Spectroscopy}",
      journal = {\apj},
         year = 2024,
        month = nov,
       volume = {976},
       number = {1},
          eid = {72},
        pages = {72},
          doi = {10.3847/1538-4357/ad7e15},
archivePrefix = {arXiv},
       eprint = {2404.17945},
 primaryClass = {astro-ph.GA},
       adsurl = {https://ui.adsabs.harvard.edu/abs/2024ApJ...976...72P}
}

@ARTICLE{Khoram2026,
       author = {{Khoram}, Amir H. and {Belli}, Sirio and {Nipoti}, Carlo and {Pascale}, Raffaele and {Newman}, Andrew B. and {Marinacci}, Federico and {Ellis}, Richard S. and {Bugiani}, Letizia and {Sapori}, Matteo and {Giunchi}, Eric},
        title = "{Death by Impact: Evidence for Merger-driven Quenching in a Collisional Ring Galaxy at Cosmic Noon}",
      journal = {\apj},
         year = 2026,
        month = feb,
       volume = {998},
       number = {1},
          eid = {59},
        pages = {59},
          doi = {10.3847/1538-4357/ae23d0},
archivePrefix = {arXiv},
       eprint = {2509.12308},
 primaryClass = {astro-ph.GA},
       adsurl = {https://ui.adsabs.harvard.edu/abs/2026ApJ...998...59K}
}

@ARTICLE{Bugiani2025,
       author = {{Bugiani}, Letizia and {Belli}, Sirio and {Park}, Minjung and {Davies}, Rebecca L. and {Mendel}, J. Trevor and {Johnson}, Benjamin D. and {Khoram}, Amir H. and {Benton}, Chlo{\"e} and {Cimatti}, Andrea and {Conroy}, Charlie and {Emami}, Razieh and {Leja}, Joel and {Li}, Yijia and {Maheson}, Gabriel and {Mathews}, Elijah P. and {Naidu}, Rohan P. and {Nelson}, Erica J. and {Tacchella}, Sandro and {Terrazas}, Bryan A. and {Weinberger}, Rainer},
        title = "{Active Galactic Nucleus Feedback in Quiescent Galaxies at Cosmic Noon Traced by Ionized Gas Emission}",
      journal = {\apj},
         year = 2025,
        month = mar,
       volume = {981},
       number = {1},
          eid = {25},
        pages = {25},
          doi = {10.3847/1538-4357/adaeaf},
archivePrefix = {arXiv},
       eprint = {2406.08547},
 primaryClass = {astro-ph.GA},
       adsurl = {https://ui.adsabs.harvard.edu/abs/2025ApJ...981...25B}
}

@ARTICLE{Lin2026,
       author = {{Lin}, Lihwai and {Wu}, Po-Feng and {Thorp}, Mallory D. and {Bluck}, Asa F.~L. and {Pan}, Hsi-An and {Ellison}, Sara L. and {Rowlands}, Kate and {Otter}, Justin Atsushi and {S{\'a}nchez}, Sebasti{\'a}n F.},
        title = "{The ALMaQUEST Survey. XVII. Unveiling Multiple Quenching Pathways in Green Valley Galaxies via Molecular Gas and Quenching Timescale Analyses}",
      journal = {\apj},
         year = 2026,
        month = mar,
       volume = {999},
       number = {2},
          eid = {263},
        pages = {263},
          doi = {10.3847/1538-4357/ae3b2b},
archivePrefix = {arXiv},
       eprint = {2601.09225},
 primaryClass = {astro-ph.GA},
       adsurl = {https://ui.adsabs.harvard.edu/abs/2026ApJ...999..263L}
}

@ARTICLE{conroy2009,
       author = {{Conroy}, Charlie and {Gunn}, James E. and {White}, Martin},
        title = "{The Propagation of Uncertainties in Stellar Population Synthesis Modeling. I. The Relevance of Uncertain Aspects of Stellar Evolution and the Initial Mass Function to the Derived Physical Properties of Galaxies}",
      journal = {\apj},
         year = 2009,
        month = jul,
       volume = {699},
       number = {1},
        pages = {486-506},
          doi = {10.1088/0004-637X/699/1/486},
archivePrefix = {arXiv},
       eprint = {0809.4261},
 primaryClass = {astro-ph},
       adsurl = {https://ui.adsabs.harvard.edu/abs/2009ApJ...699..486C}
}

@ARTICLE{strateva2001,
       author = {{Strateva}, Iskra and {Ivezi{\'c}}, {\v{Z}}eljko and {Knapp}, Gillian R. and {Narayanan}, Vijay K. and {Strauss}, Michael A. and {Gunn}, James E. and {Lupton}, Robert H. and {Schlegel}, David and {Bahcall}, Neta A. and {Brinkmann}, Jon and {Brunner}, Robert J. and {Budav{\'a}ri}, Tam{\'a}s and {Csabai}, Istv{\'a}n and {Castander}, Francisco Javier and {Doi}, Mamoru and {Fukugita}, Masataka and {Gy{\H{o}}ry}, Zsuzsanna and {Hamabe}, Masaru and {Hennessy}, Greg and {Ichikawa}, Takashi and {Kunszt}, Peter Z. and {Lamb}, Don Q. and {McKay}, Timothy A. and {Okamura}, Sadanori and {Racusin}, Judith and {Sekiguchi}, Maki and {Schneider}, Donald P. and {Shimasaku}, Kazuhiro and {York}, Donald},
        title = "{Color Separation of Galaxy Types in the Sloan Digital Sky Survey Imaging Data}",
      journal = {\aj},
         year = 2001,
        month = oct,
       volume = {122},
       number = {4},
        pages = {1861-1874},
          doi = {10.1086/323301},
archivePrefix = {arXiv},
       eprint = {astro-ph/0107201},
 primaryClass = {astro-ph},
       adsurl = {https://ui.adsabs.harvard.edu/abs/2001AJ....122.1861S}
}

@ARTICLE{baldry2004,
       author = {{Baldry}, Ivan K. and {Glazebrook}, Karl and {Brinkmann}, Jon and {Ivezi{\'c}}, {\v{Z}}eljko and {Lupton}, Robert H. and {Nichol}, Robert C. and {Szalay}, Alexander S.},
        title = "{Quantifying the Bimodal Color-Magnitude Distribution of Galaxies}",
      journal = {\apj},
         year = 2004,
        month = jan,
       volume = {600},
       number = {2},
        pages = {681-694},
          doi = {10.1086/380092},
archivePrefix = {arXiv},
       eprint = {astro-ph/0309710},
 primaryClass = {astro-ph},
       adsurl = {https://ui.adsabs.harvard.edu/abs/2004ApJ...600..681B}
}

@ARTICLE{bell2004,
       author = {{Bell}, Eric F. and {Wolf}, Christian and {Meisenheimer}, Klaus and {Rix}, Hans-Walter and {Borch}, Andrea and {Dye}, Simon and {Kleinheinrich}, Martina and {Wisotzki}, Lutz and {McIntosh}, Daniel H.},
        title = "{Nearly 5000 Distant Early-Type Galaxies in COMBO-17: A Red Sequence and Its Evolution since z\raisebox{-0.5ex}\textasciitilde1}",
      journal = {\apj},
         year = 2004,
        month = jun,
       volume = {608},
       number = {2},
        pages = {752-767},
          doi = {10.1086/420778},
archivePrefix = {arXiv},
       eprint = {astro-ph/0303394},
 primaryClass = {astro-ph},
       adsurl = {https://ui.adsabs.harvard.edu/abs/2004ApJ...608..752B}
}

@ARTICLE{kauffmann2003,
       author = {{Kauffmann}, Guinevere and {Heckman}, Timothy M. and {White}, Simon D.~M. and {Charlot}, St{\'e}phane and {Tremonti}, Christy and {Peng}, Eric W. and {Seibert}, Mark and {Brinkmann}, Jon and {Nichol}, Robert C. and {SubbaRao}, Mark and {York}, Don},
        title = "{The dependence of star formation history and internal structure on stellar mass for {}10$^{5}$ low-redshift galaxies}",
      journal = {\mnras},
         year = 2003,
        month = may,
       volume = {341},
       number = {1},
        pages = {54-69},
          doi = {10.1046/j.1365-8711.2003.06292.x},
archivePrefix = {arXiv},
       eprint = {astro-ph/0205070},
 primaryClass = {astro-ph},
       adsurl = {https://ui.adsabs.harvard.edu/abs/2003MNRAS.341...54K}
}

@ARTICLE{Hahn2017,
       author = {{Hahn}, ChangHoon and {Tinker}, Jeremy L. and {Wetzel}, Andrew},
        title = "{Star Formation Quenching Timescale of Central Galaxies in a Hierarchical Universe}",
      journal = {\apj},
         year = 2017,
        month = may,
       volume = {841},
       number = {1},
          eid = {6},
        pages = {6},
          doi = {10.3847/1538-4357/aa6d6b},
archivePrefix = {arXiv},
       eprint = {1609.04398},
 primaryClass = {astro-ph.GA},
       adsurl = {https://ui.adsabs.harvard.edu/abs/2017ApJ...841....6H}
}

@ARTICLE{rowlands2015,
       author = {{Rowlands}, K. and {Wild}, V. and {Nesvadba}, N. and {Sibthorpe}, B. and {Mortier}, A. and {Lehnert}, M. and {da Cunha}, E.},
        title = "{The evolution of the cold interstellar medium in galaxies following a starburst}",
      journal = {\mnras},
         year = 2015,
        month = mar,
       volume = {448},
       number = {1},
        pages = {258-279},
          doi = {10.1093/mnras/stu2714},
archivePrefix = {arXiv},
       eprint = {1412.6090},
 primaryClass = {astro-ph.GA},
       adsurl = {https://ui.adsabs.harvard.edu/abs/2015MNRAS.448..258R}
}

@ARTICLE{almaini2017,
       author = {{Almaini}, Omar and {Wild}, Vivienne and {Maltby}, David T. and {Hartley}, William G. and {Simpson}, Chris and {Hatch}, Nina A. and {McLure}, Ross J. and {Dunlop}, James S. and {Rowlands}, Kate},
        title = "{Massive post-starburst galaxies at z > 1 are compact proto-spheroids}",
      journal = {\mnras},
         year = 2017,
        month = dec,
       volume = {472},
       number = {2},
        pages = {1401-1412},
          doi = {10.1093/mnras/stx1957},
archivePrefix = {arXiv},
       eprint = {1708.00005},
 primaryClass = {astro-ph.GA},
       adsurl = {https://ui.adsabs.harvard.edu/abs/2017MNRAS.472.1401A}
}

@ARTICLE{Maltby2026,
       author = {{Maltby}, David T. and {Almaini}, Omar and {Wild}, Vivienne and {Taylor}, Elizabeth and {Rowlands}, Kate and {de Lisle}, Thomas and {Hewitt}, Guillaume and {Skarbinski}, Maya and {Dunlop}, James S. and {Carnall}, Adam C. and {Koekemoer}, Anton M. and {McLeod}, Derek J.},
        title = "{The multiwavelength structure of post-starburst galaxies at 0.5 < z < 3 with JWST PRIMER: compact morphologies and residual disturbances}",
      journal = {\mnras},
         year = 2026,
        month = jul,
       volume = {550},
       number = {1},
          eid = {stag987},
        pages = {stag987},
          doi = {10.1093/mnras/stag987},
archivePrefix = {arXiv},
       eprint = {2607.00085},
 primaryClass = {astro-ph.GA},
       adsurl = {https://ui.adsabs.harvard.edu/abs/2026MNRAS.550ag987M}
}

@ARTICLE{Pacifici2016,
       author = {{Pacifici}, Camilla and {Kassin}, Susan A. and {Weiner}, Benjamin J. and {Holden}, Bradford and {Gardner}, Jonathan P. and {Faber}, Sandra M. and {Ferguson}, Henry C. and {Koo}, David C. and {Primack}, Joel R. and {Bell}, Eric F. and {Dekel}, Avishai and {Gawiser}, Eric and {Giavalisco}, Mauro and {Rafelski}, Marc and {Simons}, Raymond C. and {Barro}, Guillermo and {Croton}, Darren J. and {Dav{\'e}}, Romeel and {Fontana}, Adriano and {Grogin}, Norman A. and {Koekemoer}, Anton M. and {Lee}, Seong-Kook and {Salmon}, Brett and {Somerville}, Rachel and {Behroozi}, Peter},
        title = "{The Evolution of Star Formation Histories of Quiescent Galaxies}",
      journal = {\apj},
         year = 2016,
        month = nov,
       volume = {832},
       number = {1},
          eid = {79},
        pages = {79},
          doi = {10.3847/0004-637X/832/1/79},
archivePrefix = {arXiv},
       eprint = {1609.03572},
 primaryClass = {astro-ph.GA},
       adsurl = {https://ui.adsabs.harvard.edu/abs/2016ApJ...832...79P}
}

@ARTICLE{plat2019,
       author = {{Plat}, A. and {Charlot}, S. and {Bruzual}, G. and {Feltre}, A. and {Vidal-Garc{\'\i}a}, A. and {Morisset}, C. and {Chevallard}, J. and {Todt}, H.},
        title = "{Constraints on the production and escape of ionizing radiation from the emission-line spectra of metal-poor star-forming galaxies}",
      journal = {\mnras},
         year = 2019,
        month = nov,
       volume = {490},
       number = {1},
        pages = {978-1009},
          doi = {10.1093/mnras/stz2616},
archivePrefix = {arXiv},
       eprint = {1909.07386},
 primaryClass = {astro-ph.GA},
       adsurl = {https://ui.adsabs.harvard.edu/abs/2019MNRAS.490..978P}
}

@ARTICLE{calzetti2000,
       author = {{Calzetti}, Daniela and {Armus}, Lee and {Bohlin}, Ralph C. and {Kinney}, Anne L. and {Koornneef}, Jan and {Storchi-Bergmann}, Thaisa},
        title = "{The Dust Content and Opacity of Actively Star-forming Galaxies}",
      journal = {\apj},
         year = 2000,
        month = apr,
       volume = {533},
       number = {2},
        pages = {682-695},
          doi = {10.1086/308692},
archivePrefix = {arXiv},
       eprint = {astro-ph/9911459},
 primaryClass = {astro-ph},
       adsurl = {https://ui.adsabs.harvard.edu/abs/2000ApJ...533..682C}
}

@ARTICLE{leung2024,
       author = {{Leung}, Ho-Hin and {Wild}, Vivienne and {Papathomas}, Michail and {Carnall}, Adam and {Zheng}, Yirui and {Boardman}, Nicholas and {Wang}, Cara and {Johansson}, Peter H.},
        title = "{Chemical evolution of local post-starburst galaxies: implications for the mass-metallicity relation}",
      journal = {\mnras},
         year = 2024,
        month = mar,
       volume = {528},
       number = {3},
        pages = {4029-4052},
          doi = {10.1093/mnras/stae225},
archivePrefix = {arXiv},
       eprint = {2309.16626},
 primaryClass = {astro-ph.GA},
       adsurl = {https://ui.adsabs.harvard.edu/abs/2024MNRAS.528.4029L}
}

@ARTICLE{sanchez2022,
       author = {{S{\'a}nchez}, S.~F. and {Barrera-Ballesteros}, J.~K. and {Lacerda}, E. and {Mej{\'\i}a-Narvaez}, A. and {Camps-Fari{\~n}a}, A. and {Bruzual}, Gustavo and {Espinosa-Ponce}, C. and {Rodr{\'\i}guez-Puebla}, A. and {Calette}, A.~R. and {Ibarra-Medel}, H. and {Avila-Reese}, V. and {Hernandez-Toledo}, H. and {Bershady}, M.~A. and {Cano-Diaz}, M. and {Munguia-Cordova}, A.~M.},
        title = "{SDSS-IV MaNGA: pyPipe3D Analysis Release for 10,000 Galaxies}",
      journal = {\apjs},
         year = 2022,
        month = oct,
       volume = {262},
       number = {2},
          eid = {36},
        pages = {36},
          doi = {10.3847/1538-4365/ac7b8f},
archivePrefix = {arXiv},
       eprint = {2206.07062},
 primaryClass = {astro-ph.GA},
       adsurl = {https://ui.adsabs.harvard.edu/abs/2022ApJS..262...36S}
}

@ARTICLE{maltby2016,
       author = {{Maltby}, David T. and {Almaini}, Omar and {Wild}, Vivienne and {Hatch}, Nina A. and {Hartley}, William G. and {Simpson}, Chris and {McLure}, Ross J. and {Dunlop}, James and {Rowlands}, Kate and {Cirasuolo}, Michele},
        title = "{The identification of post-starburst galaxies at z {\ensuremath{\sim}} 1 using multiwavelength photometry: a spectroscopic verification}",
      journal = {\mnras},
         year = 2016,
        month = jun,
       volume = {459},
       number = {1},
        pages = {L114-L118},
          doi = {10.1093/mnrasl/slw057},
archivePrefix = {arXiv},
       eprint = {1603.08941},
 primaryClass = {astro-ph.GA},
       adsurl = {https://ui.adsabs.harvard.edu/abs/2016MNRAS.459L.114M}
}

@ARTICLE{hopkins2008,
       author = {{Hopkins}, Philip F. and {Hernquist}, Lars and {Cox}, Thomas J. and {Dutta}, Suvendra N. and {Rothberg}, Barry},
        title = "{Dissipation and Extra Light in Galactic Nuclei. I. Gas-Rich Merger Remnants}",
      journal = {\apj},
         year = 2008,
        month = may,
       volume = {679},
       number = {1},
        pages = {156-181},
          doi = {10.1086/587544},
archivePrefix = {arXiv},
       eprint = {0802.0508},
 primaryClass = {astro-ph},
       adsurl = {https://ui.adsabs.harvard.edu/abs/2008ApJ...679..156H}
}

@ARTICLE{Remus2025,
       author = {{Remus}, Rhea-Silvia and {Kimmig}, Lucas C.},
        title = "{Relight the Candle: What Happens to High-redshift Massive Quenched Galaxies}",
      journal = {\apj},
         year = 2025,
        month = mar,
       volume = {982},
       number = {1},
          eid = {30},
        pages = {30},
          doi = {10.3847/1538-4357/ad8b4b},
archivePrefix = {arXiv},
       eprint = {2310.16089},
 primaryClass = {astro-ph.GA},
       adsurl = {https://ui.adsabs.harvard.edu/abs/2025ApJ...982...30R}
}

@ARTICLE{Forrest2020,
       author = {{Forrest}, Ben and {Marsan}, Z. Cemile and {Annunziatella}, Marianna and {Wilson}, Gillian and {Muzzin}, Adam and {Marchesini}, Danilo and {Cooper}, M.~C. and {Chan}, Jeffrey C.~C. and {McConachie}, Ian and {Gomez}, Percy and {Kado-Fong}, Erin and {La Barbera}, Francesco and {Lange-Vagle}, Daniel and {Nantais}, Julie and {Nonino}, Mario and {Saracco}, Paolo and {Stefanon}, Mauro and {van der Burg}, Remco F.~J.},
        title = "{The Massive Ancient Galaxies at z > 3 NEar-infrared (MAGAZ3NE) Survey: Confirmation of Extremely Rapid Star Formation and Quenching Timescales for Massive Galaxies in the Early Universe}",
      journal = {\apj},
         year = 2020,
        month = nov,
       volume = {903},
       number = {1},
          eid = {47},
        pages = {47},
          doi = {10.3847/1538-4357/abb819},
archivePrefix = {arXiv},
       eprint = {2009.07281},
 primaryClass = {astro-ph.GA},
       adsurl = {https://ui.adsabs.harvard.edu/abs/2020ApJ...903...47F}
}

@ARTICLE{lawrence2007,
       author = {{Lawrence}, A. and {Warren}, S.~J. and {Almaini}, O. and {Edge}, A.~C. and {Hambly}, N.~C. and {Jameson}, R.~F. and {Lucas}, P. and {Casali}, M. and {Adamson}, A. and {Dye}, S. and {Emerson}, J.~P. and {Foucaud}, S. and {Hewett}, P. and {Hirst}, P. and {Hodgkin}, S.~T. and {Irwin}, M.~J. and {Lodieu}, N. and {McMahon}, R.~G. and {Simpson}, C. and {Smail}, I. and {Mortlock}, D. and {Folger}, M.},
        title = "{The UKIRT Infrared Deep Sky Survey (UKIDSS)}",
      journal = {\mnras},
         year = 2007,
        month = aug,
       volume = {379},
       number = {4},
        pages = {1599-1617},
          doi = {10.1111/j.1365-2966.2007.12040.x},
archivePrefix = {arXiv},
       eprint = {astro-ph/0604426},
 primaryClass = {astro-ph},
       adsurl = {https://ui.adsabs.harvard.edu/abs/2007MNRAS.379.1599L}
}

@ARTICLE{Furusawa2008,
       author = {{Furusawa}, Hisanori and {Kosugi}, George and {Akiyama}, Masayuki and {Takata}, Tadafumi and {Sekiguchi}, Kazuhiro and {Tanaka}, Ichi and {Iwata}, Ikuru and {Kajisawa}, Masaru and {Yasuda}, Naoki and {Doi}, Mamoru and {Ouchi}, Masami and {Simpson}, Chris and {Shimasaku}, Kazuhiro and {Yamada}, Toru and {Furusawa}, Junko and {Morokuma}, Tomoki and {Ishida}, Catherine M. and {Aoki}, Kentaro and {Fuse}, Tetsuharu and {Imanishi}, Masatoshi and {Iye}, Masanori and {Karoji}, Hiroshi and {Kobayashi}, Naoto and {Kodama}, Tadayuki and {Komiyama}, Yutaka and {Maeda}, Yoshitomo and {Miyazaki}, Satoshi and {Mizumoto}, Yoshihiko and {Nakata}, Fumiaki and {Noumaru}, Jun'ichi and {Ogasawara}, Ryusuke and {Okamura}, Sadanori and {Saito}, Tomoki and {Sasaki}, Toshiyuki and {Ueda}, Yoshihiro and {Yoshida}, Michitoshi},
        title = "{The Subaru/XMM-Newton Deep Survey (SXDS). II. Optical Imaging and Photometric Catalogs}",
      journal = {\apjs},
         year = 2008,
        month = may,
       volume = {176},
       number = {1},
        pages = {1-18},
          doi = {10.1086/527321},
archivePrefix = {arXiv},
       eprint = {0801.4017},
 primaryClass = {astro-ph},
       adsurl = {https://ui.adsabs.harvard.edu/abs/2008ApJS..176....1F}
}

@ARTICLE{Stevenson2026,
       author = {{Stevenson}, Struan D. and {Carnall}, Adam C. and {Leung}, Ho-Hin and {Taylor}, Elizabeth and {Cullen}, Fergus and {Dunlop}, James S. and {McLeod}, Derek J. and {McLure}, Ross J. and {Begley}, Ryan and {Arellano-C{\'o}rdova}, Karla Z. and {Barrufet}, Laia and {Bondestam}, Cecilia and {Donnan}, Callum T. and {Ellis}, Richard S. and {Grogin}, Norman A. and {Koekemoer}, Anton M. and {Liu}, Feng-Yuan and {P{\'e}rez-Gonz{\'a}lez}, Pablo G. and {Rowlands}, Kate and {Sanders}, Ryan L. and {Scholte}, Dirk and {Shapley}, Alice E. and {Skarbinski}, Maya and {Stanton}, Thomas M. and {Wild}, Vivienne},
        title = "{PRIMER and JADES reveal an abundance of massive quiescent galaxies at 2 < z < 5}",
      journal = {\mnras},
         year = 2026,
        month = jan,
       volume = {545},
       number = {3},
          eid = {staf2087},
        pages = {staf2087},
          doi = {10.1093/mnras/staf2087},
archivePrefix = {arXiv},
       eprint = {2509.06913},
 primaryClass = {astro-ph.GA},
       adsurl = {https://ui.adsabs.harvard.edu/abs/2026MNRAS.545f2087S}
}

@ARTICLE{CharlotFall2000,
       author = {{Charlot}, St{\'e}phane and {Fall}, S. Michael},
        title = "{A Simple Model for the Absorption of Starlight by Dust in Galaxies}",
      journal = {\apj},
         year = 2000,
        month = aug,
       volume = {539},
       number = {2},
        pages = {718-731},
          doi = {10.1086/309250},
archivePrefix = {arXiv},
       eprint = {astro-ph/0003128},
 primaryClass = {astro-ph},
       adsurl = {https://ui.adsabs.harvard.edu/abs/2000ApJ...539..718C}
}

@ARTICLE{Bravo2023,
       author = {{Bravo}, Mat{\'\i}as and {Robotham}, Aaron S.~G. and {Lagos}, Claudia del P. and {Davies}, Luke J.~M. and {Bellstedt}, Sabine and {Thorne}, Jessica E.},
        title = "{Galaxy quenching time-scales from a forensic reconstruction of their colour evolution}",
      journal = {\mnras},
         year = 2023,
        month = jul,
       volume = {522},
       number = {3},
        pages = {4481-4498},
          doi = {10.1093/mnras/stad1234},
archivePrefix = {arXiv},
       eprint = {2301.03702},
 primaryClass = {astro-ph.GA},
       adsurl = {https://ui.adsabs.harvard.edu/abs/2023MNRAS.522.4481B}
}

@ARTICLE{Fu2025,
       author = {{Fu}, Hao and {Shankar}, Francesco and {Yuan}, Feng and {Roberts}, Daniel and {Boco}, Lumen and {Lapi}, Andrea and {Corcho-Caballero}, Pablo and {Ayromlou}, Mohammadreza and {Georgakakis}, Antonis and {Laloux}, Brivael and {Mu{\~n}oz Rodr{\'\i}guez}, Iv{\'a}n and {Peng}, Yingjie},
        title = "{The role of black hole feedback on galaxy star formation and the degeneracy with halo quenching}",
      journal = {\aap},
         year = 2025,
        month = dec,
       volume = {704},
          eid = {A244},
        pages = {A244},
          doi = {10.1051/0004-6361/202556480},
archivePrefix = {arXiv},
       eprint = {2510.26305},
 primaryClass = {astro-ph.GA},
       adsurl = {https://ui.adsabs.harvard.edu/abs/2025A&A...704A.244F}
}

@ARTICLE{Deger2026,
       author = {{Deger}, Sinan and {Peiris}, Hiranya V. and {Thorp}, Stephen and {J Mortlock}, Daniel and {Jagwani}, Gurjeet and {Alsing}, Justin and {Leistedt}, Boris and {Leja}, Joel},
        title = "{pop-cosmos: Star formation over 12 Gyr from generative modelling of a deep infrared-selected galaxy catalogue}",
      journal = {\mnras},
         year = 2026,
        month = apr,
          doi = {10.1093/mnras/stag764},
archivePrefix = {arXiv},
       eprint = {2509.20430},
 primaryClass = {astro-ph.GA},
       adsurl = {https://ui.adsabs.harvard.edu/abs/2026MNRAS.tmp..752D}
}

@ARTICLE{Li2024,
       author = {{Li}, Jiaxuan and {Melchior}, Peter and {Hahn}, ChangHoon and {Huang}, Song},
        title = "{PopSED: Population-level Inference for Galaxy Properties from Broadband Photometry with Neural Density Estimation}",
      journal = {\aj},
         year = 2024,
        month = jan,
       volume = {167},
       number = {1},
          eid = {16},
        pages = {16},
          doi = {10.3847/1538-3881/ad0be4},
archivePrefix = {arXiv},
       eprint = {2309.16958},
 primaryClass = {astro-ph.IM},
       adsurl = {https://ui.adsabs.harvard.edu/abs/2024AJ....167...16L}
}

@ARTICLE{alsing2024,
       author = {{Alsing}, Justin and {Thorp}, Stephen and {Deger}, Sinan and {Peiris}, Hiranya V. and {Leistedt}, Boris and {Mortlock}, Daniel and {Leja}, Joel},
        title = "{pop-cosmos: A Comprehensive Picture of the Galaxy Population from COSMOS Data}",
      journal = {\apjs},
         year = 2024,
        month = sep,
       volume = {274},
       number = {1},
          eid = {12},
        pages = {12},
          doi = {10.3847/1538-4365/ad5c69},
archivePrefix = {arXiv},
       eprint = {2402.00935},
 primaryClass = {astro-ph.GA},
       adsurl = {https://ui.adsabs.harvard.edu/abs/2024ApJS..274...12A}
}

@ARTICLE{thorp2024,
       author = {{Thorp}, Stephen and {Alsing}, Justin and {Peiris}, Hiranya V. and {Deger}, Sinan and {Mortlock}, Daniel J. and {Leistedt}, Boris and {Leja}, Joel and {Loureiro}, Arthur},
        title = "{pop-cosmos: Scaleable Inference of Galaxy Properties and Redshifts with a Data-driven Population Model}",
      journal = {\apj},
         year = 2024,
        month = nov,
       volume = {975},
       number = {1},
          eid = {145},
        pages = {145},
          doi = {10.3847/1538-4357/ad7736},
archivePrefix = {arXiv},
       eprint = {2406.19437},
 primaryClass = {astro-ph.CO},
       adsurl = {https://ui.adsabs.harvard.edu/abs/2024ApJ...975..145T}
}

@ARTICLE{connolly1999,
       author = {{Connolly}, A.~J. and {Szalay}, A.~S.},
        title = "{A Robust Classification of Galaxy Spectra: Dealing with Noisy and Incomplete Data}",
      journal = {\aj},
         year = 1999,
        month = may,
       volume = {117},
       number = {5},
        pages = {2052-2062},
          doi = {10.1086/300839},
archivePrefix = {arXiv},
       eprint = {astro-ph/9901300},
 primaryClass = {astro-ph},
       adsurl = {https://ui.adsabs.harvard.edu/abs/1999AJ....117.2052C}
}

@ARTICLE{wild2014,
       author = {{Wild}, Vivienne and {Almaini}, Omar and {Cirasuolo}, Michele and {Dunlop}, Jim and {McLure}, Ross and {Bowler}, Rebecca and {Ferreira}, Joao and {Bradshaw}, Emma and {Chuter}, Robert and {Hartley}, Will},
        title = "{A new method for classifying galaxy SEDs from multiwavelength photometry}",
      journal = {\mnras},
         year = 2014,
        month = may,
       volume = {440},
       number = {2},
        pages = {1880-1898},
          doi = {10.1093/mnras/stu212},
archivePrefix = {arXiv},
       eprint = {1401.7878},
 primaryClass = {astro-ph.CO},
       adsurl = {https://ui.adsabs.harvard.edu/abs/2014MNRAS.440.1880W}
}

@ARTICLE{wilkinson2021,
       author = {{Wilkinson}, Aaron and {Almaini}, Omar and {Wild}, Vivienne and {Maltby}, David and {Hartley}, William G. and {Simpson}, Chris and {Rowlands}, Kate},
        title = "{From starburst to quiescence: post-starburst galaxies and their large-scale clustering over cosmic time}",
      journal = {\mnras},
         year = 2021,
        month = jul,
       volume = {504},
       number = {3},
        pages = {4533-4550},
          doi = {10.1093/mnras/stab965},
archivePrefix = {arXiv},
       eprint = {2104.07676},
 primaryClass = {astro-ph.GA},
       adsurl = {https://ui.adsabs.harvard.edu/abs/2021MNRAS.504.4533W}
}

@ARTICLE{leja2019b,
       author = {{Leja}, Joel and {Tacchella}, Sandro and {Conroy}, Charlie},
        title = "{Beyond UVJ: More Efficient Selection of Quiescent Galaxies with Ultraviolet/Mid-infrared Fluxes}",
      journal = {\apjl},
         year = 2019,
        month = jul,
       volume = {880},
       number = {1},
          eid = {L9},
        pages = {L9},
          doi = {10.3847/2041-8213/ab2f8c},
archivePrefix = {arXiv},
       eprint = {1907.02970},
 primaryClass = {astro-ph.GA},
       adsurl = {https://ui.adsabs.harvard.edu/abs/2019ApJ...880L...9L}
}

@ARTICLE{mcleod2021,
       author = {{McLeod}, D.~J. and {McLure}, R.~J. and {Dunlop}, J.~S. and {Cullen}, F. and {Carnall}, A.~C. and {Duncan}, K.},
        title = "{The evolution of the galaxy stellar-mass function over the last 12 billion years from a combination of ground-based and HST surveys}",
      journal = {\mnras},
         year = 2021,
        month = may,
       volume = {503},
       number = {3},
        pages = {4413-4435},
          doi = {10.1093/mnras/stab731},
archivePrefix = {arXiv},
       eprint = {2009.03176},
 primaryClass = {astro-ph.GA},
       adsurl = {https://ui.adsabs.harvard.edu/abs/2021MNRAS.503.4413M}
}

@ARTICLE{Ji2026,
       author = {{Ji}, Zhiyuan and {Williams}, Christina C. and {Behroozi}, Peter and {Weibel}, Andrea and {Jespersen}, Christian Kragh and {Oesch}, Pascal A. and {Bezanson}, Rachel and {Whitaker}, Katherine E. and {Greene}, Jenny E. and {Brammer}, Gabriel and {Dayal}, Pratika and {Labb{\'e}}, Ivo and {Manning}, Sinclaire M. and {Rinaldi}, Pierluigi and {Xiao}, Mengyuan and {Zhang}, Yunchong},
        title = "{PANORAMIC: The Dawn of Massive Quiescent Galaxies I. Number Density and Cosmic Variance from 1000 arcmin$^2$ NIRCam Imaging}",
      journal = {arXiv e-prints},
         year = 2026,
        month = apr,
          eid = {arXiv:2604.05022},
        pages = {arXiv:2604.05022},
          doi = {10.48550/arXiv.2604.05022},
archivePrefix = {arXiv},
       eprint = {2604.05022},
 primaryClass = {astro-ph.GA},
       adsurl = {https://ui.adsabs.harvard.edu/abs/2026arXiv260405022J}
}

@article{cheng2025,
    author = {Cheng, Chloe M and Kriek, Mariska and Beverage, Aliza G and Slob, Martje and Bezanson, Rachel and Franx, Marijn and Leja, Joel and Mancera Piña, Pavel E and Suess, Katherine A and van der Wel, Arjen and van de Sande, Jesse and van Dokkum, Pieter G},
    title = {Ages and metallicities of quiescent galaxies: confronting broad-band (UVJ) colours with stellar absorption lines},
    journal = {\mnras},
    volume = {540},
    number = {2},
    pages = {1527-1543},
    year = {2025},
    month = {06},
    issn = {0035-8711},
    doi = {10.1093/mnras/staf806},
    url = {https://doi.org/10.1093/mnras/staf806},
    eprint = {https://academic.oup.com/mnras/article-pdf/540/2/1527/63212388/staf806.pdf},
}

@ARTICLE{narayanan2024,
       author = {{Narayanan}, Desika and {Lower}, Sidney and {Torrey}, Paul and {Brammer}, Gabriel and {Cui}, Weiguang and {Dav{\'e}}, Romeel and {Iyer}, Kartheik G. and {Li}, Qi and {Lovell}, Christopher C. and {Sales}, Laura V. and {Stark}, Daniel P. and {Marinacci}, Federico and {Vogelsberger}, Mark},
        title = "{Outshining by Recent Star Formation Prevents the Accurate Measurement of High-z Galaxy Stellar Masses}",
      journal = {\apj},
         year = 2024,
        month = jan,
       volume = {961},
       number = {1},
          eid = {73},
        pages = {73},
          doi = {10.3847/1538-4357/ad0966},
archivePrefix = {arXiv},
       eprint = {2306.10118},
 primaryClass = {astro-ph.GA},
       adsurl = {https://ui.adsabs.harvard.edu/abs/2024ApJ...961...73N}
}

@ARTICLE{Santini2021,
       author = {{Santini}, P. and {Castellano}, M. and {Merlin}, E. and {Fontana}, A. and {Fortuni}, F. and {Kodra}, D. and {Magnelli}, B. and {Menci}, N. and {Calabr{\`o}}, A. and {Lovell}, C.~C. and {Pentericci}, L. and {Testa}, V. and {Wilkins}, S.~M.},
        title = "{The emergence of passive galaxies in the early Universe}",
      journal = {\aap},
         year = 2021,
        month = aug,
       volume = {652},
          eid = {A30},
        pages = {A30},
          doi = {10.1051/0004-6361/202039738},
archivePrefix = {arXiv},
       eprint = {2011.10584},
 primaryClass = {astro-ph.GA},
       adsurl = {https://ui.adsabs.harvard.edu/abs/2021A&A...652A..30S}
}

@ARTICLE{Ni2025,
       author = {{Ni}, Yueying and {Chen}, Nianyi and {Zhou}, Yihao and {Park}, Minjung and {Yang}, Yanhui and {Di Matteo}, Tiziana and {Bird}, Simeon and {Croft}, Rupert},
        title = "{The Astrid Simulation: Evolution of Black Holes and Galaxies to z = 0.5 and Different Evolution Pathways for Galaxy Quenching}",
      journal = {\apj},
         year = 2025,
        month = sep,
       volume = {990},
       number = {2},
          eid = {120},
        pages = {120},
          doi = {10.3847/1538-4357/adf3a7},
archivePrefix = {arXiv},
       eprint = {2409.10666},
 primaryClass = {astro-ph.GA},
       adsurl = {https://ui.adsabs.harvard.edu/abs/2025ApJ...990..120N}
}

@ARTICLE{Gountanis2025,
       author = {{Gountanis}, Nicole Marcelina and {Weinberg}, David H. and {Beverage}, Aliza G. and {Sandford}, Nathan R. and {Conroy}, Charlie and {Kriek}, Mariska},
        title = "{Modeling the Ages and Chemical Abundances of Elliptical Galaxies}",
      journal = {\apj},
         year = 2025,
        month = jun,
       volume = {985},
       number = {2},
          eid = {184},
        pages = {184},
          doi = {10.3847/1538-4357/adc4e6},
archivePrefix = {arXiv},
       eprint = {2407.07971},
 primaryClass = {astro-ph.GA},
       adsurl = {https://ui.adsabs.harvard.edu/abs/2025ApJ...985..184G}
}

@ARTICLE{Zhang2026a,
       author = {{Zhang}, Yunchong and {de Graaff}, Anna and {Setton}, David J. and {Price}, Sedona H. and {Bezanson}, Rachel and {del P. Lagos}, Claudia and {Cutler}, Sam E. and {McConachie}, Ian and {Cleri}, Nikko J. and {Cooper}, Olivia R. and {Gottumukkala}, Rashmi and {Greene}, Jenny E. and {Hirschmann}, Michaela and {Khullar}, Gourav and {Labbe}, Ivo and {Leja}, Joel and {Maseda}, Michael V. and {Matthee}, Jorryt and {Miller}, Tim B. and {Nanayakkara}, Themiya and {Suess}, Katherine A. and {Wang}, Bingjie and {Whitaker}, Katherine E. and {Williams}, Christina C.},
        title = "{RUBIES Spectroscopically Confirms the High Number Density of Quiescent Galaxies from 2 < z< 5}",
      journal = {\apj},
         year = 2026,
        month = feb,
       volume = {997},
       number = {2},
          eid = {252},
        pages = {252},
          doi = {10.3847/1538-4357/ae24e1},
archivePrefix = {arXiv},
       eprint = {2508.08577},
 primaryClass = {astro-ph.GA},
       adsurl = {https://ui.adsabs.harvard.edu/abs/2026ApJ...997..252Z}
}

@ARTICLE{Kimmig2025,
       author = {{Kimmig}, Lucas C. and {Remus}, Rhea-Silvia and {Seidel}, Benjamin and {Valenzuela}, Lucas M. and {Dolag}, Klaus and {Burkert}, Andreas},
        title = "{Blowing Out the Candle: How to Quench Galaxies at High Redshift{\textemdash}An Ensemble of Rapid Starbursts, AGN Feedback, and Environment}",
      journal = {\apj},
         year = 2025,
        month = jan,
       volume = {979},
       number = {1},
          eid = {15},
        pages = {15},
          doi = {10.3847/1538-4357/ad9472},
archivePrefix = {arXiv},
       eprint = {2310.16085},
 primaryClass = {astro-ph.GA},
       adsurl = {https://ui.adsabs.harvard.edu/abs/2025ApJ...979...15K}
}

@ARTICLE{ChandroGomez2026,
       author = {{Chandro-G{\'o}mez}, {\'A}ngel and {Lagos}, Claudia del P. and {Power}, Chris and {Baker}, Willian M. and {Ben{\'\i}tez-Llambay}, Alejandro and {Chaikin}, Evgenii and {Chittenden}, Harry G. and {Correa}, Camila and {Frenk}, Carlos S. and {Hu{\v{s}}ko}, Filip and {Ito}, Kei and {McGibbon}, Robert J. and {Nanayakkara}, Themiya and {Ploeckinger}, Sylvia and {Richings}, Alexander J. and {Schaller}, Matthieu and {Schaye}, Joop and {Trayford}, James W. and {Valentino}, Francesco},
        title = "{Unveiling the population of massive quenched galaxies at $z\ge2$ in the COLIBRE simulations -- II. The role of AGN feedback and environment on their emergence}",
      journal = {arXiv e-prints},
         year = 2026,
        month = may,
          eid = {arXiv:2605.31052},
        pages = {arXiv:2605.31052},
          doi = {10.48550/arXiv.2605.31052},
archivePrefix = {arXiv},
       eprint = {2605.31052},
 primaryClass = {astro-ph.GA},
       adsurl = {https://ui.adsabs.harvard.edu/abs/2026arXiv260531052C}
}

@ARTICLE{Szpila2025,
       author = {{Szpila}, Jakub and {Dav{\'e}}, Romeel and {Rennehan}, Douglas and {Cui}, Weiguang and {Hough}, Renier T.},
        title = "{The nature and evolution of early massive quenched galaxies in the SIMBA-C simulation}",
      journal = {\mnras},
         year = 2025,
        month = feb,
       volume = {537},
       number = {2},
        pages = {1849-1868},
          doi = {10.1093/mnras/staf132},
archivePrefix = {arXiv},
       eprint = {2402.08729},
 primaryClass = {astro-ph.GA},
       adsurl = {https://ui.adsabs.harvard.edu/abs/2025MNRAS.537.1849S}
}

@ARTICLE{Lagos2025,
       author = {{Lagos}, Claudia del P. and {Valentino}, Francesco and {Wright}, Ruby J. and {de Graaff}, Anna and {Glazebrook}, Karl and {De Lucia}, Gabriella and {Robotham}, Aaron S.~G. and {Nanayakkara}, Themiya and {Chandro-Gomez}, Angel and {Bravo}, Mat{\'\i}as and {Baugh}, Carlton M. and {Harborne}, Katherine E. and {Hirschmann}, Michaela and {Fontanot}, Fabio and {Xie}, Lizhi and {Chittenden}, Harry},
        title = "{The diverse star formation histories of early massive, quenched galaxies in modern galaxy formation simulations}",
      journal = {\mnras},
         year = 2025,
        month = jan,
       volume = {536},
       number = {3},
        pages = {2324-2354},
          doi = {10.1093/mnras/stae2626},
archivePrefix = {arXiv},
       eprint = {2409.16916},
 primaryClass = {astro-ph.GA},
       adsurl = {https://ui.adsabs.harvard.edu/abs/2025MNRAS.536.2324L}
}

@ARTICLE{Mosleh2025,
       author = {{Mosleh}, Moein and {Riahi-Zamin}, Mohammad and {Tacchella}, Sandro},
        title = "{Reconstructing Star Formation Histories of High-redshift Galaxies: A Comparison of Resolved Parametric and Nonparametric Models}",
      journal = {\apj},
         year = 2025,
        month = apr,
       volume = {983},
       number = {2},
          eid = {181},
        pages = {181},
          doi = {10.3847/1538-4357/adc12e},
archivePrefix = {arXiv},
       eprint = {2503.14591},
 primaryClass = {astro-ph.GA},
       adsurl = {https://ui.adsabs.harvard.edu/abs/2025ApJ...983..181M}
}

@ARTICLE{Slob2024,
       author = {{Slob}, Martje and {Kriek}, Mariska and {Beverage}, Aliza G. and {Suess}, Katherine A. and {Barro}, Guillermo and {Bezanson}, Rachel and {Brammer}, Gabriel and {Cheng}, Chloe M. and {Conroy}, Charlie and {de Graaff}, Anna and {F{\"o}rster Schreiber}, Natascha M. and {Franx}, Marijn and {Lorenz}, Brian and {Mancera Pi{\~n}a}, Pavel E. and {Marchesini}, Danilo and {Muzzin}, Adam and {Newman}, Andrew B. and {Price}, Sedona H. and {Shapley}, Alice E. and {Stefanon}, Mauro and {van Dokkum}, Pieter and {Weisz}, Daniel R.},
        title = "{The JWST-SUSPENSE Ultradeep Spectroscopic Program: Survey Overview and Star Formation Histories of Quiescent Galaxies at 1 < z < 3}",
      journal = {\apj},
         year = 2024,
        month = oct,
       volume = {973},
       number = {2},
          eid = {131},
        pages = {131},
          doi = {10.3847/1538-4357/ad65ff},
archivePrefix = {arXiv},
       eprint = {2404.12432},
 primaryClass = {astro-ph.GA},
       adsurl = {https://ui.adsabs.harvard.edu/abs/2024ApJ...973..131S}
}

@ARTICLE{Hamadouche2026arXiv,
       author = {{Hamadouche}, Massissilia L. and {Whitaker}, Katherine E. and {Valentino}, Francesco and {Antwi-Danso}, Jacqueline and {Ito}, Kei and {Beverage}, Aliza and {Zhu}, Pengpei and {Brammer}, Gabriel and {Kokorev}, Vasily and {de Lucia}, Gabriella and {Baker}, William M. and {Farcy}, Marion and {Gallazzi}, Anna and {Gillman}, Steven and {Gottumukkala}, Rashmi and {Hirschmann}, Michaela and {Jespersen}, Christian Kragh and {Kakimoto}, Takumi and {Lee}, Minju M. and {Onodera}, Masato and {Shimakawa}, Rhythm and {Tanaka}, Masayuki and {Weaver}, John R. and {Wu}, Po-Feng},
        title = "{DeepDive: Tracing the early quenching pathways of massive quiescent galaxies at $z>3$ from their star-formation histories and chemical abundances}",
      journal = {arXiv e-prints},
         year = 2026,
        month = feb,
          eid = {arXiv:2602.02485},
        pages = {arXiv:2602.02485},
          doi = {10.48550/arXiv.2602.02485},
archivePrefix = {arXiv},
       eprint = {2602.02485},
 primaryClass = {astro-ph.GA},
       adsurl = {https://ui.adsabs.harvard.edu/abs/2026arXiv260202485H}
}

@ARTICLE{Carnall2023Abundance,
       author = {{Carnall}, A.~C. and {McLeod}, D.~J. and {McLure}, R.~J. and {Dunlop}, J.~S. and {Begley}, R. and {Cullen}, F. and {Donnan}, C.~T. and {Hamadouche}, M.~L. and {Jewell}, S.~M. and {Jones}, E.~W. and {Pollock}, C.~L. and {Wild}, V.},
        title = "{A surprising abundance of massive quiescent galaxies at 3 < z < 5 in the first data from JWST CEERS}",
      journal = {\mnras},
         year = 2023,
        month = apr,
       volume = {520},
       number = {3},
        pages = {3974-3985},
          doi = {10.1093/mnras/stad369},
archivePrefix = {arXiv},
       eprint = {2208.00986},
 primaryClass = {astro-ph.GA},
       adsurl = {https://ui.adsabs.harvard.edu/abs/2023MNRAS.520.3974C}
}

@ARTICLE{carnall2024,
       author = {{Carnall}, A.~C. and {Cullen}, F. and {McLure}, R.~J. and {McLeod}, D.~J. and {Begley}, R. and {Donnan}, C.~T. and {Dunlop}, J.~S. and {Shapley}, A.~E. and {Rowlands}, K. and {Almaini}, O. and {Arellano-C{\'o}rdova}, K.~Z. and {Barrufet}, L. and {Cimatti}, A. and {Ellis}, R.~S. and {Grogin}, N.~A. and {Hamadouche}, M.~L. and {Illingworth}, G.~D. and {Koekemoer}, A.~M. and {Leung}, H.-H. and {Lovell}, C.~C. and {P{\'e}rez-Gonz{\'a}lez}, P.~G. and {Santini}, P. and {Stanton}, T.~M. and {Wild}, V.},
        title = "{The JWST EXCELS survey: too much, too young, too fast? Ultra-massive quiescent galaxies at 3 < z < 5}",
      journal = {\mnras},
         year = 2024,
        month = oct,
       volume = {534},
       number = {1},
        pages = {325-348},
          doi = {10.1093/mnras/stae2092},
archivePrefix = {arXiv},
       eprint = {2405.02242},
 primaryClass = {astro-ph.GA},
       adsurl = {https://ui.adsabs.harvard.edu/abs/2024MNRAS.534..325C}
}

@ARTICLE{skarbinski2026,
       author = {{Skarbinski}, Maya and {Rowlands}, Kate and {Alatalo}, Katherine and {Wild}, Vivienne and {Carnall}, Adam C. and {Almaini}, Omar and {Maltby}, David and {de Lisle}, Thomas and {Heckman}, Timothy and {Begley}, Ryan and {Cullen}, Fergus and {Dunlop}, James S. and {Hewitt}, Guillaume and {Leung}, Ho-Hin and {McLeod}, Derek and {McLure}, Ross and {Otter}, Justin Atsushi and {Patil}, Pallavi and {Petric}, Andreea and {Shapley}, Alice E. and {Stevenson}, Struan and {Taylor}, Elizabeth},
        title = "{The JWST EXCELS Survey: Insights into the Nature of Quenching at Cosmic Noon}",
      journal = {\apj},
         year = 2026,
        month = apr,
       volume = {1000},
       number = {2},
          eid = {191},
        pages = {191},
          doi = {10.3847/1538-4357/ae459d},
archivePrefix = {arXiv},
       eprint = {2509.18278},
 primaryClass = {astro-ph.GA},
       adsurl = {https://ui.adsabs.harvard.edu/abs/2026ApJ..1000..191S}
}

@ARTICLE{Zhang2026b,
       author = {{Zhang}, Yunchong and {Ji}, Zhiyuan and {Bezanson}, Rachel and {Williams}, Christina C. and {Brammer}, Gabriel and {Cloonan}, Aidan P. and {de Graaff}, Anna and {Greene}, Jenny E. and {Hirschmann}, Michaela and {Jespersen}, Christian Kragh and {Khullar}, Gourav and {Lagos}, Claudia del P. and {Leja}, Joel and {Maseda}, Michael V. and {McConachie}, Ian and {Oesch}, Pascal A. and {Price}, Sedona H. and {Setton}, David J. and {Suess}, Katherine A. and {Whitaker}, Katherine E.},
        title = "{Winding Back the Clock: Recent Star Formation Histories of Massive Quiescent Galaxies Are Consistent With Their Rapid Number Density Evolution Since $\mathbf{z\sim7}$}",
      journal = {arXiv e-prints},
         year = 2026,
        month = apr,
          eid = {arXiv:2604.05024},
        pages = {arXiv:2604.05024},
          doi = {10.48550/arXiv.2604.05024},
archivePrefix = {arXiv},
       eprint = {2604.05024},
 primaryClass = {astro-ph.GA},
       adsurl = {https://ui.adsabs.harvard.edu/abs/2026arXiv260405024Z}
}

@ARTICLE{Bezanson2022,
       author = {{Bezanson}, Rachel and {Spilker}, Justin S. and {Suess}, Katherine A. and {Setton}, David J. and {Feldmann}, Robert and {Greene}, Jenny E. and {Kriek}, Mariska and {Narayanan}, Desika and {Verrico}, Margaret},
        title = "{Now You See It, Now You Don't: Star Formation Truncation Precedes the Loss of Molecular Gas by 100 Myr in Massive Poststarburst Galaxies at z   0.6}",
      journal = {\apj},
         year = 2022,
        month = feb,
       volume = {925},
       number = {2},
          eid = {153},
        pages = {153},
          doi = {10.3847/1538-4357/ac3dfa},
archivePrefix = {arXiv},
       eprint = {2111.14877},
 primaryClass = {astro-ph.GA},
       adsurl = {https://ui.adsabs.harvard.edu/abs/2022ApJ...925..153B}
}

@ARTICLE{setton2022,
       author = {{Setton}, David J. and {Verrico}, Margaret and {Bezanson}, Rachel and {Greene}, Jenny E. and {Suess}, Katherine A. and {Goulding}, Andy D. and {Spilker}, Justin S. and {Kriek}, Mariska and {Feldmann}, Robert and {Narayanan}, Desika and {Hall-Hooper}, Khalil and {Kado-Fong}, Erin},
        title = "{The Compact Structures of Massive z {\ensuremath{\sim}} 0.7 Post-starburst Galaxies in the SQuIGG$\vec{L}$E Sample}",
      journal = {\apj},
         year = 2022,
        month = may,
       volume = {931},
       number = {1},
          eid = {51},
        pages = {51},
          doi = {10.3847/1538-4357/ac6096},
archivePrefix = {arXiv},
       eprint = {2203.08835},
 primaryClass = {astro-ph.GA},
       adsurl = {https://ui.adsabs.harvard.edu/abs/2022ApJ...931...51S}
}

@ARTICLE{Tacchella2022,
       author = {{Tacchella}, Sandro and {Conroy}, Charlie and {Faber}, S.~M. and {Johnson}, Benjamin D. and {Leja}, Joel and {Barro}, Guillermo and {Cunningham}, Emily C. and {Deason}, Alis J. and {Guhathakurta}, Puragra and {Guo}, Yicheng and {Hernquist}, Lars and {Koo}, David C. and {McKinnon}, Kevin and {Rockosi}, Constance M. and {Speagle}, Joshua S. and {van Dokkum}, Pieter and {Yesuf}, Hassen M.},
        title = "{Fast, Slow, Early, Late: Quenching Massive Galaxies at z {\ensuremath{\sim}} 0.8}",
      journal = {\apj},
         year = 2022,
        month = feb,
       volume = {926},
       number = {2},
          eid = {134},
        pages = {134},
          doi = {10.3847/1538-4357/ac449b},
archivePrefix = {arXiv},
       eprint = {2102.12494},
 primaryClass = {astro-ph.GA},
       adsurl = {https://ui.adsabs.harvard.edu/abs/2022ApJ...926..134T}
}

@ARTICLE{Leung2026b,
       author = {{Leung}, Ho-Hin and {Wild}, Vivienne and {Papathomas}, Michail and {Mortlock}, Daniel J. and {Rankine}, Amy L. and {Curtis-Lake}, Emma and {Zheng}, Yirui and {Carnall}, Adam C. and {Johansson}, Peter H.},
        title = "{Spatially resolved star-formation histories of local post-starburst galaxies: Starburst and quenching spatial patterns consistent with recent mergers}",
      journal = {arXiv e-prints},
         year = 2026,
        month = feb,
          eid = {arXiv:2602.13114},
        pages = {arXiv:2602.13114},
          doi = {10.48550/arXiv.2602.13114},
archivePrefix = {arXiv},
       eprint = {2602.13114},
 primaryClass = {astro-ph.GA},
       adsurl = {https://ui.adsabs.harvard.edu/abs/2026arXiv260213114L}
}

@ARTICLE{Ditrani2024,
       author = {{Ditrani}, F.~R. and {Longhetti}, M. and {Fossati}, M. and {Wolter}, A.},
        title = "{Understanding the unusual life of the Cartwheel galaxy using stellar populations}",
      journal = {\aap},
         year = 2024,
        month = aug,
       volume = {688},
          eid = {A89},
        pages = {A89},
          doi = {10.1051/0004-6361/202349070},
archivePrefix = {arXiv},
       eprint = {2405.19403},
 primaryClass = {astro-ph.GA},
       adsurl = {https://ui.adsabs.harvard.edu/abs/2024A&A...688A..89D}
}

@ARTICLE{Nersesian2026,
       author = {{Nersesian}, Angelos and {Kaushal}, Yasha and {Martorano}, Marco and {van der Wel}, Arjen and {Wu}, Po-Feng and {Bezanson}, Rachel and {Bell}, Eric F. and {D'Eugenio}, Francesco and {Gallazzi}, Anna R. and {Leja}, Joel and {Zibetti}, Stefano and {Tacchella}, Sandro},
        title = "{The LEGA-C galaxy survey: Multiple quenching channels for quiescent galaxies at z {\ensuremath{\sim}} 1}",
      journal = {\aap},
         year = 2026,
        month = jan,
       volume = {705},
          eid = {A153},
        pages = {A153},
          doi = {10.1051/0004-6361/202557293},
archivePrefix = {arXiv},
       eprint = {2512.10383},
 primaryClass = {astro-ph.GA},
       adsurl = {https://ui.adsabs.harvard.edu/abs/2026A&A...705A.153N}
}

@ARTICLE{sandles2022,
       author = {{Sandles}, L. and {Curtis-Lake}, E. and {Charlot}, S. and {Chevallard}, J. and {Maiolino}, R.},
        title = "{Bayesian hierarchical modelling of the M$_{*}$-SFR relation from 1 {\ensuremath{\lesssim}} z {\ensuremath{\lesssim}} 6 in ASTRODEEP}",
      journal = {\mnras},
         year = 2022,
        month = sep,
       volume = {515},
       number = {2},
        pages = {2951-2969},
          doi = {10.1093/mnras/stac1999},
archivePrefix = {arXiv},
       eprint = {2207.06322},
 primaryClass = {astro-ph.GA},
       adsurl = {https://ui.adsabs.harvard.edu/abs/2022MNRAS.515.2951S}
}

@ARTICLE{Leung2026a,
       author = {{Leung}, Ho-Hin and {Carnall}, Adam C. and {Taylor}, Elizabeth and {Stevenson}, Struan D. and {Beverage}, Aliza G. and {Cullen}, Fergus and {Dunlop}, James S. and {McLeod}, Derek J. and {McLure}, Ross J. and {Begley}, Ryan and {Almaini}, Omar and {Antonogiannaki}, Stella and {Arellano-C{\'o}rdova}, Karla Z. and {Barrufet}, Laia and {Bondestam}, Cecilia and {Donnan}, Callum T. and {Holst}, Isaac J.~B. and {Liu}, Feng-Yuan F. and {Rowlands}, Kate and {Sanders}, Ryan L. and {Scholte}, Dirk and {Skarbinski}, Maya and {Stanton}, Thomas M. and {Wild}, Vivienne},
        title = "{The JWST EXCELS survey: The ages and abundances of 3 < z < 5 massive quiescent galaxies show that downsizing was already in place by z ≃ 4}",
      journal = {\mnras},
         year = 2026,
        month = may,
          doi = {10.1093/mnras/stag827},
archivePrefix = {arXiv},
       eprint = {2602.05934},
 primaryClass = {astro-ph.GA},
       adsurl = {https://ui.adsabs.harvard.edu/abs/2026MNRAS.tmp..787L}
}

@ARTICLE{baker2025,
       author = {{Baker}, William M. and {Valentino}, Francesco and {Lagos}, Claudia del P. and {Ito}, Kei and {Jespersen}, Christian Kragh and {Gottumukkala}, Rashmi and {Hjorth}, Jens and {Langeroodi}, Danial and {Sedgewick}, Aidan},
        title = "{Exploring over 700 massive quiescent galaxies at z = 2─7: Demographics and stellar mass functions}",
      journal = {\aap},
         year = 2025,
        month = oct,
       volume = {702},
          eid = {A270},
        pages = {A270},
          doi = {10.1051/0004-6361/202555829},
archivePrefix = {arXiv},
       eprint = {2506.04119},
 primaryClass = {astro-ph.GA},
       adsurl = {https://ui.adsabs.harvard.edu/abs/2025A&A...702A.270B}
}

@ARTICLE{weaver2023,
       author = {{Weaver}, J.~R. and {Davidzon}, I. and {Toft}, S. and {Ilbert}, O. and {McCracken}, H.~J. and {Gould}, K.~M.~L. and {Jespersen}, C.~K. and {Steinhardt}, C. and {Lagos}, C.~D.~P. and {Capak}, P.~L. and {Casey}, C.~M. and {Chartab}, N. and {Faisst}, A.~L. and {Hayward}, C.~C. and {Kartaltepe}, J.~S. and {Kauffmann}, O.~B. and {Koekemoer}, A.~M. and {Kokorev}, V. and {Laigle}, C. and {Liu}, D. and {Long}, A. and {Magdis}, G.~E. and {McPartland}, C.~J.~R. and {Milvang-Jensen}, B. and {Mobasher}, B. and {Moneti}, A. and {Peng}, Y. and {Sanders}, D.~B. and {Shuntov}, M. and {Sneppen}, A. and {Valentino}, F. and {Zalesky}, L. and {Zamorani}, G.},
        title = "{COSMOS2020: The galaxy stellar mass function. The assembly and star formation cessation of galaxies at 0.2< z {\ensuremath{\leq}} 7.5}",
      journal = {\aap},
         year = 2023,
        month = sep,
       volume = {677},
          eid = {A184},
        pages = {A184},
          doi = {10.1051/0004-6361/202245581},
archivePrefix = {arXiv},
       eprint = {2212.02512},
 primaryClass = {astro-ph.GA},
       adsurl = {https://ui.adsabs.harvard.edu/abs/2023A&A...677A.184W}
}

@ARTICLE{Iyer2020,
       author = {{Iyer}, Kartheik G. and {Tacchella}, Sandro and {Genel}, Shy and {Hayward}, Christopher C. and {Hernquist}, Lars and {Brooks}, Alyson M. and {Caplar}, Neven and {Dav{\'e}}, Romeel and {Diemer}, Benedikt and {Forbes}, John C. and {Gawiser}, Eric and {Somerville}, Rachel S. and {Starkenburg}, Tjitske K.},
        title = "{The diversity and variability of star formation histories in models of galaxy evolution}",
      journal = {\mnras},
         year = 2020,
        month = oct,
       volume = {498},
       number = {1},
        pages = {430-463},
          doi = {10.1093/mnras/staa2150},
archivePrefix = {arXiv},
       eprint = {2007.07916},
 primaryClass = {astro-ph.GA},
       adsurl = {https://ui.adsabs.harvard.edu/abs/2020MNRAS.498..430I}
}

@ARTICLE{Iyer2025,
       author = {{Iyer}, Kartheik G. and {Starkenburg}, Tjitske K. and {Bryan}, Greg L. and {Somerville}, Rachel S. and {Alfonzo}, Juan Pablo and {Angl{\'e}s-Alc{\'a}zar}, Daniel and {Cooray}, Suchetha and {Dav{\'e}}, Romeel and {Gabrielpillai}, Austen and {Genel}, Shy and {Hassan}, Sultan and {Hernquist}, Lars and {Jespersen}, Christian Kragh and {Lovell}, Christopher C. and {Oh}, Boon Kiat and {Pacifici}, Camilla and {Perez}, Lucia A. and {Sommovigo}, Laura and {Speagle}, Joshua S. and {Tacchella}, Sandro and {Tillman}, Megan T. and {Villaescusa-Navarro}, Francisco and {Wu}, John F.},
        title = "{How Does Feedback Affect the Star Formation Histories of Galaxies?}",
      journal = {\apj},
         year = 2025,
        month = dec,
       volume = {994},
       number = {2},
          eid = {174},
        pages = {174},
          doi = {10.3847/1538-4357/ae0334},
archivePrefix = {arXiv},
       eprint = {2508.21152},
 primaryClass = {astro-ph.GA},
       adsurl = {https://ui.adsabs.harvard.edu/abs/2025ApJ...994..174I}
}

@ARTICLE{Diemer2017,
       author = {{Diemer}, Benedikt and {Sparre}, Martin and {Abramson}, Louis E. and {Torrey}, Paul},
        title = "{Log-normal Star Formation Histories in Simulated and Observed Galaxies}",
      journal = {\apj},
         year = 2017,
        month = apr,
       volume = {839},
       number = {1},
          eid = {26},
        pages = {26},
          doi = {10.3847/1538-4357/aa68e5},
archivePrefix = {arXiv},
       eprint = {1701.02308},
 primaryClass = {astro-ph.GA},
       adsurl = {https://ui.adsabs.harvard.edu/abs/2017ApJ...839...26D}
}

@ARTICLE{Cloudy2025,
       author = {{Gunasekera}, C.~M. and {van Hoof}, P.~A.~M. and {Dehghanian}, M. and {Chakraborty}, P. and {Shaw}, G. and {Bianchi}, S. and {Chatzikos}, M. and {Tsujimoto}, M. and {Ferland}, G.~J.},
        title = "{The 2025 release of Cloudy}",
      journal = {\rmxaa},
         year = 2025,
        month = nov,
       volume = {61},
        pages = {120-133},
          doi = {10.22201/ia.01851101p.2025.61.03.01},
archivePrefix = {arXiv},
       eprint = {2508.01102},
 primaryClass = {astro-ph.GA},
       adsurl = {https://ui.adsabs.harvard.edu/abs/2025RMxAA..61..120G}
}

@ARTICLE{williams2009,
       author = {{Williams}, Rik J. and {Quadri}, Ryan F. and {Franx}, Marijn and {van Dokkum}, Pieter and {Labb{\'e}}, Ivo},
        title = "{Detection of Quiescent Galaxies in a Bicolor Sequence from Z = 0-2}",
      journal = {\apj},
         year = 2009,
        month = feb,
       volume = {691},
       number = {2},
        pages = {1879-1895},
          doi = {10.1088/0004-637X/691/2/1879},
archivePrefix = {arXiv},
       eprint = {0806.0625},
 primaryClass = {astro-ph},
       adsurl = {https://ui.adsabs.harvard.edu/abs/2009ApJ...691.1879W}
}

@ARTICLE{trayford2016,
       author = {{Trayford}, James W. and {Theuns}, Tom and {Bower}, Richard G. and {Crain}, Robert A. and {Lagos}, Claudia del P. and {Schaller}, Matthieu and {Schaye}, Joop},
        title = "{It is not easy being green: the evolution of galaxy colour in the EAGLE simulation}",
      journal = {\mnras},
         year = 2016,
        month = aug,
       volume = {460},
       number = {4},
        pages = {3925-3939},
          doi = {10.1093/mnras/stw1230},
archivePrefix = {arXiv},
       eprint = {1601.07907},
 primaryClass = {astro-ph.GA},
       adsurl = {https://ui.adsabs.harvard.edu/abs/2016MNRAS.460.3925T}
}

@ARTICLE{Walters2022,
       author = {{Walters}, Dan and {Woo}, Joanna and {Ellison}, Sara L.},
        title = "{Quenching time-scales in the IllustrisTNG simulation}",
      journal = {\mnras},
         year = 2022,
        month = apr,
       volume = {511},
       number = {4},
        pages = {6126-6142},
          doi = {10.1093/mnras/stac283},
archivePrefix = {arXiv},
       eprint = {2202.00015},
 primaryClass = {astro-ph.GA},
       adsurl = {https://ui.adsabs.harvard.edu/abs/2022MNRAS.511.6126W}
}

@ARTICLE{Wright2019,
       author = {{Wright}, Ruby J. and {Lagos}, Claudia del P. and {Davies}, Luke J.~M. and {Power}, Chris and {Trayford}, James W. and {Wong}, O. Ivy},
        title = "{Quenching time-scales of galaxies in the EAGLE simulations}",
      journal = {\mnras},
         year = 2019,
        month = aug,
       volume = {487},
       number = {3},
        pages = {3740-3758},
          doi = {10.1093/mnras/stz1410},
archivePrefix = {arXiv},
       eprint = {1810.07335},
 primaryClass = {astro-ph.GA},
       adsurl = {https://ui.adsabs.harvard.edu/abs/2019MNRAS.487.3740W}
}

@ARTICLE{daCunha2008_magphys,
       author = {{da Cunha}, Elisabete and {Charlot}, St{\'e}phane and {Elbaz}, David},
        title = "{A simple model to interpret the ultraviolet, optical and infrared emission from galaxies}",
      journal = {\mnras},
         year = 2008,
        month = aug,
       volume = {388},
       number = {4},
        pages = {1595-1617},
          doi = {10.1111/j.1365-2966.2008.13535.x},
archivePrefix = {arXiv},
       eprint = {0806.1020},
 primaryClass = {astro-ph},
       adsurl = {https://ui.adsabs.harvard.edu/abs/2008MNRAS.388.1595D}
}

@ARTICLE{skelton2012,
       author = {{Skelton}, Rosalind E. and {Bell}, Eric F. and {Somerville}, Rachel S.},
        title = "{Modeling the Red Sequence: Hierarchical Growth yet Slow Luminosity Evolution}",
      journal = {\apj},
         year = 2012,
        month = jul,
       volume = {753},
       number = {1},
          eid = {44},
        pages = {44},
          doi = {10.1088/0004-637X/753/1/44},
archivePrefix = {arXiv},
       eprint = {1112.1077},
 primaryClass = {astro-ph.CO},
       adsurl = {https://ui.adsabs.harvard.edu/abs/2012ApJ...753...44S}
}

@ARTICLE{Naab2017,
       author = {{Naab}, Thorsten and {Ostriker}, Jeremiah P.},
        title = "{Theoretical Challenges in Galaxy Formation}",
      journal = {\araa},
         year = 2017,
        month = aug,
       volume = {55},
       number = {1},
        pages = {59-109},
          doi = {10.1146/annurev-astro-081913-040019},
archivePrefix = {arXiv},
       eprint = {1612.06891},
 primaryClass = {astro-ph.GA},
       adsurl = {https://ui.adsabs.harvard.edu/abs/2017ARA&A..55...59N}
}

@ARTICLE{Schreiber2018,
       author = {{Schreiber}, C. and {Glazebrook}, K. and {Nanayakkara}, T. and {Kacprzak}, G.~G. and {Labb{\'e}}, I. and {Oesch}, P. and {Yuan}, T. and {Tran}, K.-V. and {Papovich}, C. and {Spitler}, L. and {Straatman}, C.},
        title = "{Near infrared spectroscopy and star-formation histories of 3 {\ensuremath{\leq}} z {\ensuremath{\leq}} 4 quiescent galaxies}",
      journal = {\aap},
         year = 2018,
        month = oct,
       volume = {618},
          eid = {A85},
        pages = {A85},
          doi = {10.1051/0004-6361/201833070},
archivePrefix = {arXiv},
       eprint = {1807.02523},
 primaryClass = {astro-ph.GA},
       adsurl = {https://ui.adsabs.harvard.edu/abs/2018A&A...618A..85S}
}

@ARTICLE{Sherman2020,
       author = {{Sherman}, Sydney and {Jogee}, Shardha and {Florez}, Jonathan and {Stevans}, Matthew L. and {Kawinwanichakij}, Lalitwadee and {Wold}, Isak and {Finkelstein}, Steven L. and {Papovich}, Casey and {Ciardullo}, Robin and {Gronwall}, Caryl and {Cora}, Sof{\'\i}a A. and {Hough}, Tom{\'a}s and {Vega-Mart{\'\i}nez}, Cristian A.},
        title = "{Investigating the growing population of massive quiescent galaxies at cosmic noon}",
      journal = {\mnras},
         year = 2020,
        month = dec,
       volume = {499},
       number = {3},
        pages = {4239-4260},
          doi = {10.1093/mnras/staa3167},
archivePrefix = {arXiv},
       eprint = {2010.04741},
 primaryClass = {astro-ph.GA},
       adsurl = {https://ui.adsabs.harvard.edu/abs/2020MNRAS.499.4239S}
}

\appendix

% make the hyper-refs work for appendix figures.
% solution from here: https://tex.stackexchange.com/questions/560474/hyperlinks-to-tables-that-are-in-appendix-do-not-work-well
\setcounter{figure}{0}
\renewcommand\theHfigure{Appendix.\thefigure}

\section{Mock parameter recover tests}
\label{app:A}

\begin{figure*}[h]
    \centering
    \includegraphics[width=\linewidth]{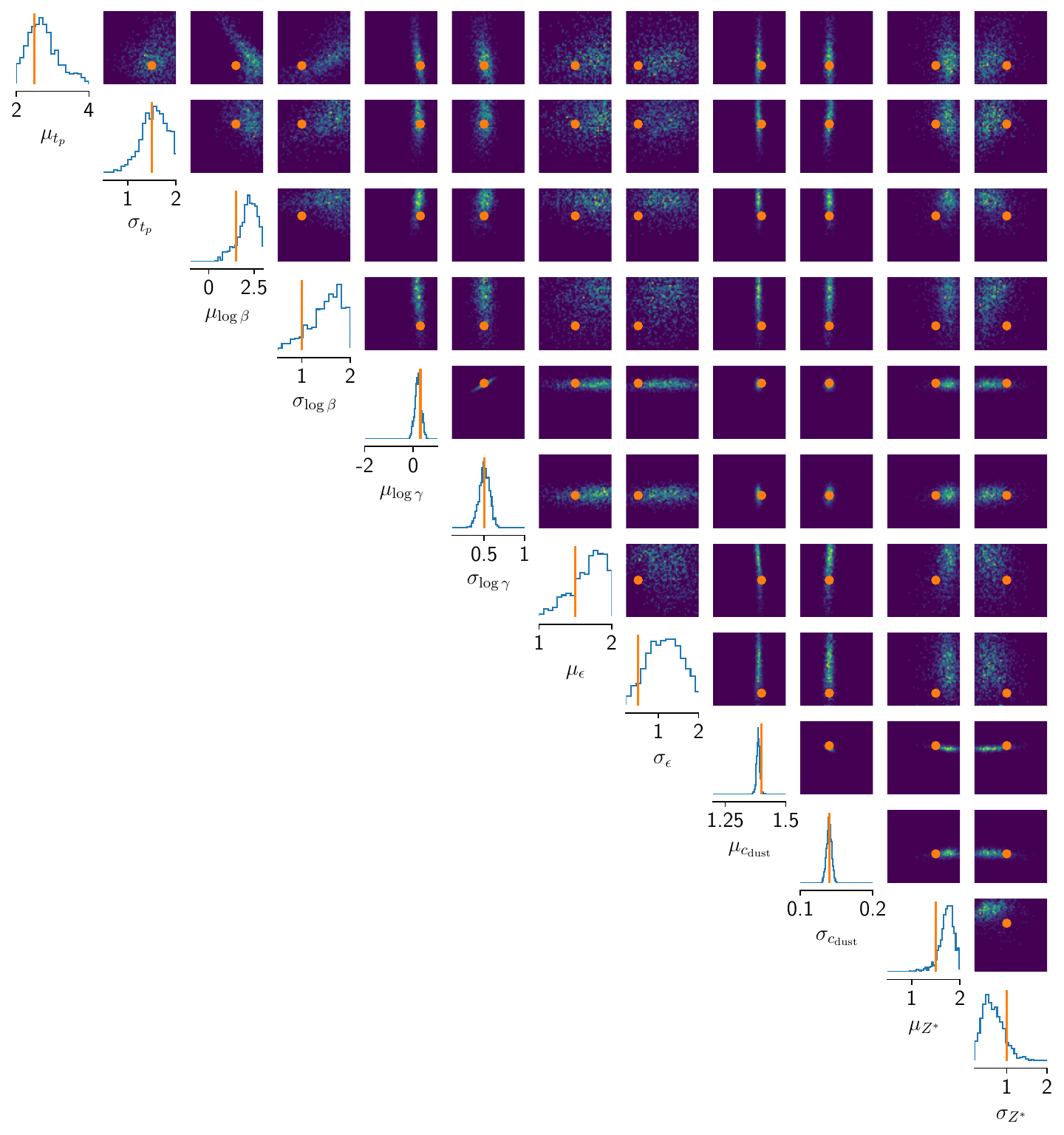}
    \caption{A mock recovery test for the mock shown in Fig~\ref{fig:mockposterior} showing the joint posterior distributions of the model population parameters as described in Table \ref{tab:theta}, with the orange dots and lines marking the true mock parameters. }
    \label{fig:mockcorners}
\end{figure*}

\begin{figure*}
    \centering
    \includegraphics[width=\linewidth]{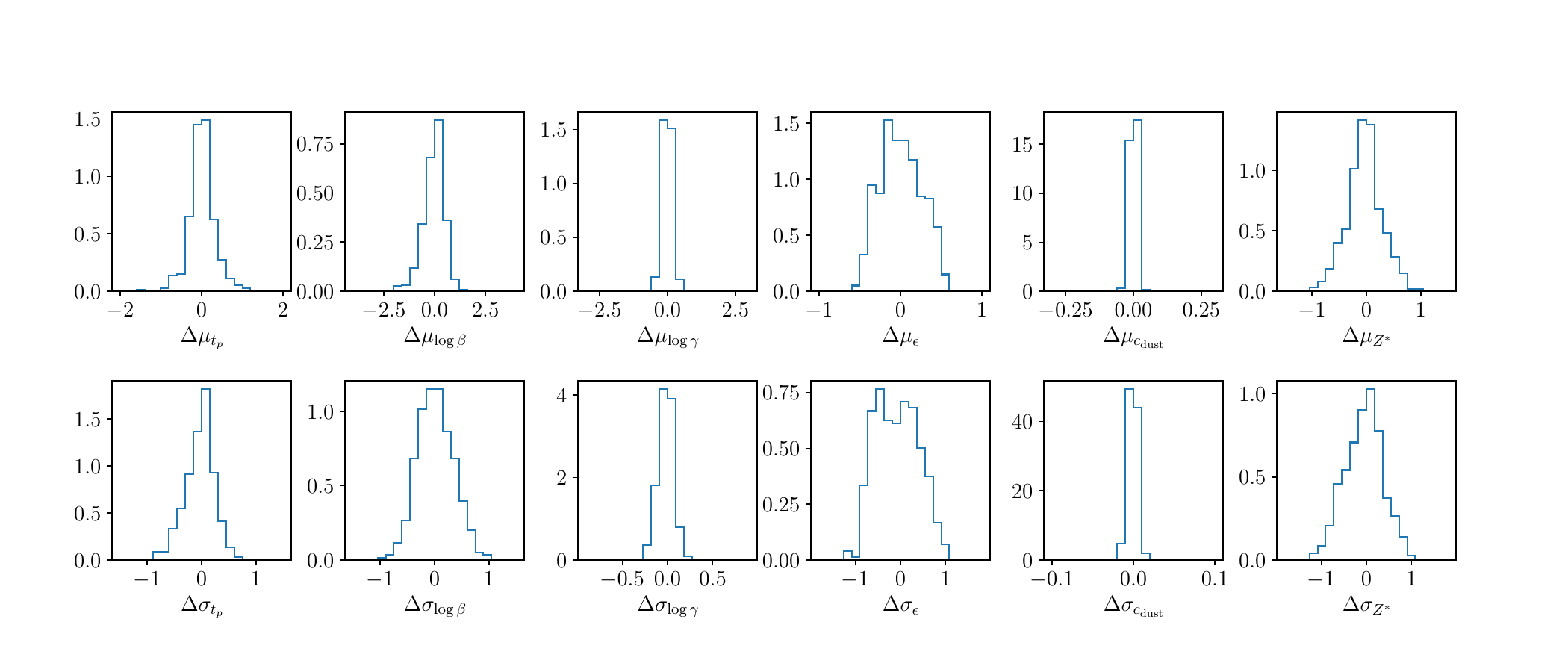}
    \caption{The differences between input and posterior median values for 500 mock simulations drawn from a new test dataset. Each panel's x-axis range matches the possible range given the upper and lower limits on the flat prior. We can see that the population parameters that describe the quenching timescale distribution ($\gamma$) and amplitude of dust attenuation ($c_{dust}$) are well constrained by the super-colour image, while those that describe the metallicity distribution, width of the SFR rise time distribution, and additional dust attenuation factor for young stars are not.}
    \label{fig:mockrecovery}
\end{figure*}

Fig.~\ref{fig:mockcorners} shows the corner plot distribution of 1000 samples from the posterior for the mock dataset presented in Section \ref{sec:methods:mocks}. The true parameters are marked by orange dots and lines. Fig.~\ref{fig:mockrecovery} shows the distributions of median posterior minus true population parameter values for 500 mock simulations in a new test dataset.

\clearpage
\section{Corner plot for model fit to data}
\label{app:B}
\begin{figure*}[h]
    \centering
    \includegraphics[width=\linewidth,trim={3cm 2cm 2cm 3cm},clip]{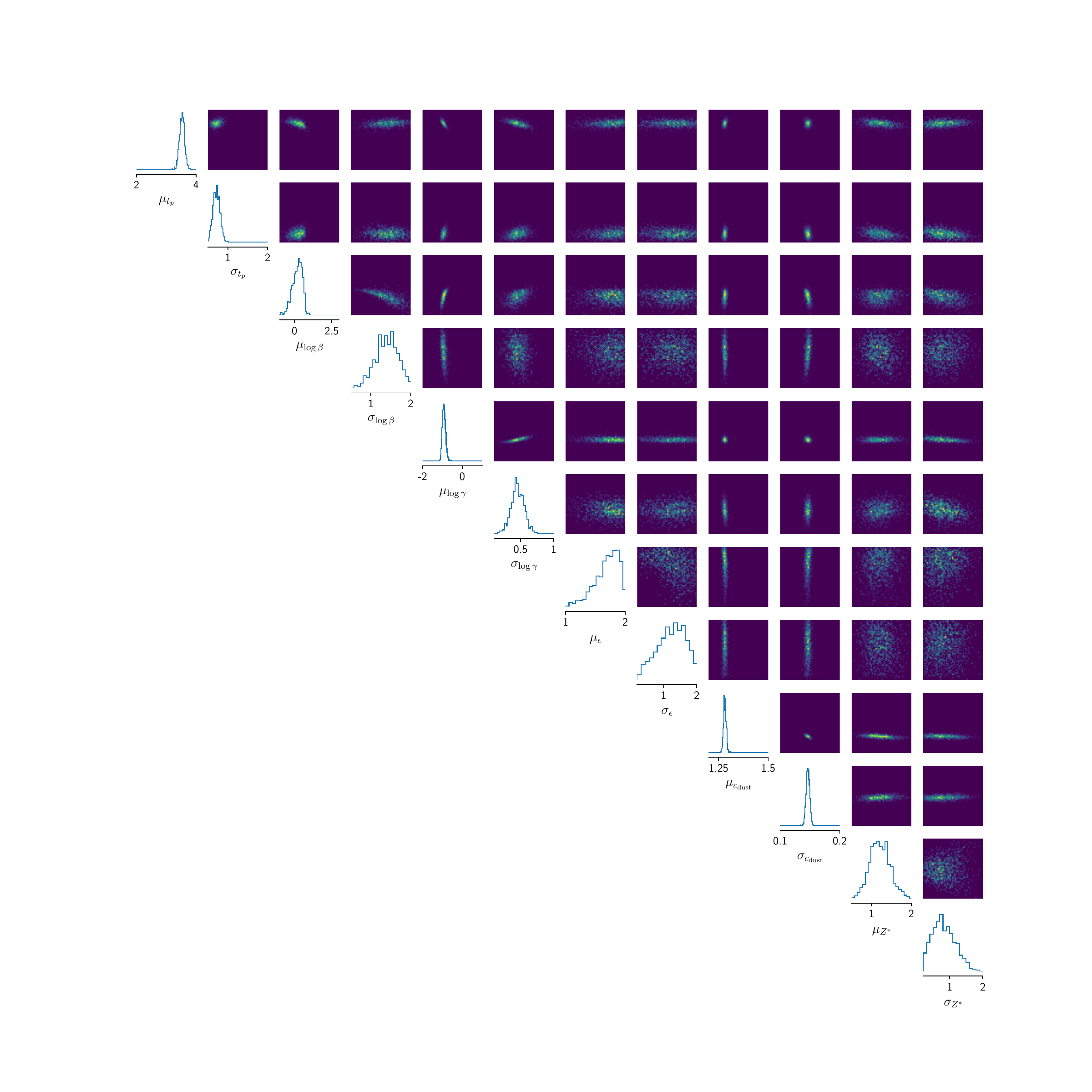}
    \caption{The joint posterior distributions of the population parameters fit to the UDS galaxies, depicted by 1000 samples from the posterior. The parameters are described in Tables \ref{tab:theta} and the median and percentiles of these distributions are given in Table \ref{tab:popparams}.}
    \label{fig:datacorners}
\end{figure*}

Fig.~\ref{fig:datacorners} shows the corner plot distribution of each of the population parameters for 1000 samples from the posterior of the UDS dataset analysed in Section \ref{sec:results}. 

\vspace{1cm}
\vfill

\end{document}